\documentclass[longauth]{aa}

\usepackage{graphicx}

\usepackage{amsmath}
\usepackage{amssymb}
\usepackage{dsfont}

\usepackage{natbib}

\usepackage{lipsum}
\usepackage[draft]{todonotes}

\newcommand{\refstar}{\mathrm{star}}
\newcommand{\onstar}{\mathrm{onstar}}
\newcommand{\planet}{\mathrm{planet}}
\newcommand{\refplanet}{\mathrm{planet}} 
\newcommand{\onplanet}{\mathrm{onplanet}}

\newcommand{\RA}{\alpha}
\newcommand{\DEC}{\delta}

\newcommand{\mat}[1]{{\bf#1}}
\newcommand{\cmat}[1]{\complex{\mat{#1}}}

\newcommand{\myunderline}[2]{\mkern#2mu\underline{\mkern-#2mu#1\mkern-#2mu}\mkern#2mu}
\newcommand{\complex}[1]{\myunderline{#1}{3}}
\newcommand{\conj}[1]{#1^{*}}

\newcommand{\cmathbb}[1]{\complex{\mathbb{#1}}}

\newcommand{\transp}[1]{#1^{T}}
\newcommand{\adj}[1]{{#1^\dagger}}

\newcommand{\textcommand}[1]{\mkern2mu\mathrm{#1}\mkern-2mu}
\newcommand{\real}[1]{\textcommand{Re}\left(#1\right)}
\newcommand{\imag}[1]{\textcommand{Im}\left(#1\right)}

\newcommand{\cov}[1]{\textcommand{cov}\left(#1\right)}
\newcommand{\pcov}[1]{\textcommand{pcov}\left(#1\right)}

\begin{document}

\title{Peering into the formation history of beta Pictoris b with VLTI/GRAVITY long-baseline interferometry}
\titlerunning{Peering into the formation history of beta Pictoris b}

\author{GRAVITY Collaboration\thanks{ Corresponding author \email{mcn35@cam.ac.uk}.}:
M. Nowak \inst{\ref{lesia}, \ref{ioa}}        
\and S.~Lacour\inst{\ref{lesia}, \ref{eso}, \ref{mpe}}
\and P.~Mollière   \inst{\ref{leiden},\ref{cologne}}    
\and J.~Wang \inst{\ref{caltech},}\thanks{51 Pegasi b Fellow.} 
\and B.~Charnay\inst{\ref{lesia}}
\and E.F.~van~Dishoeck   \inst{\ref{mpe},\ref{leiden}}
\and R.~Abuter\inst{\ref{eso}}
\and A.~Amorim\inst{\ref{lisboa},\ref{centra}}
\and J.P.~Berger\inst{\ref{grenoble}}
\and H.~Beust  \inst{\ref{grenoble}}
\and M.~Bonnefoy   \inst{\ref{grenoble}}
\and H.~Bonnet\inst{\ref{eso}}
\and W.~Brandner\inst{\ref{mpia}}
\and A.~Buron\inst{\ref{mpe}}
\and F.~Cantalloube\inst{\ref{mpia}}
\and C.~Collin\inst{\ref{lesia}}
\and F.~Chapron\inst{\ref{lesia}}
\and Y.~Cl\'{e}net\inst{\ref{lesia}}
\and V.~Coud\'e~du~Foresto\inst{\ref{lesia}}
\and P.T.~de~Zeeuw\inst{\ref{mpe},\ref{leiden}}
\and R.~Dembet\inst{\ref{lesia}}
\and J.~Dexter\inst{\ref{mpe}}
\and G.~Duvert\inst{\ref{grenoble}}
\and A.~Eckart\inst{\ref{cologne},\ref{mpifr}}
\and F.~Eisenhauer \inst{\ref{mpe}}
\and N.M.~Förster~Schreiber\inst{\ref{mpe}}
\and P.~Fédou\inst{\ref{lesia}}
\and R.~Garcia~Lopez\inst{\ref{dublin},\ref{mpia}}
\and F.~Gao\inst{\ref{mpe}}
\and E.~Gendron\inst{\ref{lesia}}
\and R.~Genzel\inst{\ref{mpe},\ref{berkeley}}
\and S.~Gillessen\inst{\ref{mpe}}
\and F.~Haußmann\inst{\ref{mpe}}
\and T.~Henning\inst{\ref{mpia}}
\and S.~Hippler\inst{\ref{mpia}}
\and Z.~Hubert\inst{\ref{lesia}}
\and L.~Jocou\inst{\ref{grenoble}}
\and P.~Kervella\inst{\ref{lesia}}
\and A.-M.~Lagrange   \inst{\ref{grenoble}}
\and V.~Lapeyr\`ere\inst{\ref{lesia}}
\and J.-B.~Le~Bouquin\inst{\ref{grenoble}}
\and P.~L\'ena\inst{\ref{lesia}}
\and A.-L.~Maire\inst{\ref{star},\ref{mpia}}
\and T.~Ott\inst{\ref{mpe}}
\and T.~Paumard\inst{\ref{lesia}}
\and C.~Paladini\inst{\ref{eso}}
\and K.~Perraut\inst{\ref{grenoble}}
\and G.~Perrin\inst{\ref{lesia}}
\and L.~Pueyo  \inst{\ref{stsci}} 
\and O.~Pfuhl  \inst{\ref{mpe},\ref{eso}}      
\and S.~Rabien\inst{\ref{mpe}}
\and C.~Rau\inst{\ref{mpe}}
\and G.~Rodr\'iguez-Coira\inst{\ref{lesia}}
\and G.~Rousset\inst{\ref{lesia}}
\and S.~Scheithauer\inst{\ref{mpia}}
\and J.~Shangguan\inst{\ref{mpe}}
\and O.~Straub\inst{\ref{lesia},\ref{mpe}}
\and C.~Straubmeier\inst{\ref{cologne}}
\and E.~Sturm\inst{\ref{mpe}}
\and L.J.~Tacconi\inst{\ref{mpe}}
\and F.~Vincent\inst{\ref{lesia}}
\and F.~Widmann\inst{\ref{mpe}}
\and E.~Wieprecht\inst{\ref{mpe}}
\and E.~Wiezorrek\inst{\ref{mpe}}
\and J.~Woillez\inst{\ref{eso}}
\and S.~Yazici\inst{\ref{mpe},\ref{cologne}}
\and D.~Ziegler\inst{\ref{lesia}}
}

\institute{
LESIA, Observatoire de Paris, Universit\'e PSL, CNRS, Sorbonne Universit\'e, Univ. Paris Diderot,
Sorbonne Paris Cit\'e, 5 place Jules Janssen, 92195 Meudon, France
\label{lesia}
\and European Southern Observatory, Karl-Schwarzschild-Straße 2, 85748 Garching, Germany
\label{eso}
\and Max Planck Institute for extraterrestrial Physics, Giessenbachstraße~1, 85748 Garching, Germany
\label{mpe}
\and Max Planck Institute for Astronomy, K\"onigstuhl 17, 69117 Heidelberg, Germany
\label{mpia}
\and Max Planck Institute for Radio Astronomy, Auf dem H\"ugel 69, 53121 Bonn, Germany
\label{mpifr}
\and $1^{\rm st}$ Institute of Physics, University of Cologne, Z\"ulpicher Straße 77, 50937 Cologne, Germany
\label{cologne}
\and Univ. Grenoble Alpes, CNRS, IPAG, 38000 Grenoble, France
\label{grenoble}
\and Universidade de Lisboa - Faculdade de Ci\^encias, Campo Grande, 1749-016 Lisboa, Portugal
\label{lisboa}
\and Sterrewacht Leiden, Leiden University, Postbus 9513, 2300 RA Leiden, The Netherlands
\label{leiden}
\and Departments of Physics and Astronomy, Le Conte Hall, University of California, Berkeley, CA 94720, USA
\label{berkeley}
\and CENTRA - Centro de Astrof\'{\i}sica e Gravita\c c\~ao, IST, Universidade de Lisboa, 1049-001 Lisboa, Portugal
\label{centra}
\and
Space Telescope Science Institute, Baltimore, MD 21218, USA
\label{stsci}
\and
STAR Institute, Universit\'e de Li\`ege, All\'ee du Six Ao\^ut 19c, B-4000 Li\`ege, Belgium
\label{star}
\and Department of Astronomy, California Institute of Technology, Pasadena, CA 91125, USA
\label{caltech}
\and Institute of Astronomy, University of Cambridge, Cambridge CB3 0HA, United Kingdom
\label{ioa}
}

\date{\today}
  \abstract
      {Beta Pictoris is arguably one of the most studied stellar systems outside of our own. Some 30 years of observations have revealed a highly-structured circumstellar disk, with rings, belts, and a giant planet:
        $\beta$ Pictoris b. However very little is known about how this system came into being.} 
      {Our objective is to estimate the C/O ratio in the atmosphere of $\beta$ Pictoris b and obtain an estimate of the dynamical mass of the planet, as well as to refine its orbital parameters using high-precision astrometry.}
      {We used the GRAVITY instrument with the four 8.2 m telescopes of the Very Large Telescope Interferometer to obtain K-band spectro-interferometric data on $\beta$ Pic b. We extracted a medium resolution
        (R=500) K-band spectrum of the planet and a high-precision astrometric position. We estimated the planetary C/O ratio using two different approaches (forward modeling and free retrieval) from two different codes (ExoREM and petitRADTRANS, respectively). Finally, we used a simplified model of two formation scenarios (gravitational collapse and core-accretion) to determine which can best explain the measured C/O ratio.}
      {Our new astrometry disfavors a circular orbit for $\beta$ Pic b ($e=0.15^{+0.05}_{-0.04}$). Combined with previous results and with Hipparcos/GAIA measurements, this astrometry points to a planet mass of $M = 12.7\pm{}2.2\,M_\mathrm{Jup}$. This value is compatible with the mass derived with the free-retrieval code petitRADTRANS using spectral data only. The forward modeling and free-retrieval approches yield very similar results regarding the atmosphere of $\beta$ Pic b. In particular, the C/O ratios derived with the two codes are identical ($0.43\pm{}0.05$ vs $0.43^{+0.04}_{-0.03}$). We argue that if the stellar C/O in $\beta$ Pic is Solar, then this combination of a very high mass and a low C/O ratio for the planet suggests a formation through core-accretion, with strong planetesimal enrichment. 
      }
        {}
        
   \keywords{Exoplanets -- Instrumentation: interferometers
   -- Techniques: high angular resolution               }

   \maketitle

\section{Introduction}



The ever-increasing number of exoplanet detections (over 4000, at the time of this writing\footnote{http://exoplanets.eu}) proves that our instrumental capabilities are getting better and better at discovering these other worlds. But even though exoplanets are now routinely being observed, determining their physical properties (temperature, mass, composition), let alone the history of their formation, remains extremely challenging. And yet, these measurements are key to understanding the details of planetary formation processes.

Among all measurable quantities, element abundance ratios are emerging as some of the most promising for understanding planetary formation. The question of the supersolar abundances of heavy elements in the atmosphere of Jupiter is probably what motivated the first attempts to link abundance ratios to planetary formation, and several studies have been carried out to understand how planetesimal accretion can lead to heavy element enrichment \citep{Helled2006, Helled2009, Owen1999, Alibert2005}. On the exoplanet front, the work of \cite{Oberg2011} was the first general attempt to show that element ratios in an exoplanet atmosphere can be an imprint of its formation history. This idea has since been investigated further by several authors \citep[e.g.,][]{Ali-Dib2014,Thiabaud2014,Helling2014,Marboeuf2014b,Marboeuf2014,Madhusudhan2014,Madhusudhan2017,Mordasini2016,Oberg2016,Cridland2016,Eistrup2016,Eistrup2018}. While \cite{Oberg2011} highlighted how gas disk abundances can influence the atmospheric composition, the importance of icy planetesimals for the atmospheric enrichment is stressed in \citet{Mordasini2016}, where exoplanet spectra are derived from modeling full formation in the core-accretion paradigm.

Measuring the element ratios is not easy, and requires high-quality data. \cite{Madhusudhan2011} used a free retrieval method on a set of Spitzer and ground-based photometric data in 7 different bands to obtain the first exoplanetary C/O ratio on the hot Jupiter WASP-12b. But the value of $\mathrm{C/O}>1$ they obtained has since been ruled out by \cite{Kreidberg2016}, showing the difficulty of obtaining reliable abundance ratios. \cite{Konopacky2013} used a different approach in their study of HR~8799~c. They obtained K-band spectroscopic observations of the planet with the spectrograph OSIRIS on the Keck II telescope, and were able to extract an estimate of the C/O ratio using model grid fitting. They found a value of $\mathrm{C/O}=0.65\pm{}0.15$. Looking at the same planetary system, \cite{Lavie2017} estimated the C/O ratio for four planets (HR~8799~b, c, d, and e), using a retrieval analysis method. In their analysis, they notably emphasized the importance of high-quality K-band spectroscopic data, which they found to be critical for a reliable measurement of the C/O and C/H ratios.

With the recent direct detection of the giant planet HR~8799~e with the GRAVITY instrument \citep{GravityLacour2019} on the Very Large Telescope Interferometer (VLTI), optical interferometry has become a new arrow in the quiver of exoplanet observers. By taking advantage of the angular-resolution offered by 100+ meter baselines, optical interferometers can separate a dim exoplanet from the overwhelming residual starlight, leading to accurate measurements of the astrometric position \citep[up to 10\,$\mu$as,][]{GravityCollaboration2018}, and high signal-to-noise spectroscopic data with absolute calibration of the continuum.

In this paper, we present observations of the giant planet $\beta$ Pic~b obtained with GRAVITY and we investigate the possibility of using this K-band spectro-interferometric data to determine the C/O ratio of the planet. The observations are presented in Section~\ref{sec:observations}, together with a brief summary of the data reduction (a complete explanation is given in Appendix~\ref{sec:reduction_appendix}). Section~\ref{sec:orbit_mass} focuses on the orbit and mass of $\beta$ Pic b. We show in this section how the new GRAVITY astrometric data impacts the best orbital estimate currently available and we provide a new estimate of the dynamical mass of the planet. Section~\ref{sec:atmosphere} is devoted to the measurement of atmospheric properties and, in particular, to the determination of the C/O ratio, using two different approaches: forward modeling and free retrieval. In Section~\ref{sec:co_ratio}, we discuss the C/O ratio obtained in the case of a formation of $\beta$ Pic b through gravitational accretion and then through core-accretion. Our general conclusions can be found in Section~\ref{sec:conclusion}.

\section{Observations and data reduction}
\label{sec:observations}
\subsection{Observations}

Observations of $\beta$ Pictoris b were obtained on September, 22, 2018, using the GRAVITY instrument \citep{GravityFirstLight}, with the four 8~m Unit Telescopes (UTs) of the VLT. The instrument was set up in its medium resolution mode ($\mathrm{R}=500$), and observations were conducted in on-axis/dual-field mode.

The observing strategy was similar to the one described in \cite{GravityLacour2019}:  the fringe-tracker \citep{Lacour2019} was using the flux from the central star during the observing sequence, while the position of the science fiber was changed at each exposure, alternating between the central star and the position of the planet. Since the planet was not visible on the acquisition camera, the position used to center the fiber during the planet exposures was a theoretical position, based on predictions from previous monitoring \citep{Wang2016, Lagrange2018}.

A total of 16 exposures (resp. 17) were acquired on the star (resp. the planet). Each star exposure was made of 50 individual 0.3~s integrations. For the planet, which is $\sim{}10~\mathrm{mag}$ fainter than the star, the integration time was initially set to 30~s, with 10 integrations per exposure, and reduced to 10~s with 30 integrations at mid-course, since the observing conditions were good (seeing $< 0.8''$). {The complete dataset contains 1.4~hr of integration on the planet (and 0.35~hr of associated background exposures), and 4~min~30~s of integration on the central star (plus 1~min~15~s of sky background). The observing log is given in Table~\ref{tab:log}.}



\begin{table*}
  \begin{center}
    \begin{tabular}{l c c c c c c c c c}
      \hline
      \hline
      Target & Start Time & End Time & EXP & DIT & NDIT & Seeing & $\tau_0$ & Airmass & Parallactic angle \\
      & (UTC) & (UTC) & & (s) & & ('') & (ms) & & (deg) \\
      \hline      
      $\beta$ Pictoris b & 07:37:40 & 08:31:40 & 7 & 30.0 & 10 & 0.4/0.9 & 4.7\,/\,10.4 & 1.33\,/\,1.21 & -66.4\,/\,-50.1\\
	SKY & 07:50:30 & 08:24:56 & 2 & 30.0 & 10 &  0.4\,/\,0.9 & 4.7\,/\,10.4 & 1.33\,/\,1.21 & N/A\\
       $\beta$ Pictoris b  & 08:38:31 & 09:51:49 & 10 & 10.0 & 30 & 0.6\,/\,1.2 & 5.9\,/\,8.4 & 1.20\,/\,1.12 & -47.7\,/\,-16.6\\
	SKY & 08:50:41& 09:25:03 & 2 & 10.0 & 30 &  0.6\,/\,1.2 & 5.9\,/\,8.4 & 1.20\,/\,1.12 & N/A\\
	 $\beta$ Pictoris A  & 07:43:55 & 09:58:31& 18 & 0.3 & 50 & 0.4\,/\,1.2 & 4.7\,/\,10.4 & 1.31\,/\,1.12 & -64.7\,/\,-13.2\\
SKY & 07:57:14 & 09:59:20 & 5& 0.3 & 50 & 0.4\,/\,1.2 & 4.7\,/\,10.4 & 1.31\,/\,1.12 & N/A\\
      \hline
    \end{tabular}
    \caption[Observing log for the $\beta$ Pic DDT program]{Observing log for the DDT $\beta$ Pic b program, carried out on September, 22, 2018.}
    \label{tab:log}
  \end{center}
\end{table*}

\subsection{General data reduction}

During planet exposures, the science fiber at each telescope is kept at an offset position with respect to the star, reducing significantly the star to fiber coupling ratio. But even though most of the stellar flux is rejected, speckle noise can still couple to the science fiber and dominate the exposures, hence the need for careful data reduction to disentangle the planet signal from the remaining coherent stellar flux.

The general data reduction method used to reduce the VLTI/GRAVITY observations of $\beta$ Pic b is presented in details in Appendix~\ref{sec:reduction_appendix}. It can be divided into different parts: pipeline reduction (common to all GRAVITY observations), astrometric extraction, and spectrum extraction. These steps are described in Appendix~\ref{sec:app_pipeline}, \ref{sec:app_astro}, and \ref{sec:app_spectrum}. The end products are an astrometric position for the planet with respect to the star ($\Delta\alpha, \Delta\delta$), and a planet-to-star contrast spectrum $C(\lambda)=S_\mathrm{P}(\lambda)/S_\star(\lambda)$ which is the ratio between the spectra of the planet and of the star.

\subsection{K-band spectrum}

\begin{figure*}
  \includegraphics[width=\linewidth, clip = True, trim = 0cm 0.5cm 0cm 0cm]{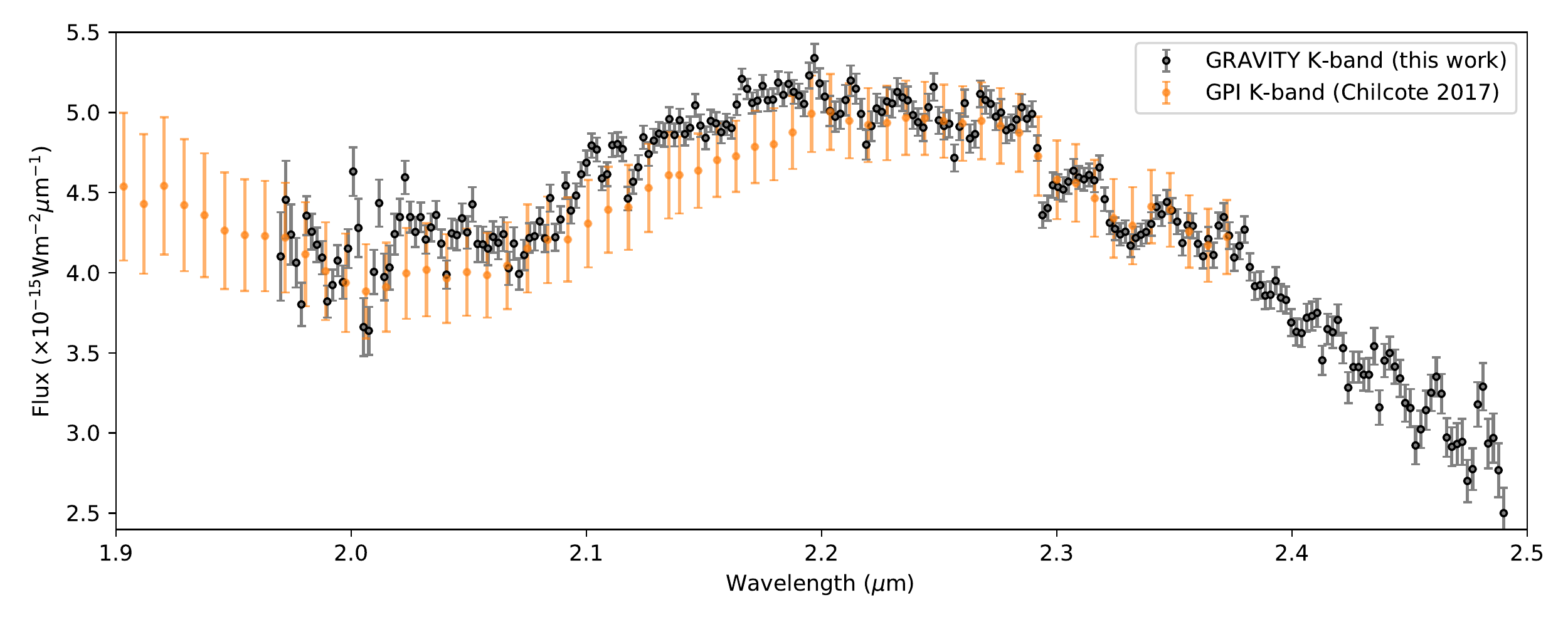}
  \caption{Calibrated K-band spectrum of $\beta$ Pictoris b, at $\mathrm{R}=500$, extracted from the VLTI/GRAVITY observations (gray points). For comparison, the K-band part of the GPI spectrum from \cite{Chilcote2017} ($\mathrm{R}\simeq{}70$) is also overplotted (orange points). The error bars plotted for the GRAVITY spectrum only represent the diagonal part of the full covariance matrix.}
  \label{fig:spectrum}
\end{figure*}

The contrast spectrum of $\beta$ Pic b was converted to an absolute spectrum of the planet using a model of the stellar spectrum: $S_\mathrm{P}(\lambda) = C(\lambda)\times{}S_\star(\lambda)$. We used a BT-NextGen model \citep{Hauschildt1999}, with a temperature of 8000 K, a surface gravity of $\log(g/g_0) = 4$, and a Solar metallicity, as close as possible to the measured value for this star \citep{Lanz1995, Gray2006}.  We scaled this synthetic spectrum to an ESO K-band magnitude of 3.495, taking into account the correct filter \citep{VanDerBliek1996}. {This strategy, based on the extraction of a contrast spectrum and the use of a model  for the star, helps to reduce the impact of Earth's atmosphere on the final planet spectrum}. The result is given in Figure~\ref{fig:spectrum}.

\subsection{Astrometry}

Using the data reduction method described in Appendix~\ref{sec:app_astro}, we found a mean relative planet to star astrometry on all the exposure files of:
\begin{equation}
  \begin{cases}
    \Delta\mathrm{RA} = 68.48~\mathrm{mas} \\
    \Delta\mathrm{DEC} = 126.31~\mathrm{mas}
  \end{cases}
\end{equation}
The $1\,\sigma$ confidence interval is given by the covariance matrix of all the 17 exposure files:
\begin{align*}
  \mathrm{Covar}\left(\Delta{}\mathrm{RA}, \Delta\mathrm{DEC}\right) & = \begin{bmatrix}0.0027 & -0.0035 \\ -0.0035 & 0.0045 \end{bmatrix}~\mathrm{mas}^2
\end{align*}
This GRAVITY measurement is shown in the inset plot of Figure \ref{fig:orbitsky}.

In its dual-field mode, GRAVITY is limited to observations of planets above the diffraction limit of a single telescope (to separate the planet from the central star), but the relative astrometry derived from these observations still fully benefits from the length of the telescope array.



\section{Orbit and dynamical mass}
\label{sec:orbit_mass}
\subsection{Orbital parameters \label{sec:orbit_regular}}

\begin{figure}
  \includegraphics{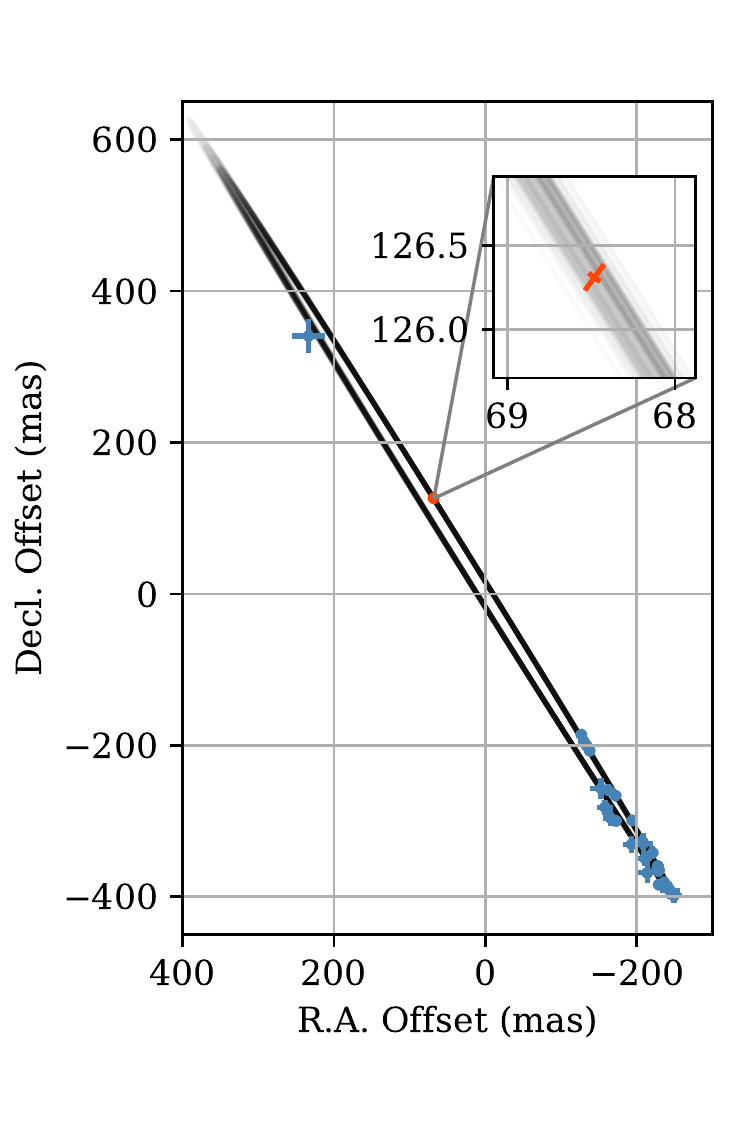}
  \caption{Visual orbit of $\beta$ Pic b. Plotted in black are possible orbits randomly drawn from the posterior using only relative astrometry (Section \ref{sec:orbit_regular}). Previous astrometric measurements used in the orbit fit are in blue. The GRAVITY measurement from this work is in red, with an inset plot that is zoomed in by a factor of $\sim$2000 to display the uncertainties on this measurement. }
  \label{fig:orbitsky}
\end{figure}

We fit a Keplerian orbit to the visual astrometry of the planet to characterize its dynamics. As our new GRAVITY point is more than an order of magnitude more precise than any other published astrometric point on the northeastern half of its orbit \citep[c.f.,][]{Lagrange2019}, we expected a better constraint on the eccentricity of the planet's orbit. We used the published astrometry from \citet{Chauvin2012}, \citet{Nielsen2014}, and \citet{Wang2016} in this analysis. The orbit was fit using the open-source Python orbit fitting package \texttt{orbitize!} \citep{Blunt2019}. We included a custom likelihood to fit the GRAVITY measurement along the two principal axes of the error ellipse. We fit for the same eight parameters as \cite{Wang2016}: semi-major axis ($a$), eccentricity ($e$), inclination ($i$), argument of periastron ($\omega$), position angle of the ascending node ($\Omega$), the first periastron passage after MJD = 55,000 in units of fractional orbital period ($\tau$), system parallax, and total system mass ($M_{tot}$). We generally used relatively unconstrained priors for most of the orbital parameters (see Table \ref{tab:orbitparams}). For $\Omega$, we constrained it to between $\pi/10$ and $\pi/2$ to account for the fact that \citet{Snellen2014} detected the RV signal of the planet. However, we chose not to explicitly include the RV in the fit as there could be systematics in the reported uncertainties. For the parallax, we used a normal distribution to represent the parallax of $51.44 \pm 0.12$~mas measured by Hipparcos \citep{VanLeeuwen2007}. We sampled the posterior using the parallel-temperature affine-invariant sampler in \texttt{ptemcee} \citep{Foreman-Mackey2013,Vousden2016} with 20 temperatures, 1000 walkers per temperature. We discarded the first 15,000 steps to allow the walkers to converge. We assessed convergence using the autocorrelation time and by visual inspection of the samples. We then ran each walker for 5000 steps, keeping only every tenth sample to mitigate correlations in the samples produced by any given walker. 
 
\begin{table*}
\begin{center}
\begin{tabular}{c|c|cc|cc|cc}
\hline
\hline
Orbital Element  & Prior & \multicolumn{2}{c}{\shortstack[c]{Only Relative\\Astrometry}} & \multicolumn{2}{c}{\shortstack[c]{Hipparcos IAD\\and Gaia DR2}} & \multicolumn{2}{c}{\shortstack[c]{Brandt (2018) HGCA\\and Stellar RVs}}  \\
& & 68\% CI & Best Fit & 68\% CI & Best Fit & 68\% CI & Best Fit \\
\hline
 $a$ (au)               & LogUniform(1, 100) & $10.6 \pm 0.5$ & 10.9 & $11.0^{+0.3}_{-0.4}$ & 11.2         & $10.0^{+0.6}_{-0.5}$ & 10.2 \\
 $e$                    & Uniform(0, 1) & $0.15^{+0.04}_{-0.05}$ & 0.18 & $0.19^{+0.02}_{-0.03}$ & 0.21        & $0.11 \pm 0.05$ & 0.13 \\
 $i$ (\degr)            & $\sin(i)$ & $89.04 \pm 0.03$ &89.05 & $89.06 \pm 0.02$ & 89.07                   & $88.99^{+0.03}_{-0.04}$ & 89.00 \\
 $\omega$ (\degr)       & Uniform(0, 2$\pi$) & $196^{+3}_{-4}$ & 196 & $197 \pm 2$ & 197                   & $202 \pm 5 $ & 202 \\
 $\Omega$ (\degr)       & Uniform($\pi/10$, $\pi/2$) & $31.88 \pm 0.05$ & 31.90 & $31.90 \pm 0.05$ & 31.92 & $31.87 \pm 0.05$ & 31.88 \\
 $\tau$                 & Uniform(0, 1) & $0.159 \pm 0.009$ & 0.157 & $0.155^{+0.008}_{-0.006}$ & 0.152    & $0.185^{+0.019}_{-0.016}$ & 0.185 \\
 Parallax (mas)         & $\mathcal{N}$(51.44, 0.12) & $51.44 \pm 0.12$ & 51.45 & $51.44 \pm 0.12$ & 51.49 & $51.44 \pm 0.12$ & 51.47 \\
 $M_{tot}$ ($M_\odot$)  & Uniform(1.4, 2) & $1.82 \pm 0.03$ & 1.82 & $1.83 \pm 0.03$ & 1.81                & $1.79 \pm 0.03$ & 1.78 \\ 
 $M_b$ ($M_{\rm{Jup}}$) & Uniform(1, 100) & - & - & $12.7 \pm 2.2$ & 13.8                                  & $14.2^{+3.7}_{-3.9}$ & 15.1 \\ 
 \hline
\end{tabular}
\caption{Orbital Parameters of $\beta$ Pic b. Listed are fits using just astrometry of the planet (Section \ref{sec:orbit_regular}) and also including measurements of the stellar orbit for dynamical mass estimates of the planet (Section \ref{sec:orbit_withmass}). For each fit, the first column lists the 68\% credible interval centered about the median. The second column lists the fit with the maximum posterior probability. We note that this the best fit orbit is generally not the best estimate of the true orbit. However, it is useful as a valid representative orbit, whereas using the median of all of the orbital parameters often is not a valid orbit due to complex covariances. }
\label{tab:orbitparams}
\end{center}
\end{table*}

Our constraints on the orbit of $\beta$ Pic b using just astrometry of the planet are collected in Table \ref{tab:orbitparams} and plotted in Figure \ref{fig:orbitsky}. We find that $< 2\%$ of allowed orbits have $e < 0.05$ and $< 0.5\%$ of orbits have $e < 0.03$, although there are still some allowed circular orbits. \citet{Dupuy2019} also proposed an $e \approx 0.25$ when including astrometric and radial velocity data on the system. To statistically assess whether eccentric orbits are preferred, we refit the orbit fixing $e = 0$ and $\omega = 0$ resulting in a fit with two less parameters. Similar to \citep{Wang2018} in assessing the coplanarity of the HR 8799 planets, we compared the Bayesian Information Criterion (BIC) of the fit that allowed eccentric orbits with the fit that fixed the orbit to be circular, and found that the BIC disfavors the circular orbit by 9.9. The reduction in model parameters for a purely circular orbit does not compensate for an increase in fitting residuals, so we disfavor circular orbits for a single planet model. However, additional confusion on this measurement could be due to a second planet in the system \citep{Lagrange2019b}. The second planet $\beta$ Pic c would induce epicycles in the apparent orbit of $\beta$ Pic b around the star due to the gravitational influence of the second planet on the orbit of the host star. Using parameters for $\beta$ Pic c from \citet{Lagrange2019b}, the magnitude of these epicycles are several hundred $\mu$as, so well detectable by GRAVITY, but hidden beneath the uncertainty of previous astrometry. Thus, they would also bias this single GRAVITY measurement, and continued astrometric monitoring is required to separate out the signal of the separate planet from a possibly eccentric orbit of $\beta$ Pic b. 

However, a moderate eccentricity would fit nicely in the dynamics of the system. An $e \approx 0.15$ is consistent with the picture of an eccentric $\beta$ Pic b launching small bodies towards the star, causing spectroscopic and transiting signatures of exo-comets in observations of the star \citep{Thebault2001, Zieba2019}. An interesting question is how such a massive planet acquired a significant eccentricity. The obvious conclusion would point to a second massive planet in the system, such as the radial velocity detected $\beta$ Pic c \citep{Lagrange2019b}. 
Otherwise, \citet{Dupuy2019} proposed that if the planet had formed further out and migrated inwards, resonant interactions with the circumstellar disk could pump up its eccentricity to the values we observe today. Characterizing the detailed structure of the circumstellar dust in the system as well as the chemical composition of $\beta$ Pic b could test this theory.  

Generally, the other orbital parameters of $\beta$ Pic b have already been sufficiently well constrained previously that out results agree with the conclusions drawn in previous works \citep{MillarBlanchaer2015,Wang2016,Lagrange2019,Dupuy2019}. We still find that the planet did not transit the star in 2017, and that the Hill sphere of the planet did transit. Assuming a planet mass of $12.9 \pm 0.2$ $M_{\rm{Jup}}$, we find a Hill sphere ingress at MJD $57852 \pm 2$ (2017 April 8) and a Hill sphere egress at MJD $58163 \pm 2$ (2018 February 13). The closest approach, which does not require an assumption on the planet's mass, is at MJD $58008 \pm 1$ (2017 September 11), with the planet passing $8.57 \pm 0.13$~mas from the star ($0.166 \pm 0.003$ au in projection). The precise astrometry of the GRAVITY epoch post conjunction has significantly improved the transit ephemeris from \citet{Wang2016}.

\subsection{Dynamical mass determination \label{sec:orbit_withmass}}
A significant astrometric acceleration for the star $\beta$ Pic was detected when comparing its average velocity over the course of the Hipparcos mission and the average velocity inferred by the change in position of the star between the Hipparcos and Gaia missions \citep{Snellen2018, Kervela2019}. Assuming this acceleration is due entirely to $\beta$ Pic b, \citet{Snellen2018} and \cite{Dupuy2019} used it in conjunction with the visual orbit to measure a dynamical mass for the planet. \citet{Snellen2018} used the Hipparcos intermediate astrometric data \citep[IAD;][]{VanLeeuwen2007} and Gaia DR2 position \citep{Gaia2018} to fit the position, proper motion, and orbital motion of the host star to derive the mass of the planet. \cite{Dupuy2019} used the re-calibrated Hipparcos and Gaia proper motions from the Hipparcos-Gaia Catalog of Accelerations \citep[HGCA;][]{Brandt2018} and the stellar radial velocities from \citet{Lagrange2012} to derive the mass of the planet. Being agnostic to which method is more accurate, we repeated both analyses here, now with the new GRAVITY epoch providing strong constraints on $a$ and $e$, which are otherwise degenerate with the mass of $\beta$ Pic b, $M_b$. To repeat the \citet{Snellen2018} orbit fit, we include five additional parameters in the fit: the position and proper motion of the star in RA and DEC, as well as the mass of the planet. We also switch the prior on parallax to a uniform prior between 50.24 and 52.64~mas, since the Hipparcos intermediate astrometric data now constrains this parallax. To repeat the \citet{Dupuy2019} analysis, we only fit for changes in the tangential velocity of the host star, so we do not need to fit for its actual position and proper motion. We only include a RV offset and RV jitter term for the stellar RV data. We modified the \texttt{orbitize!} custom likelihood function to include these measurements of the host star, and repeated the orbit fit. 

\begin{figure}
  \includegraphics[width = \linewidth]{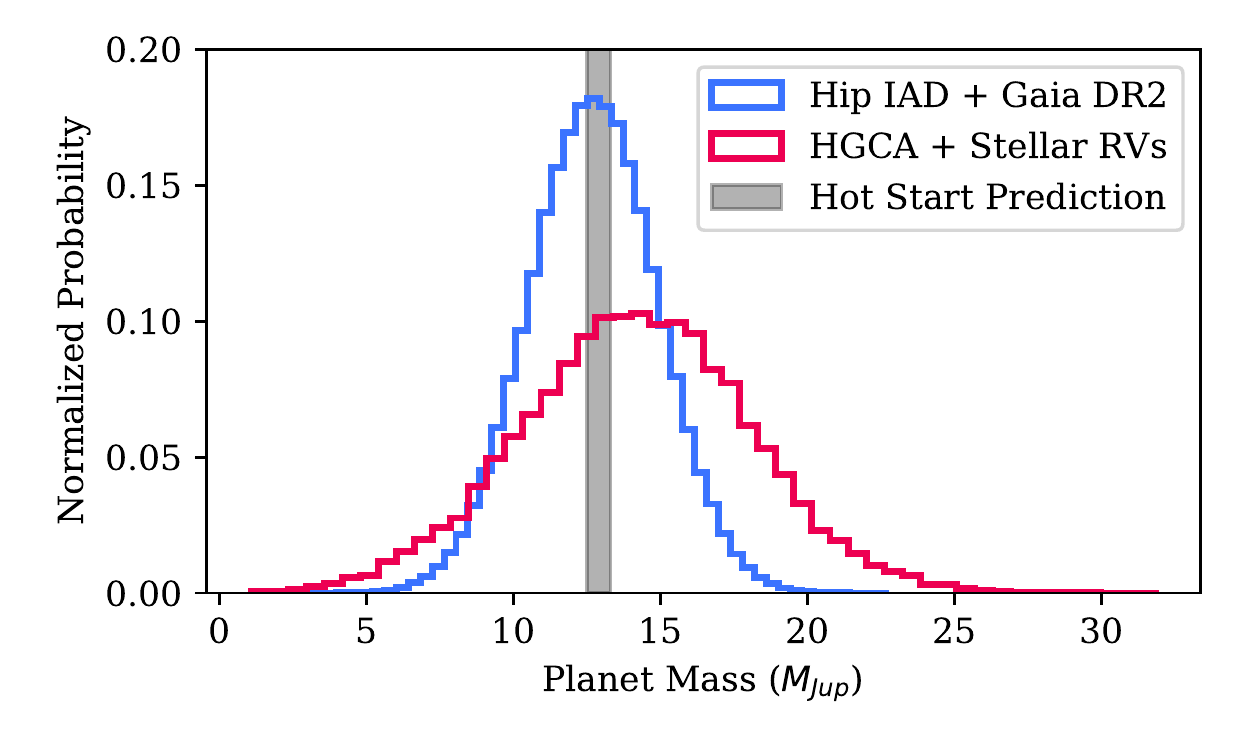}
  \caption{Dynamical mass estimates of $\beta$ Pic b using the two different methods described in Section \ref{sec:orbit_withmass}. The shaded grey region is the 2$\sigma$ uncertainty on the hot-start derived mass from \citet{Chilcote2017}.  }
  \label{fig:masshist}
\end{figure}

We list the orbital and mass constraints in Table \ref{tab:orbitparams}, marginalizing over astrometric parameters of the host star and stellar RV calibration numbers in the two fits. We also plot the posterior probabilties for the mass of $\beta$ Pic b in Figure \ref{fig:masshist}. In the fit using the Hipparcos IAD, the semi-major axis and eccentricity posteriors now favor slightly higher values by 1$\sigma$. We find a dynamical mass of $\beta$ Pic b of $12.7 \pm 2.2$~$M_{\rm{Jup}}$, which is consistent with the values from \citet{Snellen2018}. Conversely, using the recalibrated stellar astrometry from the \citet{Brandt2018} HGCA catalog and the stellar RVs, we find a slightly lower semi-major axis and eccentricity by 1$\sigma$ than the relative astrometry only fit. Despite these minor differences, all three fits considered in this work favor an eccentricity between 0.1-0.2. We do not find orbital solutions with $e > 0.25$ as has been suggested by \citet{Dupuy2019}. In the HGCA fit, we also find a weaker dynamical mass constraint for $\beta$ Pic b of $14.2^{+3.7}_{-3.9}$~$M_{\rm{Jup}}$, which is consistent with \citet{Dupuy2019}. The Hipparcos IAD method provides more stringent constraints on the planet mass, likely because it has smaller uncertainties. It is unclear whether this better constrain is unbiased, or if the uncertainties are underestimated due to calibration systematics or effects of other planets on the stellar astrometry. However, as seen in Figure \ref{fig:masshist}, both fits agree with each other, and both dynamical masses are consistent with hot-start derived masses of $12.7 \pm 0.3$~$M_{\rm{Jup}}$ from \citet{Morzinski2015} and $12.9 \pm 0.2$~$M_{\rm{Jup}}$ from \citet{Chilcote2017}. More accurate stellar astrometry or RVs are necessary to test hot-start evolutionary models more stringently, given that the model-dependent hot-start masses have an order of magnitude better precision than the dynamical masses.

\section{The atmosphere of beta Pic b}
\label{sec:atmosphere}

\subsection{Previous work}

Physical parameters of $\beta$ Pictoris b have been reported in a number of previous studies (see Table~15 of \cite{Morzinski2015}, Table~2 in \cite{Chilcote2017} for a summary of these results). The temperature of the planet has been estimated by several authors, using atmospheric or evolutionary model grid fitting, on photometric and/or spectroscopic data. The most extensive study to date was performed by \cite{Chilcote2017}, who obtained GPI spectroscopic data at $R\simeq{}50$ in Y, J, H, and K-band, as well as photometric points in different bands ranging from 1 to $5~\mu\mathrm{m}$. Using different atmospheric models (BT-Settl: \citeauthor{Allard2012}, \citeyear{Allard2012}; DRIFT-PHOENIX: \citeauthor{Woitke2003}, \citeyear{Woitke2003}, \citeauthor{Helling2006}, \citeyear{Helling2006}; AMES-DUSTY: \citeauthor{Chabrier2000}, \citeyear{Chabrier2000}; \citeauthor{Allard2001}, \citeyear{Allard2001}), they obtained values ranging from 1650~K to 1800~K for the temperature, and 3.0 to 4.5 for $\log(g/g_0)$. These values are similar to what is reported in \cite{Bonnefoy2013, Bonnefoy2014}, \cite{Chilcote2015}, and \cite{Morzinski2015}, with the same models.

The lower limit of the range of temperatures estimated comes from \cite{Baudino2015}. Using their Exo-REM model grid, and a set of photometric data only (the GPI spectrum was not available at the time), they derived a temperature of 1550~K, and a surface gravity of $\log(g/g_0) = 3.5$.

\subsection{ExoREM atmospheric grid fitting}
\label{subsect:exorem}

Using either the GRAVITY K-band spectrum only, or the GRAVITY K-band and GPI YJH bands spectra, we performed a grid model fitting using the newest ExoREM grid \citep{Charnay2018}.

We performed a $\chi^2$-based grid model fitting on the GRAVITY K-band only data, using the same ExoREM model grid as used to fit the GRAVITY HR\,8799\,e spectrum in \cite{GravityLacour2019}, ranging from 400 to 1800~K in temperature, with a step-size of 50~K, from  3.0 to 5.0 in $\log(g/g_0)$, with a step-size of 0.2, for a metalicity of $[\mathrm{Fe}/\mathrm{H}]=-0.5$, 0, and 0.5, and with a Solar C/O ratio. The best fit was obtained for a Solar metallicity, a temperature of 1750~K, and a $\log(g/g_0)$ of 3.30. However, this best fit also leads to a mass of $1.3\,M_\mathrm{Jup}$, more than $5\,\sigma$ away from our estimate given in Table~\ref{tab:orbitparams}. To force the result of the fit to be in agreement with our mass estimate, we added a mass prior in the $\chi^2$ calculation. We used a weight for the prior similar to the weight of the entire GRAVITY spectrum:
\begin{footnotesize}
\begin{equation}
    \chi^2 =  n_\lambda\frac{\left(m - 12.7\,M_\mathrm{Jup}\right)^2}{\left(2.2\,M_\mathrm{Jup}\right)^2} + \sum \frac{\left(F_\mathrm{data}(\lambda_k) - F_\mathrm{model}(\lambda_k)\right)^2}{\sigma_{F}(\lambda_k)^2}
  \label{eq:biased_chi2}
\end{equation}
\end{footnotesize}

\noindent{}in which $F_\mathrm{data}$ and $F_\mathrm{model}$ represent the flux from the data and from the model at the different wavelengths, $\sigma_F$ the error on the data, and $m$ the mass derived from the flux level.

With this new definition of the $\chi^2$, the same ExoREM grid led to a best fit at $T=1500~\mathrm{K}$, $\log(g/g_0) = 4.0$, for a Solar metallicity. The corresponding planet radius is $1.9\,R_\mathrm{Jup}$, and the mass is $14\,M_\mathrm{Jup}$, compatible with the estimate of Section~\ref{sec:orbit_mass}. However, we find that the fit itself was not very good, with a $\chi^2_\mathrm{red}{}$ value of 6.8. The CO region around $2.3~\mu\mathrm{m}$ was particularly poorly fitted.



\begin{figure*}
  \includegraphics[width = \linewidth]{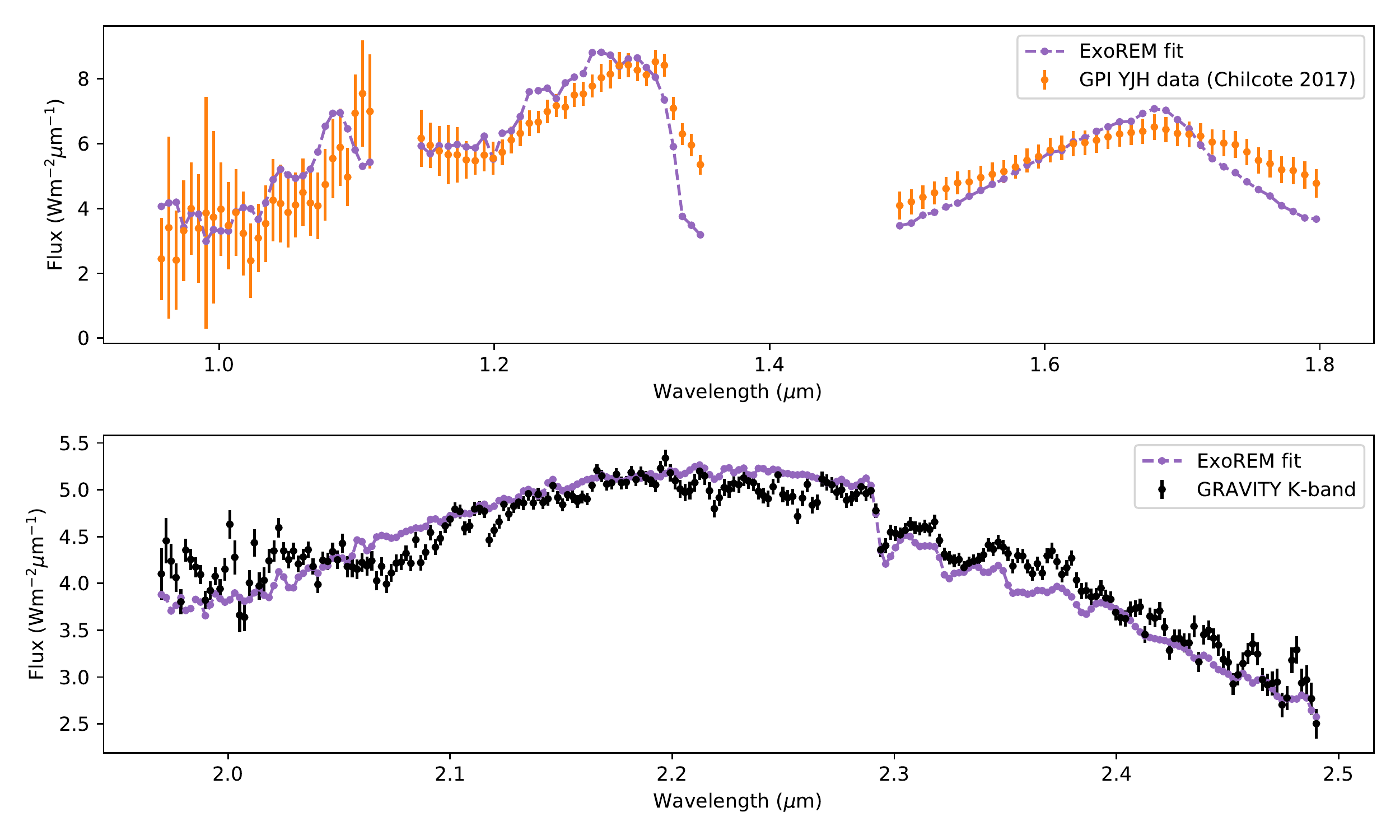}
  \caption{Best fit obtained with the ExoREM atmospheric model \citep{Charnay2018} using GPI Y, J, H + GRAVITY K bands, and a mass prior.}
  \label{fig:exorem_fit}
\end{figure*}

The fit was improved by generating a second ExoREM grid, which included the C/O ratio as an additional parameter. The grid was generated on the same range of temperature, surface gravity, and metallicity, for C/O values ranging from 0.3 to 0.8, with a step of 0.05.

Without the mass prior, the new grid yielded a best fit corresponding to a temperature of $1700\pm{}50~\mathrm{K}$, a surface gravity of $\log(g/g_0) = 3.5$, a metallicity of -0.5 (the lowest value available in our grid), and a C/O ratio $<0.3$. The resulting planet mass remained too low, at $2\,M_\mathrm{Jup}$. Adding the mass prior in the definition of the $\chi^2$, as in Equation~\ref{eq:biased_chi2} led to a temperature of $1550\pm{}20~\mathrm{K}$, a surface gravity $\log(g/g_0)$ of 4.0, a metallicity of $[\mathrm{Fe}/\mathrm{H}]=0.5$ (highest value from our grid), and a C/O ratio of $0.41\pm{}0.05$, for a planet mass of $11.5\,M_\mathrm{Jup}$, in very good agreement with the result of Section~\ref{sec:orbit_mass}. Contraining the fit to Solar metallicity resulted in a very low C/O ratio of 0.3, with similar temperature and surface gravity.

Including the GPI Y, J, and H band data from \cite{Chilcote2017} and allowing for a multiplicative scaling factor between GPI and GRAVITY resulted in a temperature of $T=1590\pm{}20~\mathrm{K}$, with a C/O of $0.43\pm{}0.05$, for a metallicity of $[\mathrm{Fe}/\mathrm{H}]$ = 0.5. {For reference, the typical multiplicative factors needed to scale the GPI spectra on the ExoREM grid were $\simeq{}0.85$ for the Y band, and $\simeq{}0.9$ for J and H bands}.

The results of these different fits are summarized in Table~\ref{tab:exorem_results}, and the best fit obtained using GRAVITY+GPI and a mass prior is shown in Figure~\ref{fig:exorem_fit}.

\subsection{Free retrieval with petitRADTRANS}
\label{sec:pRT_fit}

\subsubsection{Retrieval forward model}

In addition to fitting a model grid to the $\beta$~Pic~b observation, we carried out a free retrieval. To this end, the spectra were compared to the predictions of a spectral synthesis code, where the atmospheric structure was parametrized. In such an approach more weight is given on atmospheric conditions as constrained by the data, while principles such as radiative-convective equilibrium do not have to be strictly fulfilled. This approach was motivated by the work of \citet{Line2015,Line2017,Zalesky2019} for clear, and \citet{Burningham2017} for cloudy brown dwarfs, in which the power of free retrievals to constrain condensation and cloud processes has been demonstrated.

Our ``forward model'', used for predicting the spectra, was constructed using petitRADTRANS \citep{Molliere2019b}. Because the atmosphere of $\beta$~Pic~b is expected to be cloudy, we added scattering to petitRADTRANS. We verified the calculations by comparing to spectra of self-consistent models for cloudy, self-luminous planets obtained with petitCODE \citep{Molliere2015,Molliere2017}, which agreed excellently.

One benefit of using a free retrieval is that one of the most uncertain physical processes, namely the formation of clouds, can be parametrized, letting the observations constrain the cloud mass fraction and particle size distribution. A related approach was taken by \citet{Burningham2017}, who carried out free retrievals for cloudy brown dwarfs for the first time. Here, we assume that our clouds consist of iron and silicate particles, which fixes the location of the cloud base for a given temperature profile. The cloud parameterization of \citet{Burningham2017} was even more general. One of them retrieved the cloud location (where it becomes optically thick), scale height, the single scattering albedo, as well as the power law slope of the opacity.

For the fits presented here, we parametrized the clouds using the \citet{Ackerman2001} cloud model. However, in contrast to the usual treatment in grid models \citep[see, e.g.][]{Ackerman2001,Marley2012,Morley2014,Molliere2017,Samland2017,Charnay2018}, we retrieved all of its three parameters. First, the settling parameter $f_{\rm sed}$, which is the mass-averaged ratio between the settling and mixing velocity of the cloud particles. This determines the decrease of the cloud mass fraction with altitude, which we set to be $\propto P^{f_{\rm sed}}$. Second, the atmospheric mixing coefficient $K_{zz}$, which sets the average particle size, once $f_{\rm sed}$ is fixed. In grid models using the \citet{Ackerman2001} cloud model, this parameter is usually fixed by mixing length theory (with overshooting) or held constant. Third, the width of the log-normal particle size distribution $\sigma_g$, which is normally also kept constant. The cloud mass fraction at the bottom of the cloud was a free parameter, whereas the position of the cloud base was found by intersecting the $P$-$T$ profile with the saturation vapor pressure curves \citep[taken from][in the corrected pressure units]{Ackerman2001} of the cloud species we considered, Fe and MgSiO$_3$.

In the future we plan to also test the \citet{Burningham2017} models of clouds, as they are more general, and do not assume the prevalence of a certain cloud species. Moreover, retrieving the power law slope and albedo of the cloud opacities may represent a better choice: for us this is encoded in our choice of cloud species, particle sizes and width of the log-normal particle size distribution, in a non-trivial way. Based on their findings, \citet{Burningham2017} suggest that a log-normal particle size distribution may not be the ideal choice, and that a Hansen distribution \citep{Hansen1971} may be better.

While carrying out verification retrievals of cloudy petitCODE spectra, we found that we had to be very careful with how the temperature was parametrized. If the temperature model was too flexible \citep[e.g., independent layers + p-spline interpolation, as used in][]{Line2015}, test retrievals of cloudy synthetic spectra lead to clear, hot atmospheres with shallow temperature gradients, that well matched the synthetic input spectrum, but were inconsistent with the input temperature and cloud structure. This could indicate that the cloud-free solutions occupied a larger prior volume, and were thus favored when using a Markov Chain Monte Carlo (MCMC) retrieval.

\begin{table*}
  \begin{center}
    \begin{tabular}{l c c c c c c c}
      \hline
      \hline
      Fit performed & $T$ & $\log(g/g_0)$ & metallicity & C/O ratio & Mass & $\chi^2_\mathrm{red}$ \\ 
      & (K) &             &   $[\mathrm{Fe}/\mathrm{H}]$   &           &  ($M_\mathrm{Jup}$) \\
      \hline
      ExoREM \\
      GRAVITY data only & $1700\pm{}50$ & 3.5 & $-0.5$ & $\le 0.30$ & 2.0 & 3.4 \\ 
      GRAVITY + GPI YJH band data & $1590\pm{}20$ & 4.0 & $0.5$ & $0.43\pm{}0.05$ & $12.4^{(*)}$ & 2.4 \\ 
      \hline
      petitRADTRANS \\ GRAVITY data only
      &  $1847\pm 55$ & $3.3_{-0.42}^{+0.54}$ & $-0.53_{-0.34}^{+0.28}$ &
      $0.35_{-0.09}^{+0.07}$ & $1.4_{-0.87}^{+3.94}$   & 2.6$^{(a)}$   \\
      GRAVITY + GPI YJH band data & $1742\pm 10$ & $4.34_{-0.09}^{+0.08}$ &$0.68_{-0.08}^{+0.11}$ &
      $0.43_{-0.03}^{+0.04}$ & $15.43_{-2.79}^{+2.91}$ & 2.1$^{(b)}$ \\
      \hline
    \end{tabular}\\
    \caption{Results obtained with the ExoREM model grid and free parameter retrieval petitRADTRANS. (*) Using a mass prior in the fit. (a) Mean value of 100 posterior samples, assuming 17 free parameters, using the GRAVITY covariance matrix. (b) Mean value of 100 posterior samples, assuming 21 free parameters, using the GRAVITY covariance matrix.}
    \label{tab:exorem_results}
  \end{center}
\end{table*}

Specifically, we found it to be necessary to impose a temperature profile in the photospheric region that follows the Eddington approximation, that is
\begin{equation}
T_{\rm phot}^4 = \frac{3}{4}T_{\rm 0}^4\left(\frac{2}{3}+\tau\right) ,
\end{equation}
where $T_{\rm 0}$ is normally the internal temperature (taken to be a free nuisance parameter here) and $\tau$ the optical depth. This shape was used from $\tau=0.1$ to the radiative-convective boundary, below which we forced the atmosphere onto a moist adiabat. The optical depth was modeled via
\begin{equation}
\tau = \delta P^{\rm \alpha},    
\end{equation}
where $\delta$ and $\alpha$ are free parameters. A quite strict prior was imposed on $\alpha$. We rejected all models where $|\alpha-\Tilde{\alpha}|>0.1$, where $\Tilde{\alpha}$ is the power law index measured from the opacity structure of a given forward model realization. It was obtained from estimating the Rosseland mean opacity using the non-gray opacity of the atmosphere, across the spectral range of the observations. These altitude-dependent values were then used to calculate an optical depth $\Tilde{\tau}$, and from this
\begin{equation}
\Tilde{\alpha}=\left<\frac{d {\rm log}\Tilde{\tau}}{d {\rm log}P}\right>.
\end{equation}
Here $\left<\right>$ denotes the average over the photospheric region.
This prior ensures that the parametrized, pressure-dependent opacity is consistent with the atmosphere's non-gray opacity structure. In future applications of the parametrized $P$-$T$ we will test to not downright reject models with too large $|\alpha-\Tilde{\alpha}|$. Instead one could adapt the log-likelihood by adding
\begin{equation}
L_\alpha = -\frac{(\alpha-\Tilde{\alpha})^2}{2\sigma_\alpha^2} - \frac{1}{2} {\rm log}\left(2\pi \sigma_\alpha^2\right)
\end{equation}
and fitting for $\sigma_\alpha$ as a free parameter.
Moreover, other P-T parametrizations, for example that of \citet{Madhusudhan2009}, should be tested. This parametrization was also used in \citet{Burningham2017}.

In order to prevent the location of the Eddington photosphere to be unrealistically deep in the atmosphere, we also rejected models where 
\begin{equation}
P(\tau=1)>5P(\Tilde{\tau}=1).
\end{equation}
Above the photosphere the temperature was freely variable. We modeled these high altitudes by retrieving the temperature of three locations spaced equidistantly in ${\rm log}(P)$ space, and spline interpolating between them.

The chemical abundances and moist adiabat of the atmosphere were found by interpolating in a chemical equilibrium table which contained these quantities as a function of $T$, $P$, C/O and [Fe/H]. This table was calculated with the equilibrium chemistry code described in \citep{Molliere2017}. In addition, we also retrieved a quench pressure $P_{\rm quench}$. At pressures smaller than $P_{\rm quench}$ the abundances of CH$_4$, H$_2$O and CO were held constant, so as to model the effect of chemical quenching in regions where the chemical reaction timescales become longer than the mixing timescales \citep[see, e.g.,][]{Zahnle2014}. 

The following absorption opacity sources where included: CO, H$_2$O, CH$_4$, NH$_3$, CO$_2$, H$_2$S, Na, K, PH$_3$, FeH, VO, TiO, H$_2$-H$_2$ (CIA), H$_2$-He (CIA), Fe clouds (crystalline particles, irregularly shaped), MgSiO$_3$ clouds (crystalline particles, irregularly shaped). The following scattering opacity sources where included: H$_2$ Rayleigh scattering, He Rayleigh scattering, Fe clouds, MgSiO$_3$ clouds. The opacity references can be found in \citet{Molliere2019b}.

Using the setup described above, we were able to successfully retrieve the spectrum and atmospheric parameters for a synthetic observation of a cloudy, self-consistent model obtained with petitCODE. The implementation of the retrieval forward model presented here will be described in detail in an upcoming paper. It will contain a description of how the scattering was added, and the verification thereof, as well as the verification retrieval test.

\begin{figure*}
  \includegraphics[width = \linewidth]{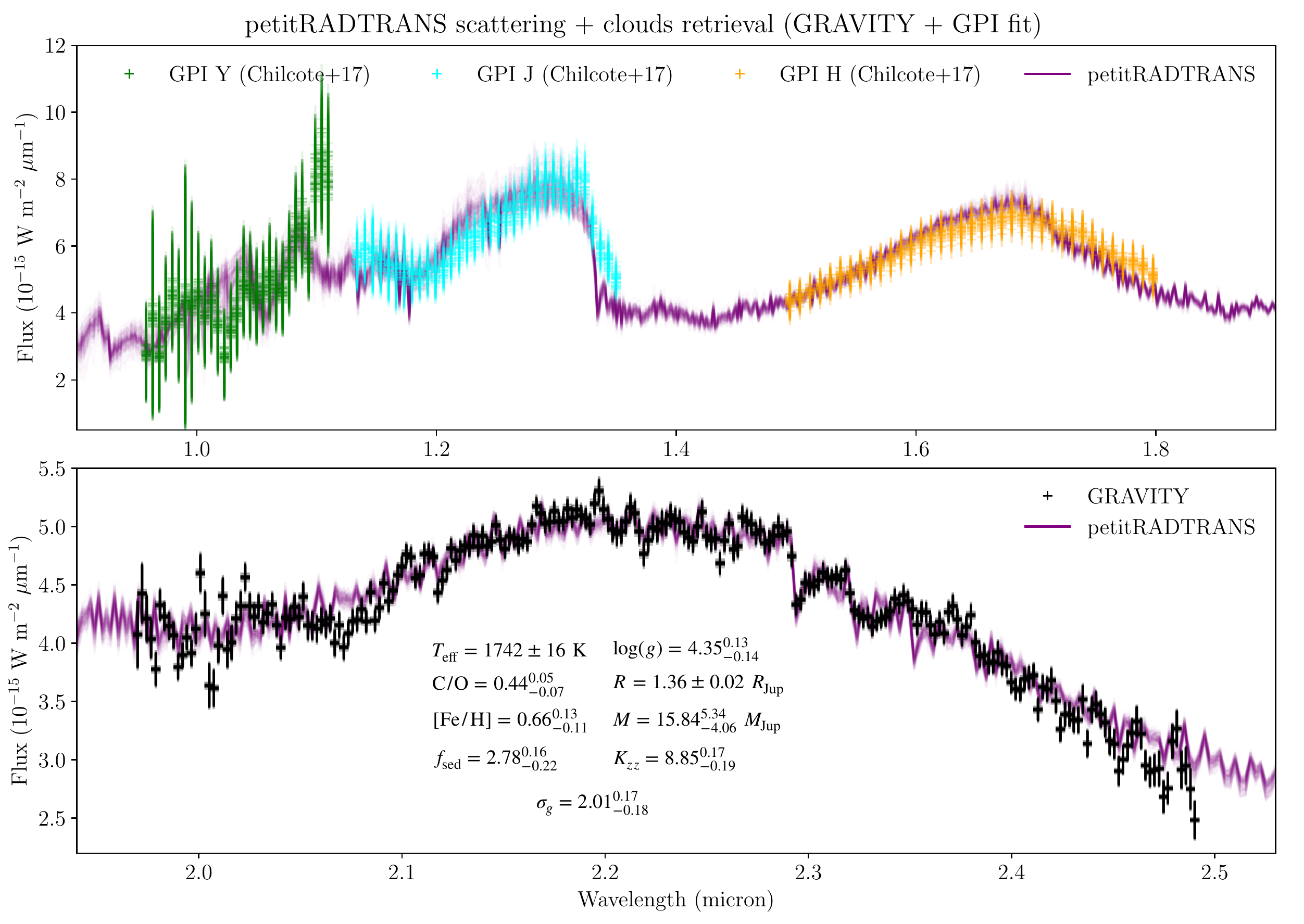}
  \caption{Results of the combined (GRAVITY+GPI) fit of the $\beta$~Pic~b spectrum with petitRADTRANS. No prior on the mass was used in the fit, and the spectroscopically retrieved mass is consistent with the astrometric value. For producing this plot, 100 samples were drawn from the posterior distribution, for both the model and the data scaling. The 2-d projection of the posterior can be found in Appendix \ref{sec:pRT_posterior_appendix}. {\it Top panel:} the GPI Y, J and H-band data of \citet{Chilcote2017} are plotted as green, cyan, and orange points with error bars, respectively, the petitRADTRANS models are plotted as purple solid lines. The fit is dominated by the high S/N of the GRAVITY data (shown in the {\it bottom panel}), leading to a worse fit in the GPI bands, see text. {\it Right panel:} the GRAVITY data are shown as black points with errorbars, the petitRADTRANS models are plotted as purple solid lines.
  }
  \label{fig:pRT_fit_spec}
\end{figure*}

The parameter estimation was carried out using emcee \citep{Foreman-Mackey2013}. Due to the complex priors resulting from the temperature parametrization, the high dimensionality, and a potentially multimodal posterior, the acceptance fraction is low (of the order of 1-2~\%) such that one million models were drawn (started around  the best-fit position of a pre-burn) to obtain results where the walker positions had converged. As is common for all parameter estimations using an MCMC method, we cannot guarantee that the retrieval results have converged to the true global maximum of the log-probability. While the multi-modality of our model is an inherent property, the acceptance fraction can be improved by setting up the chain closely around the best-fit position of the pre-burn-in run \citep{Foreman-Mackey2013}. In addition, we are currently working on implementing a parameter estimation using nested sampling, which also applies a clustering algorithm for the parameter estimation. This should alleviate the low acceptance rate problem, and lead to a complete sampling of the parameter space \citep{Feroz2008,Feroz2009}. With these current limitations in mind, we note that we retrieved similar values as in the grid retrieval with Exo-REM in Section \ref{subsect:exorem}, and could successfully retrieve self-consistent cloudy input models when testing our method.

\subsubsection{Retrieval parameter results}

Our forward model has 17 free parameters: 6 for the temperature model described above, 3 for the abundances (C/O, [Fe/H], $P_{\rm quench}$), 5 for the clouds (the cloud mass fraction of the MgSiO$_3$ and Fe at the cloud base, the settling parameter $f_{\rm sed}$, the eddy diffusion coefficient $K_{zz}$, the width of the log-normal particle size distribution $\sigma_g$), the gravity $\log(g)$, the planetary radius $R_{\rm Pl}$, and the abundance of FeH (currently not included in the chemical table). We used uniform or log-uniform priors for all parameters. In addition to the parameters above we allowed for an individual scaling of the GPI (Y, J, H) bands by up to $\pm$50~\%, and by up to $\pm$2.5~\% for the GRAVITY data. A large value for the scaling of the GPI data was chosen because a similar scaling value was found when comparing the GPI and SPHERE J-band for 51~Eri~b in \citet{Samland2017}. The maximum scaling we retrieve for $\beta$~Pic~b is 13~\% in the GPI Y-band, see below.

Similar to the Exo-REM analysis in Section \ref{subsect:exorem}, we ran retrievals for a GRAVITY only and GRAVITY + GPI case. In Figure \ref{fig:pRT_fit_spec} we show the results when fitting the GRAVITY and GPI data together, {but without imposing a prior on the mass of $\beta$~Pic~b}. For producing this plot, we sampled the posterior distribution 100 times, and plot both the spectra and the accordingly scaled data points. We give the median, and 16 and 84 percentile values of some of the free parameters in the Figure. The full posterior and resulting temperature confidence envelopes are shown in Appendix \ref{sec:pRT_posterior_appendix}. The effective temperature was obtained from integrating the flux of the sampled spectra from 0.5 to 20 microns. The mass was calculated from the $\log(g)$ and $R_{\rm P}$ values of the posterior sample.

As can be seen in Figure \ref{fig:pRT_fit_spec}, the GRAVITY data can be well fit. At least two CO bandheads at $\sim$2.3 micron are visible in the data. 
The GPI data is less well fit, which may be partially due to the high S/N of the GRAVITY data, which dominates the fit. We found that the fit of the GPI data improved when increasing the error bars of the GRAVITY data. Likewise, we found that the slope in the red part of the GRAVITY spectrum can be fit better if the GPI data are neglected.
The retrieved parameters are presented in Table~\ref{tab:exorem_results}, together with the ExoRem results for comparison.
Most interestingly, petitRADTRANS retrieves a mass which is consistent with the values from the astrometric measurement in Section \ref{sec:orbit_mass}, {without the need of imposing a prior on the mass}. Here we find $M_{\rm P} = 15.43_{-2.79}^{+2.91}~M_{\rm Jup}$, which is consistent with the values presented in Table~\ref{tab:orbitparams}.

From the retrievals presented here, it appears that the GRAVITY K-band data are useful for constraining the planetary C/O ratio, while adding the GPI Y, J, and H-bands is important for obtaining better constraints on the planet's gravity, and hence mass (petitRADTRANS retrieves a too low mass if the GPI data are neglected, see Section \ref{sect:fit_comp} below). This is   consistent with the sensitivity of the NIR YJH bands to gravity (atomic and molecular features, as well as the H band shape), while the CO absorption in the K band is not expected to probe the surface gravity very well \citep[see, e.g.,][]{Bejar2018}.
As demonstrated here, estimating the planetary mass from a planet's spectrum alone may become feasible when applying free retrievals over a broad spectral range, such as carried out here with petitRADTRANS.


\subsection{Comparison between the grid and free retrieval}
\label{sect:fit_comp}

In agreement with the Exo-REM fit for GRAVITY+GPI, petitRADTRANS obtains a cloudy atmosphere, and a slightly sub-solar C/O of $0.43_{-0.03}^{+0.04}$ (Exo-REM found $0.43\pm 0.05$). The free retrieval obtains a metallicity of $0.68_{-0.08}^{+0.11}$. This is higher than Exo-REM, where $0.5$ was found, but this was at the boundaries of the Exo-REM grid, and could likely be higher. This could also be the reason for the slightly higher $\log(g/g_0)$ value ($4.34_{-0.09}^{+0.08}$, and 4 for Exo-REM), due to the gravity-metallicity correlation. petitRADTRANS finds an effective temperature which is higher than in the Exo-REM fit by about 150~K ($1742\pm 10$~K, compared to 1590~K for Exo-REM). The larger radius found by Exo-REM is likely due to the lower temperature it retrieved, so as to conserve the total amount of flux. At the estimated age of $\beta$~Pic \citep[$24\pm 3$~Myr; see][]{Bell2015}, a radius of 1.7~$R_{\rm Jup}$ (the value Exo-REM retrieved) requires masses in excess of 20~$M_{\rm Jup}$, and effective temperatures of around 2500~K, when considering hot start models \citep{Spiegel2011}. Core accretion models under the warm\footnote{These models are somewhat warmer than the classical cold start assumption \citep{Marley2007}, because the planetesimal accretion is not shut off after the isolation mass is reached.} start assumption, which include deuterium burning, require similarly large masses and temperatures, but potentially somewhat younger ages \citep{Molliere2012,Mordasini2017}, and would put the planet firmly into the mass regime of brown dwarfs currently undergoing deuterium burning. Hence, the large radius retrieved by Exo-REM is likely to be inconsistent with the retrieved mass and temperature. The values of the mass, temperature, and radius (1.36~$R_{\rm Jup}$) retrieved by petitRADTRANS agree with both the cold and hot start predictions, given the age of $\beta$~Pic.

Also when fitting only the GRAVITY data, the petitRADTRANS results are mostly consistent with Exo-REM. Without a mass prior we find C/O = $0.35_{-0.09}^{+0.07}$ (Exo-REM found $\lesssim 0.3$), [Fe/H] = $-0.53_{-0.34}^{+0.28}$ (Exo-REM found -0.5), $\log(g/g_0)=3.3_{-0.42}^{+0.54}$ (Exo-REM found 3.5), $M=1.4_{-0.87}^{+3.94}$~$M_{\rm Jup}$ (Exo-REM found 2 $M_{\rm Jup}$). Only the temperature is larger again, at $1847\pm 55$~K (petitRADTRANS), compared to 1700~K (Exo-REM).

In summary, a free retrieval approach gives similar results to a more classical retrieval from a grid of forward models. We note that here a free retrieval appears to lead to physically more consistent results when constraining radii and effective temperatures. Another possible cause for the differences could be how the opacities of gas and clouds are treated. For the gas opacities we note that petitRADTRANS uses the opacity database of petitCODE, the latter of which was successfully benchmarked with Exo-REM in \citet{Baudino2017}. Small remaining differences, identified to stem from the use of different line lists in \citet{Baudino2017}, have since been removed by updating the opacity database of petitCODE/petitRADTRANS in 2017.

\subsection{Comparison to \citet{Chilcote2017}}

A substantial analysis of the NIR spectrocopy of $\beta$~Pic~b was carried out in \citet{Chilcote2017}, using GPI YJHK band spectra. The data were compared to low gravity and field brown dwarf spectra, the derived bolometric luminosity was compared to evolutionary models, and spectral fits were carried out with four different model grids.

Comparing their bolometric luminosity to evolutionary models \citep[hot start models of][]{Baraffe2003}, \citet{Chilcote2017} found a mass of $12.9\pm0.2$~$M_{\rm Jup}$, an effective temperature of $1724\pm15$~K, a surface gravity of $\log(g/g_0)=4.18\pm0.01$ and a radius of $1.46\pm 0.01$~$R_{\rm Jup}$. Their mass measurement is consistent both with our astrometrically and spectroscopically inferred mass values. Moreover, the other values inferred from the the YJHK fit of petitRADTRANS are close to the evolutionary values of \citet{Chilcote2017}, but not within the uncertainties of one another (e.g., $1742\pm 10$~K vs. $1724\pm15$~K). As noted in \citet{Chilcote2017}, these uncertainties do not contain a contribution of the model uncertainties, and the true uncertainties must be larger. The same holds for our retrievals and fits carried out here. The Exo-REM fits with mass prior lead to a similar agreement in gravity, but the radii and temperatures are further away from the  \citet{Chilcote2017} values, with the planet being cooler, and thus larger, in the Exo-REM fits.

The best grid model fit of the combined photometry and spectroscopy in \citet{Chilcote2017} was obtained with Drift-PHOENIX \citep[e.g., ][]{Helling2008}, where $T_{\rm eff}=1651$~K, ${\rm log}(g)=3$ and $R=1.58$~$R_{\rm Jup}$ was found, leading to a mass of $\sim$1~$M_{\rm Jup}$. The AMES-Dusty \citep[e.g.,][]{Allard2001} fit gave the highest mass (most consistent with our astrometric, spectroscopic and \citealt{Chilcote2017}'s evolutionary mass), namely 17~$M_{\rm Jup}$. The AMES-Dusty best-fit values are also closer to the evolutionary parameters derived in \citet{Chilcote2017}, at $T_{\rm eff}=1706$~K, ${\rm log}(g)=4.5$ and $R=1.18$~$R_{\rm Jup}$, but at an overall worse $\chi^2$ than the Drift-PHOENIX fit.

In summary, the results of our spectral characterization compare well with the evolutionary values inferred for $\beta$~Pic~b in \citet{Chilcote2017}. Especially the parameter values of the free retrieval carried out with petitRADTRANS are close to the evolutionary values. It is also noteworthy that our masses, inferred indepentently with astrometry or spectral retrieval, are consistent with the evolutionary mass of \citet{Chilcote2017}.

\section{C/O ratio and the formation of beta Pic b}
\label{sec:co_ratio}

\subsection{Stellar and planetary C/O ratio}

\cite{Holweger1997} have shown that the abundances of several elements (C, Ca, Ti, Cr, Fe, Sr, Ba) on the surface of $\beta$ Pictoris are Solar. But measuring the abundance of oxygen in stars is a notoriously difficult task, due to line blending, deviations from local thermal equilibrium predictions, or sensitivity to the 3D temperature structure of the star \citep{Asplund2005}. As a consequence, to our knowledge, the abundance of oxygen -- and hence the C/O ratio -- has not yet been reported in the literature. We note, though, that a subsolar C/O ratio (i.e., $\simeq{} 0.4$) would invalidate most of the following discussion.

In our atmosphere analysis, both the ExoREM grid model fitting and the petitRADTRANS free retrieval point to the same result: the C/O number ratio in $\beta$ Pictoris b is $\simeq{}0.43\pm{}0.05$, which is subsolar (the solar C/O ratio is 0.55, see \citeauthor{Asplund2009}, \citeyear{Asplund2009}).

A number of studies have been done in the recent years to try to find links between planetary formation processes and element abundances. In particular, \cite{Oberg2011} first attempted to relate the C/O ratio to the position of the different icelines in a protoplanetary system, and to the proportion of gas and solid material accreted by a young planet. They concluded that substellar C/O ratio was a sign of a formation by either gravitational collapse or core-accretion, followed by icy planetesimal enrichment. The objective of this section is to show that the C/O ratio can possibly be used to disentangle the two formation scenarios.

\subsection{General model for the evolution of the C/O ratio}

In a similar fashion as to \cite{Oberg2011}, we assume that the main sources of carbon and oxygen in the protoplanetary disk in which $\beta$ Pic b formed were $\mathrm{CO}$, $\mathrm{CO}_2$, $\mathrm{H}_2\mathrm{O}$, silicates and carbon grains. Assuming a solar C/O ratio for the star, the table of relative abundances given in \cite{Oberg2011} is valid, and we use it as a baseline to set the abundances of each species (see Table~\ref{tab:abundances}).

\begin{table}
  \begin{center}  
    \begin{tabular}{l c}
      \hline
      \hline
      Species & $n_\mathrm{species}/n_{\mathrm{H}_2\mathrm{O}}$ \\
      \hline
      $\mathrm{H}_2\mathrm{O}$ & 1 \\
      $\mathrm{CO}$ & 1.67 \\
      $\mathrm{CO}_2$ & 0.33 \\
      C (grains) & 0.67 \\
      O (silicates) & 1.54 \\
      \hline
    \end{tabular}
    \caption{Relative abundances of the different species taken from Table~1 of \cite{Oberg2011}, and used in our young $\beta$ Pic protoplanetary disk. All values are given relative to $\mathrm{H}_2\mathrm{O}$.}
    \label{tab:abundances}
  \end{center}
\end{table}

In the framework developed by \cite{Oberg2011}, the C/O ratio in the atmosphere of a planet can be calculated from the amount of solid and gaseous material entering its composition.
We denote $n_{X,s}$ (resp. $n_{X,g}$) the abundance of element $X$ in the solid phase (resp. gas phase) of the disk, given in number of atoms per unit of disk mass. We also write $M_\mathrm{solid}$ (resp. $M_\mathrm{gas}$) the total mass of solid (resp. gas) entering the composition of the atmosphere of the planet, and $f_\mathrm{s/g}$ the dust-to-gas fraction in the disk, which we assume to be equal to 0.01. With these notations, the total number of elements $X$ in the atmosphere of the planet is given by:
\begin{equation}
  N_X = \frac{n_{X,s}}{f_\mathrm{s/g}}\times{}M_\mathrm{solid} + \frac{n_{X,g}}{1-f_\mathrm{s/g}}\times{}M_\mathrm{gas}
\end{equation}
And the C/O number ratio is then:
\begin{equation}
  \mathrm{C/O} = \frac{n_{\mathrm{C},s}{f_\mathrm{s/g}}^{-1}M_\mathrm{solid}+n_{\mathrm{C},g}(1-{f_\mathrm{s/g}})^{-1}M_\mathrm{gas}}{n_{\mathrm{O},s}{f_\mathrm{s/g}}^{-1}M_\mathrm{solid}+n_{\mathrm{O},g}(1-{f_\mathrm{s/g}})^{-1}M_\mathrm{gas}}
  \label{eq:c/o}
\end{equation}
Note that both the numerator and the denominator can be given relative to a reference species without affecting the validity of this Eq.~(\ref{eq:c/o}). In Table~\ref{tab:abundances} and in all the following, we implictly use abundances relative to $\mathrm{H}_2\mathrm{O}$.

The exact values of $n_\mathrm{C,s}$, $n_\mathrm{C,g}$, $n_\mathrm{O,s}$, and $n_\mathrm{O,g}$ depends on the abundances given in Table~\ref{tab:abundances}, and on the state (solid or gaseous) of each species. and hence on the location of the forming planet with respect to the different icelines.

Using ALMA observations, \cite{Qi2015} have shown that the CO iceline in the disk around HD~163296 was likely to be located at $\simeq{}90~\mathrm{AU}$ from the star. Other observations of the same system, also performed with ALMA, led \cite{Notsu2019} to conclude that the water iceline was located at a distance of $\le 20~\mathrm{AU}$. Since HD~163296 is also an A-type star, these two values give an idea of the possible location of the $\mathrm{H}_2\mathrm{O}$ and $\mathrm{CO}$ icelines in the $\beta$ Pic system. However, little is known about the relationship between the current orbit of $\beta$ Pic b and its exact formation location, and about possible variations of the locations of these icelines between systems. Thus, no definitive assumption can be made as to where the planet formed in comparison to the water iceline, and the two options must be considered: a formation within the water iceline, and a formation between the water and the $\mathrm{CO}_2$ icelines. 



From there, the terms $n_{\mathrm{C}, s}, n_{\mathrm{C}, g}, n_{\mathrm{O}, s}$, and $n_{\mathrm{O}, g}$ from Eq.~(\ref{eq:c/o}) can be determined from the values listed in Table~\ref{tab:abundances}. For a planet forming within the water iceline, we have:
\begin{equation}
  \begin{cases}
  n_{\mathrm{O}, g} = n_{\mathrm{H}_2\mathrm{O}} + n_{\mathrm{CO}} + 2\times{}n_{\mathrm{C}\mathrm{O}_2} = 3.33 \\
  n_{\mathrm{O}, s} =  n_{\mathrm{O}~\mathrm{(silicates)}} = 2.12 \\
  n_{\mathrm{C}, g} = n_{\mathrm{CO}} + n_{\mathrm{C}\mathrm{O}_2} = 2.0 \\
  n_{\mathrm{C}, s} = n_{\mathrm{C}~\mathrm{(grains)}} = 0.67
  \label{eq:c_o_abundances_1}  
  \end{cases}
\end{equation}

\noindent{}And for a planet forming between the water and $\mathrm{CO}_2$ icelines:
\begin{equation}
  \begin{cases}
  n_{\mathrm{O}, g} = n_{\mathrm{CO}} + 2\times{}n_{\mathrm{C}\mathrm{O}_2} = 2.33\\
  n_{\mathrm{O}, s} = n_{\mathrm{H}_2\mathrm{O}} + n_{\mathrm{O}~\mathrm{(silicates)}} = 3.12\\
  n_{\mathrm{C}, g} = n_{\mathrm{CO}} + n_{\mathrm{C}\mathrm{O}_2} = 2.0\\
  n_{\mathrm{C}, s} = n_{\mathrm{C}~\mathrm{(grains)}} = 0.67
  \end{cases}
  \label{eq:c_o_abundances_2}
\end{equation}


\subsection{C/O ratio in the gravitational collapse paradigm}

\begin{figure*}
  \begin{center}
    \includegraphics[width=\linewidth]{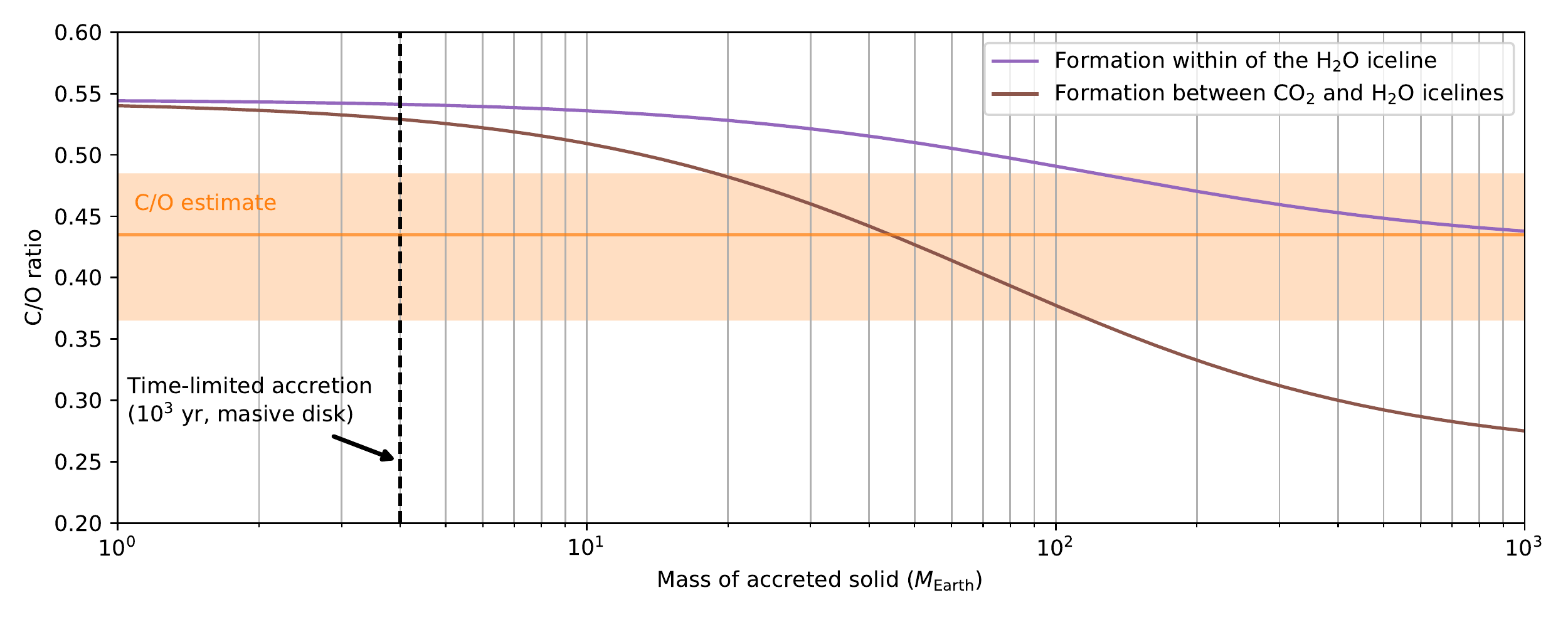}
    \caption[C/O versus mass of accreted solid in the gravitational collapse paradigm]{Gravitational collapse scenario: evolution of the C/O ratio as a function of the total mass of solid accreted after the initial formation of the protoplanet. The purple curve corresponds to a formation within the $\mathrm{H}_2\mathrm{O}$ iceline, and the brown curve to a formation between the $\mathrm{H}_2\mathrm{O}$ and $\mathrm{CO}_2$ icelines. The orange area gives the 68\% confidence interval for the value of the C/O ratio. Dashed vertical lines corresponds to different solid accretion limits discussed in the text.} 
    \label{fig:co_accretion}
  \end{center}
\end{figure*}

We consider the case of a formation through gravitational collapse \citep{Bodenheimer1974}, a violent mechanism which shares similarities with star formation. In this scenario, an entire region of the circumstellar disk becomes unstable, and rapidly collapses to form a protoplanet, which then slowly contracts and cools down. 

The total mass of solid entering in the composition of the atmosphere of a planet formed through gravitational collapse can be separated in two terms: the mass of solid initially contained in the disk fragment which collapsed to create the protoplanet, and the mass of solid planetesimals later accreted by the protoplanet. The solid mass contained in the initial clump is directly related to the dust-to-gas ratio of the disk, and we can write:
\begin{equation}
  M_\mathrm{solid} = f_\mathrm{s/g}M_\mathrm{planet} + M_\mathrm{accreted}
  \label{eq:msolid=fmplanet+maccreted}
\end{equation}
\noindent{}This equation assumes that no core has formed in the young protoplanet, which, for a planet as massive as $\beta$ Pic b is reasonable \citep{Helled2008}. For a planet less massive, for which a core could form, sedimentation of a fraction of the initial solid mass on the core should be taken into account.

Injecting the definition of $M_\mathrm{solid}$ into Eq.~(\ref{eq:c/o}), and using $f_\mathrm{s/g}=0.01$, $M_\mathrm{planet} = 12.7\,M_\mathrm{Jup}$, as well as the values for C and O abundances given in Eq.~(\ref{eq:c_o_abundances_1}) or (\ref{eq:c_o_abundances_2}), it is possible to determine the C/O ratio as a function of the mass of accreted planetesimals $M_\mathrm{accreted}$ in the gravitational collapse paradigm. The results is given in Figure~\ref{fig:co_accretion}, for two possible formation locations: within the water iceline, and between the water and carbon dioxide icelines. We have also added the $1\,\sigma$ confidence intervals of our ExoREM and petitRADTRANS measurements on this graph.
This figure shows that a formation bewteen the $\mathrm{H}_2\mathrm{O}$ and $\mathrm{CO}_2$ icelines is more favorable to a large deviation from the stellar C/O ratio, mainly due to the injection of oxygen coming from solid water ice during planetesimal accretion.   

The formation of a planet by gravitational instability can be separated in a few different steps \citep{Bodenheimer1974}: formation of the initial clump in the disk, quasi-equilibrium contraction, hydrodynamic collapse, and a new hydrostatic quasi-equilibrium phase. Accretion of planetesimals is thought to be efficient only during the pre-collapse phase \citep{Helled2009}. The duration of this phase decreases with increasing planet mass, and typical values ranges from a few $10^5$ years for a Jupiter mass planet, to less than $10^3$ years for more massive planets \citep{DeCampli1979, Bodenheimer1980}. Using the model proposed by \cite{Helled2009}, the mass of planetesimal accreted during the pre-collapse phase of $\beta$ Pic b can be estimated using:
\begin{equation}
  M_\mathrm{accreted} = \int_0^{t_\mathrm{collapse}}\pi R_\mathrm{capture}^2(t)\sigma(a, t)\Omega(a)\mathrm{d}t
\end{equation}
\noindent{}Where $t_\mathrm{collapse}$ is the time of collapse, $R_\mathrm{capture}$ the protoplanet's capture radius, $\sigma$ the surface density of solids in the disk at the location of the protoplanet, and $\Omega$ the orbital frequency.

\cite{Andrews2005} presented a large survey of 153 young stellar objects in the Taurus-Auriga star forming region. Among all these objects, AB Aur and V892 Tau are two A-type stars, for which they give an estimate of the mass: $0.004~M_\odot$ and $0.009~M_\odot$. Considering all stellar types, the median disk-to-star mass ratio they found is 0.5\%. More recent studies of protoplanetray disk demographics based on ALMA observations yielded similar results, with typical dust to star mass ratios of $\simeq{}10^{-4.5}$ \citep{Pascucci2016, Ansdell2017}, i.e. disk-to-star mass ratios of $\simeq{}0.3\%$, assuming a dust-to-gas ratio of $1\%$.

Considering the upper limit of an extremely massive disk ($M_\mathrm{disk} = 0.1~M_\odot$), and using a power-law for the surface density ($\sigma = \sigma_0\,(a/5~\mathrm{AU})^{-\alpha}$, with $\alpha = 1.00$), the solid density at $a = 11~\mathrm{AU}$ is:
\begin{equation}
  \sigma(11~\mathrm{AU}) \simeq{} 6~\mathrm{g}/\mathrm{cm}^2
\end{equation}
\noindent{}The orbital period of the planet is $\sim 20~\mathrm{yr}$ \citep[Section~\ref{sec:orbit_mass} of this work]{Wang2016, Lagrange2018}. The capture radius decreases with the contraction of the planet, but an optimistic value would be $ 2~\mathrm{to}~3\times{}10^{12}~\mathrm{cm}$ for a $1~M_\mathrm{Jup}$ planet \citep{Helled2006}. For a planet 10 times more massive, the effective radius could be $\simeq{}5\times{}10^{12}~\mathrm{cm}$. This yields:
\begin{equation}
  M_\mathrm{accreted} \simeq 4\times{}M_\mathrm{Earth}\times{}\frac{t_\mathrm{collapse}}{1000~\mathrm{yr}}
\end{equation}
The corresponding accretion limit has been added to Figure~\ref{fig:co_accretion}, for a reasonable assumption of $t_\mathrm{collapse} = 10^3~\mathrm{yr}$.

Taking into account the effective time available for efficient planetesimal accretion during the pre-collapse stage, the low C/O ratio measured with GRAVITY is difficult to explaine, even in the case of a planet forming oustide the $\mathrm{H}_2\mathrm{O}$ iceline. For the C/O ratio to reach a value of $\simeq{}0.43$, we need to assume a massive protoplanetary disk and an unusually long time for the pre-collapse phase, or an extremely efficient accretion (with an accretion rate of $4\times{}10^{-3}\,M_\mathrm{Earth}/\mathrm{yr}$).

\subsection{C/O ratio in the core-accretion paradigm}


\begin{figure*}
  \begin{center}
    \includegraphics[width=\linewidth]{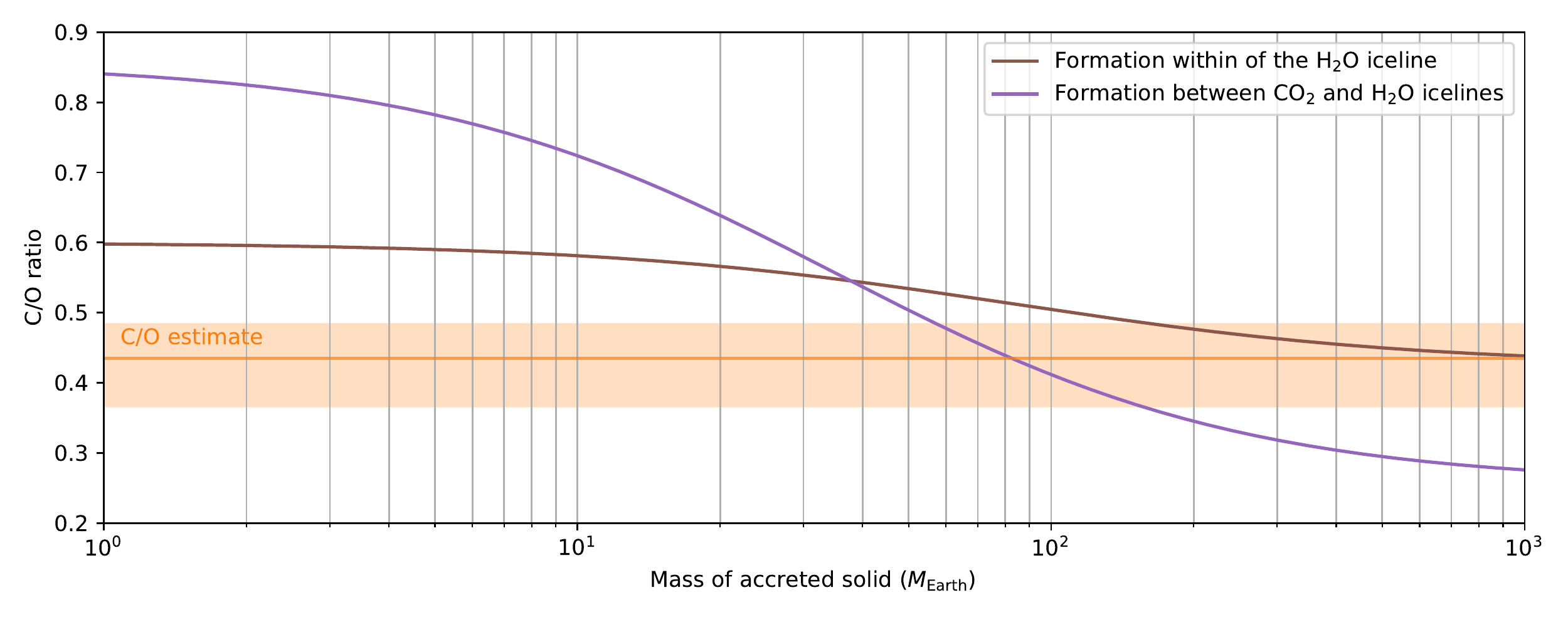}
    \caption[C/O versus mass of solid in the core-accretion paradigm]{Core-accretion scenario: evolution of the C/O ratio in the atmosphere of $\beta$ Pic b as a function of the total mass of solid accreted by the protoplanet, for a formation between the $\mathrm{H}_2\mathrm{O}$ and the $\mathrm{CO}_2$ icelines, or within the $\mathrm{H}_2\mathrm{O}$ iceline. The orange area gives the 68\% confidence interval for the value of the C/O ratio.}
    \label{fig:co_core_accretion}
  \end{center}
\end{figure*}

Core-accretion is another formation mechanism, in which an initial solid core forms, and slowly accretes gas from the disk. When the mass of gas is roughly the same as the mass of the core, the protoplanet enters a phase of ``runaway gas accretion'', during which it gains significant amount of gas over a short time \citep{GiantPlanetFormation}. In this scenario, the formation of a planet is a much longer process than with gravitational instability, which gives more time to enrich the proto-atmosphere in solid material and to lower its C/O ratio.



\cite{Mordasini2016} explored the effect of planetesimal enrichment coupled with disk composition, in a core-accretion scenario. They focused on the case of Jupiter mass planets migrating to short period orbits (``hot Jupiters''), which is a different archetype than $\beta$ Pic b. But the general sequence of events they use to form their planets in the core-accretion paradigm can still be applied to $\beta$ Pic b, only leaving out the inward migration part. First, the core of the planet forms from the accretion of solid material. Then, once the core has formed, the protoplanet starts accreting a gaseous envelope which, during its formation, is enriched by the accretion of disintegrating planetesimals. When the planet reaches a critical mass, runaway accretion occurs, and the mass of the planet significantly increases. This runaway gas accretion clears a gap in the disk, and ends the formation of the planet.

In the gravitational instability scenario, because the formation of the planet happens so quickly compared to typical timescales of disk evolution, the gas and solid making the atmosphere necessarily have a stellar combined C/O. If the solid and gas in the atmosphere are in the same proportion as they are in the disk ($M_\mathrm{solid} = f_\mathrm{s/g}M_\mathrm{gas}$), the C/O of the atmosphere is stellar. A deviation of the solid to gas proportion in the atmosphere is required to alter the C/O ratio.

In the case of core-accretion, the situation is different. Witout planetismal enrichment before the runaway gas accretion phase, the atmosphere of the planet would not be made of a mixture of gas and solid material, but purely of gas. Thus, without planetesimal enrichment, the atmospheric C/O ratio in the core-accretion paradigm can be expected to be close to the C/O ratio of the gas in the disk, that is, superstellar.

In this core-accretion paradigm, it is still possible to use Eq.~(\ref{eq:c/o}) to calculate how the final C/O ratio of the atmosphere is impacted by the mass of solid material accreted before the runaway accretion phase. But in this case, all of the solid mass $M_\mathrm{solid}$ corresponds to accreted material: $M_\mathrm{solid} = M_\mathrm{accreted}$, as opposed to Eq.~(\ref{eq:msolid=fmplanet+maccreted}).

In Figure~\ref{fig:co_core_accretion}, we show the evolution of the C/O ratio as a function of the mass of accreted planetesimals, for a formation thourgh core-accretion, within the water iceline, or between the water and $\mathrm{CO}_2$ icelines. In this scenario, it is possible to reach C/O values compatible with our GRAVITY measurement with accretion of $\simeq{}80~M_\mathrm{Earth}$, if the planet formed between the water and $\mathrm{CO}_2$ icelines. A formation within the water iceline is more diffcult to explain, as it would require at least $150~M_\mathrm{Earth}$ of solid material enrichment to reach the upper limit of the $1\,\sigma$ interval on the C/O measurement, and up to several $10^2\,M_\mathrm{Earth}$ to reach a value of 0.43.

\section{Summary and conclusions}
\label{sec:conclusion}
In this work, we presented the first VLTI/GRAVITY spectro-interferometric observation of the giant planet $\beta$ Pictoris b. Using an adequate data reduction technique detailed in the appendix of this paper, we extracted a high quality K-band spectrum of the planet, at a resolution of $R = 500$. We also derived the most precise relative astrometry obtained to date on this object, with an error of $\simeq{}40~\mu\mathrm{as}$.

We find that the astrometry disfavors circular orbits for $\beta$ Pic b, with a value of $e\simeq{}0.15^{+0.05}_{-0.04}$. It remains unclear how a massive planet like $\beta$ Pic b can acquire such a significant eccentricity.
Using this new astrometric datapoint together with previously published visual astrometry and Hipparcos/Gaia data, we were able to derive an estimate of the dynamical mass of $\beta$ Pictoris b, in a similar fashion as to what \cite{Snellen2018} and \cite{Dupuy2019} did. Our value is compatible with these previous studies, with a best estimate of $12.7\pm{}2.2\,M_\mathrm{Jup}$.

We were also able to retrieve a similar mass, albeit with larger error bars, using only the spectral data. Using a free retrieval, including the effect of scattering and clouds, with petitRADTRANS \citep{Molliere2019} to fit the spectrum of $\beta$ Pic b in Y, J, H, and K bands \citep[Y, J, H from][K from this work]{Chilcote2017}, we obtained a mass of $15.43^{+2.91}_{-2.79}\,M_\mathrm{Jup}$. This constitutes a rare case of validation of an atmospheric model with a model-independent measurement.

We performed an in-depth analysis of the K-band spectrum extracted from our GRAVITY observation using two different approaches: forward modeling with the ExoREM code \citep{Charnay2018}, and free-retrieval with petitRADTRANS. We found that both approaches point to a C/O ratio of $\mathrm{C/O}=0.43\pm{}0.05$.

We showed that, if the C/O ratio of the host star $\beta$ Pictoris is Solar, it is difficult to explain this C/O ratio with a gravitational collapse formation scenario. This is mainly due to the high mass of $\beta$ Pictoris b, which has the dual consequence of requiring large amount of planetesimal enrichment to lower the initial C/O ratio, while at the same time making the whole formation process extremely short. In this case, it appears that a slower formation via core-accretion, somewhere between the $\mathrm{H}_2\mathrm{O}$ and $\mathrm{CO}_2$ icelines, is more likely. This scenario can potentially explain the subsolar C/O ratio if the planet was enriched in oxygen by icy planetesimal accretion.

The high metal enrichment we retrieve from the spectral fits appears to corroborate this assessment, with the exact value being quite high and at the edge of what is expected from classical core accretion \cite{Mordasini2016}.

This model still comes with several important limitations. One of them is that the exact compositition of the initial protoplanetary disk around $\beta$ Pic remains largely unknown. Another major issue is the efficiency of the planetesimal enrichment, which we have assumed to be of $100\%$ (i.e., all the solid material accreted by the planet is disintegrated in the atmosphere). This is unlikely to be the case, as fraction of this material can be deposited into the planetary core, or can stay at the bottom of the atmosphere. This is particularly true for the core-accretion scenario, in which the solid material is accreted before most of the gas \citep{Mordasini2016}. Strong vertical mixing can potentially mitigate this problem, but further studies are required to be able to take into account these phenomena. Finally, disk chemistry may also play a role. For example, \cite{Eistrup2018} have shown that a large fraction of water molecules can be transformed into dioxygen ($\mathrm{O}_2$) over a few Myr, along a chemical pathway detailed in \cite{Walsh2015}. Oustide of the water iceline, such a chemical evolution can potentially deplete the solid material from its oxygen, while enriching the gas.




The observations of $\beta$ Pictoris b presented in this paper show the potential of long-baseline optical interferometry with VLTI/GRAVITY for exoplanet science. The instrument gives access to medium resolution spectroscopy in K-band and high-precision astrometry, which are both extremely useful to characterise giant exoplanets and to start peering into their formation history.

\begin{acknowledgements} 
  Based on observations collected at the European Southern Observatory under ESO programme 0101.C-0912(A) and 2101.C-5050(A). 
   M.N. acknowledges funding for his PhD from the European Research Council (ERC), under the European Union’s Horizon 2020 research and innovation programme (Grant agreement No. 639248).
   P.M thanks M. Line for insightful discussions. P.M acknowledges support from the European Research Council under the European Union's Horizon 2020 research and innovation programme (Grant agreement No. 694513 and 832428).
   J.W. thanks R. De Rosa for helpful discussions on Gaia and Hipparcos data. J.W. is supported by the Heising-Simons Foundation 51 Pegasi b postdoctoral fellowship.
   R.G.L. received financial support of Science Foundation Ireland under Grant number 18/SIRG/5597.
   A-M.L acknowledges support from the French CNRS and from the Agence Nationale de la Recherche (ANR grant GIPSE ANR-14-CE33-0018).
\end{acknowledgements}

\clearpage 
\appendix
\section{Reduction of the GRAVITY dataset}
\label{sec:reduction_appendix}

\subsection{Nomenclature and pipeline errors}
\label{sec:complexError}

The data reduction used to extract the beta Pictoris b signal from the GRAVITY observations makes heavy use of complex linear algebra, complex error formalism, and maximum likelihood estimation. To avoid confusion and mistakes, complex numbers in this appendix are underlined (e.g., $\complex{V}$), whereas real numbers are not (e.g., $X$).

Most of the GRAVITY data manipulated are quantities which depends on the wavelength $\lambda$. These quantities can be represented as vectors of size $n_\lambda$ (the number of wavelength channels) by concatenating the individual values. These vectors are denoted using a bold font. For example, in the case of the complex visibility obtained on baseline $b$ at time $t$, we denote:
\begin{equation}
\cmat{V}_{b, t} = \begin{pmatrix}\complex{V}(b, t, \lambda_1) \\ \complex{V}(b, t, \lambda_2) \\ \vdots{} \\ \complex{V}(b, t, \lambda_{n_\lambda})\end{pmatrix}
\end{equation}

For a given DIT, it is also possible to concatenate all baselines to create a vector of size $n_\mathrm{b}\times{}n_\lambda$, where  $n_\mathrm{b} = 6$ is the number of baselines. In this case, the subscript $b$ is dropped:
\begin{equation}
\cmat{V}_t = \begin{pmatrix} \cmat{V}_{b_1, t} \\ \vdots{} \\ \cmat{V}_{b_{n_b}, t} \end{pmatrix}
\end{equation}

The complex-conjugate of a complex number $\complex{V}$ is denoted $\conj{\complex{V}}$, and the complex-transpose of a vector or matrix $\cmat{A}$ is denoted $\adj{\cmat{A}}$. It is defined by: $\adj{\cmat{A}} = \transp{{\conj{\cmat{A}}}}$ where $\transp{}$ is the transpose operator.

All the $\lambda$-vectors are understood as elements of a $n_\lambda$-dimension complex linear space (i.e. a linear space for which the scalar field is the set of complex numbers $\mathbb{C}$, rather than the set of real numbers $\mathbb{R}$). Adding the natural scalar product operator (i.e. $\langle V_1, V_2\rangle = \adj{V_1}V_2$) makes this linear space an Euclidean space. This mathematical structure allows for several useful concepts: it is possible to compute othogonal projections, to use projector matrices, to define othogonal and/or orthonormal basis, etc.

The data set can be subdivided into two parts: the observations taken with the science fiber on the planet, and the observations taken on the star (see observing log in Table~\ref{tab:log}). On-planet and on-star phase-referenced visibilities are calculated from the coherent fluxes measured by GRAVITY, called VISDATA in the FITS files generated by the pipeline. The VISDATA are complex numbers, affected by noise. The GRAVITY pipeline reports these errors in another set of complex numbers, called VISERR. The real part of VISERR contains the uncertainties on the real part of VISDATA, and the imaginary part of VISERR contains the uncertainty on the imaginary part of VISDATA. These errors do not take into account any possible correlation between different spectral channels, or between the real and imaginary parts of the visibility. To take into account such correlations, it is necessary to use the covariance/pseudo-covariance formalism of complex random variables.

In our data reduction algorithm, the GRAVITY pipeline errors are systematically replaced by an empirical estimate of the covariance and pseudo-covariance matrices of the visibilities. We assume that the noise affecting the measurements does not vary significantly over the individual DITs of a single exposure file ($\sim{}5~\mathrm{min}$), but can vary from file to file. We also allow for correlations between different spectral channels and/or between different baselines. Under these assumptions, the errors on the coherent fluxes are best represented by a set of $n_\mathrm{EXP}$ (the number of exposure files) covariance matrices $\mat{W}_k$ and $n_\mathrm{EXP}$ pseudo-covariance matrices $\mat{Z}_k$, both of size $n_b\times{}n_\lambda$, where $n_b=6$ is the number of baselines and $n_\lambda = 235$ is the number of wavelength channels. The covariance and pseudo-covariance matrices for each exposure file are estimated directly from the DITs sequence:
\begin{small}
\begin{align*}
\cmat{W}_k = \frac{1}{n_{\rm DIT}-1}\left(\sum_{t=1}^{n_{\rm DIT}}\cmat{V}_{t}\adj{\cmat{V}_{t}} - \frac{1}{n_{\rm DIT}}\left(\sum_{t=1}^{n_{\rm DIT}}\cmat{V}_{t}\right)\adj{\left(\sum_{t=1}^{n_{\rm DIT}}{\cmat{V}_{t}}\right)}\right)\\
\cmat{Z}_k = \frac{1}{n_{\rm DIT}-1}\left(\sum_{t=1}^{n_{\rm DIT}}\cmat{V}_{t}\transp{\cmat{V}_{t}} - \frac{1}{n_{\rm DIT}}\left(\sum_{t=1}^{n_{\rm DIT}}\cmat{V}_{t}\right)\transp{\left(\sum_{t=1}^{n_{\rm DIT}}{\cmat{V}_{t}}\right)}\right)
\end{align*}
\end{small}
where the dummy $t$ runs over the $n_{\rm DIT}$ DITs of the $k$-th exposure. 

The covariance and pseudo-covariance matrices are always related to the covariance of the real and imaginary parts by the following equations:
\begin{align}
  \cov{\real{\cmat{V}}, \real{\cmat{V}}} &= \frac{1}{2}\,\real{\cmat{W} + \cmat{Z}}\\
  \cov{\imag{\cmat{V}}, \imag{\cmat{V}}} &= \frac{1}{2}\,\real{\cmat{W}-\cmat{Z}}\\
  \cov{\real{\cmat{V}}, \imag{\cmat{V}}} &= \frac{1}{2}\,\imag{-\cmat{W}+\cmat{Z}}\\
  \cov{\imag{\cmat{V}}, \real{\cmat{V}}} &= \frac{1}{2}\,\imag{\cmat{W}+\cmat{Z}}
  \label{eq:cov_complex_to_real}  
\end{align}

The covariance and pseudo-covariance matrices can be propagated during the data reduction algorithm by using the complex error propagation equations:
\begin{align}
  \cov{\cmat{A}\cmat{V}} &= \cmat{A}\cmat{W}\adj{\cmat{A}} \label{eq:cov_propagation}\\
  \pcov{\cmat{A}\cmat{V}} &= \cmat{A}\cmat{Z}\transp{\cmat{A}} \label{eq:pcov_propagation} 
\end{align}
\noindent{}with $\cmat{A}$ any complex matrix of appropriate size.

The $\cmat W$ and $\cmat Z$ matrices can also be used to resolve linear equations involving complex data. In the case of an unknown real parameter vector \mat{X}, the solution of the linear problem $\cmat{V} = \cmat{A}\mat{X}$ (in the sense of maximum likelihood) is:
\begin{equation}
\hat{\mat{X}} = \transp{{\left(\real{\adj{\cmat{V}_{2}}\cmat{W}_2^{-1}\cmat{A}_2}\left[\real{\adj{\cmat{A}_2}\cmat{W}_2^{-1}\cmat{A}_2}\right]^{-1}\right)}}
\label{eq:solutionComplex}
\end{equation}
where 
\begin{eqnarray}
\cmat{V}_{2}&=& \begin{pmatrix}\cmat{V}\\ \cmat{V}^* \end{pmatrix} \\
\cmat{W}_{2}&=&\begin{pmatrix} \cmat{W} & \cmat{Z}  \\  \cmat{Z}^\dag & \cmat{W}^*   \end{pmatrix} \label{eq:covaMatrix}\\
\cmat{A}_{2}&=&\begin{pmatrix} \cmat{A}\\ \cmat{A}^* \end{pmatrix}  \label{eq:covaMatrix2}
\end{eqnarray}

For a complete mathematical derivation of Eq.~(\ref{eq:solutionComplex}), we refer the reader to Appendix~B of \cite{NowakPhd}.

\subsection{Pipeline reduction and phase referencing}
\label{sec:app_pipeline} 


The initial step uses the pipeline reduction and is common to all VLTI/GRAVITY observations. It consists in extracting the complex visibilities from the raw data, using the ESO pipeline \citep{Lapeyrere2014}. The pipeline takes care of the background subtraction, flat-field correction, bad-pixel interpolation and P2VM multiplication \citep{Tatulli2007}. It also corrects the phase of the visibilities using the metrology data, and combines all DITs within each exposure.
This last step performed by the pipeline (averaging of all DITs within each exposure file) is unwanted for exoplanet observations (see Section~\ref{sec:complexError}). Thus, for the $\beta$ Pic b observations, an intermediate file product generated by the pipeline is used: the ``astrored'' files, in which all DITs are kept separate. The complex visibilities contained in the ``astrored'' files are not corrected for the metrology and fringe-tracker zero-point, and thus the correction must be applied manually (see recipe in \citeauthor{NowakPhd},~\citeyear{NowakPhd}).

For each baseline $b$, and each time $t$ (i.e. for each DIT), the wavelength-dependent complex visibility $\mathrm{VISDATA}_\onstar$ and $\mathrm{VISDATA}_\onplanet$
 extracted by the pipeline (and with the above-mentionned corrections) are then ``phase-referenced'' to the star:
\begin{align}
\complex{V}_\onstar &= |\mathrm{VISDATA}_\onstar|\\
\complex{V}_\onplanet &= \mathrm{VISDATA}_\onplanet \times{}e^{-i\arg{(\mathrm{VISDATA}_\onstar)}} \label{eq:VISDATA}
\end{align}
where $\arg{(\mathrm{VISDATA}_\onstar)}$ is the phase of the stellar complex visibility as measured by GRAVITY when the science fiber is positioned on the star.

This phase-referencing step is performed both on the star exposures, and on the planet exposures. When dealing with a star exposure, phase-referencing the visibility is mathematically equivalent to extracting the modulus of the visibility. But when dealing with on-planet exposures, a problem arises: the instrument does not simultaneously observe both the planet and the star. Thus, the quantity $\arg{(\mathrm{VISDATA}_\onstar)}$ must be estimated from the available star exposures. In the observing strategy used for acquiring the $\beta$ Pictoris data reported here, a star exposure was performed before and after each on-planet exposure. For each on-planet exposure, the phase reference is then simply estimated by taking the phase of the stellar complex visibily averaged on these two star exposures (before and after the on-planet observation).

\subsection{A model for the on-planet visibility}

In the absence of stellar flux, the on-planet visibility measured by the instrument and phase-referenced to the star can be written:
\begin{equation}
  \complex{V}_\onplanet(b, t, \lambda) = \complex{G}(b, t, \lambda)\complex{V}_\refplanet(b, t, \lambda)
  \label{eq:naive_model}  
\end{equation}
\noindent{}in which $V_\refplanet$ is the planet astrophysical visibility phase-referenced to the star, and $G$ is the instrumental response. 
We note that the visibilities are not calibrated, meaning that the visibility at zero frequency is not 1, but the un-normalized flux. Therefore, as long as the planet remains unresolved by the instrument, its astrophysical visibility is given by:
\begin{equation}
  \complex{V}_\refplanet(b, t, \lambda) = S_\planet(\lambda)\times{}e^{-i\,\frac{2\pi}{\lambda}\left(\Delta\RA\times{}U+\Delta\DEC\times{}V\right)}
  \label{eq:refplanet}
\end{equation}
\noindent{}in which $(U, V)$ are the coordinates of baseline $b$ in the UV-plane, ($\Delta\RA$, $\Delta\DEC$) the sky-coordinates of the planet relative to the star, and $S_\planet(\lambda)$ the spectrum of the planet.

Given the typical VLTI baseline lengths (between 45 and $130\,\mathrm{m}$ with the UTs), and the expected $\beta$ Pic b planet-to-star separation at time of observation ($\simeq{}140~\mathrm{mas}$), the exponential term in the above equation should produce significant oscillations of the complex visibility over the GRAVITY wavelength range (1.9 to 2.35~$\mu\mathrm{m}$). But the phase-referenced on-planet visibilities extracted from our $\beta$ Pic b observations show no such oscillations. The reason is that the data are dominated by remaining starlight, which needs to be taken into account.

To take into account the coherent starlight leaking into the fiber, Eq.~(\ref{eq:naive_model}) must be modified with an additional term, proportional to the stellar phase-referenced visibility $V_\refstar$. In practice, since the leaking starlight does not originate in the direct coupling of the star to the fiber, but rather in the coupling of speckle noise to the fiber, this term needs to be multiplied by a polynomial in $\lambda$ to account for its chromaticity. The model is now given by:
\begin{equation}
\begin{aligned}
  \complex{V}_\onplanet(b, t, \lambda) =&~ \complex{Q}(b, t, \lambda)\complex{G}(b, t, \lambda)V_\refstar(b, t, \lambda) \\
  & +\complex{G}(b, t, \lambda) \complex{V}_\refplanet(b, t, \lambda)  \label{eq:onplanet}  
\end{aligned}
\end{equation}
\noindent{}with $\lambda\rightarrow{}\complex{Q}(b, t, \lambda)$ a polynomial function in $\lambda$, whose coefficients vary with baseline $b$ and time $t$.

If the star is not resolved by the instrument, its astrophysical phase-referenced visibility corresponds to its spectrum. If the star is partially resolved by the instrument, the spectrum needs to be multiplied by a term accounting for the resulting drop in visibility, which depends on the angular size of the star, limb-darkening model, etc. Explicitly separating these two terms, the referenced stellar astrophysical visibility can be written using the following equation, in which $S_\star(\lambda)$ is the star spectrum, and $J$ a function accounting for the visibility drop due to the star geometry (typically, $J$ is a bessel function of first order):
\begin{equation}
  V_\refstar(b, t, \lambda) = S_\star(\lambda)J(b, t, \lambda)
  \label{eq:refstar}
\end{equation}

Going back to Eq.~(\ref{eq:onplanet}), the planet term in the right-hand side can be factored by $V_\star$ by introducing the planet-to-star contrast spectrum $C(\lambda)=S_\planet(\lambda)/S_\star(\lambda)$:
\begin{equation}
  \begin{aligned}
    \complex{V}_\onplanet &= \complex{Q}\complex{G}V_\refstar\\&\;+J^{-1}\complex{G}V_\refstar C(\lambda)e^{-i\frac{2\pi}{\lambda}\left(\Delta\RA U+\Delta\DEC V\right)}
\label{eq:V_model}
\end{aligned}
\end{equation}

The on-star equivalent of Eq.~(\ref{eq:onplanet}) is simpler, as the reference visibility observed on-star only depends on the stellar referenced visibility and the instrumental response:
\begin{equation}
  \begin{aligned}
    \complex{V}_\onstar(b, t, \lambda) &= \complex{G}(b, t, \lambda)V_\refstar(b, t, \lambda)
  \end{aligned}
  \label{eq:onstar}
\end{equation}

This provides a natural way to estimate the term $\complex{G}V_\refstar$ in Eq.~(\ref{eq:V_model}), and thus to calibrate $\complex{V}_\onplanet$:
\begin{equation}
\begin{aligned}
\complex{U}(b, t, \lambda) &= \frac{\complex{V}_\onplanet}{\complex{V}_\onstar}(b, t, \lambda) \\ &= \complex{Q} + J^{-1}C(\lambda)e^{-i\frac{2\pi}{\lambda}\left(\Delta\RA U+\Delta\DEC V\right)}
\label{eq:V_cal}
\end{aligned}
\end{equation}

Equation~(\ref{eq:V_cal}) shows how the physical quantities of interest (i.e. the contrast spectrum $C(\lambda)$ and the planet separation $\Delta\RA, \Delta\DEC$) are encoded in the on-planet data. Even taking into account the filtering of the starlight by the off-axis fiber, as well as the only partly coherent nature of the speckle noise, the polynomial $\complex{Q}$ in the right-hand side of Eq.~(\ref{eq:V_cal}), which model the stellar residuals, is still a factor 20 to 30 superior to the planet signal. It is only because of the phase modulation naturally introduced by the planet separation that this planet signal can be retrieved. To do so, we proceed in two steps: we first extract the star-planet separation vector under some hypothesis on the contrast spectrum, and we then extract the contrast spectrum using the estimated separation vector. The two steps are iterated on time, to check the consistency of the results.

In matrix notations, the multiplications by $\complex{G}$, $J^{-1}$, the exponential, or even the polynomial $\complex{Q}$ can all be represented by diagonal-matrix multiplications. We write:
\begin{equation}
\cmat{U}_{b, t} = \sum_{k=0}^{m}\complex{a}_{b, t, k} \mat{\Lambda}^k\mathds{1} + \mat{J}_{b, t}^{-1}\mat{\Phi}_{b, t}^{\Delta\alpha,\Delta\delta}\mat{C}\,,
\label{eq:V_cal_mat}
\end{equation}
where $m$ is the order of the polynomial $\complex{Q}$, and the $\complex{a}_k$s are complex coefficients used to describe the polynomial. The vector $\mat{C}$ is defined from $C(\lambda)$ using the notations introduced in Section~\ref{sec:complexError}, $\mathds{1}$ is a column vector filled with 1's, and the matrices $\mat{\Lambda}$, $\mat{J}^{-1}$, and $\cmat{\Phi}$ are all diagonal matrices of size $n_\lambda\times{}n_\lambda$ defined by:
\begin{equation}
\begin{aligned}
\cmat{\Lambda} &= \mathrm{diag}\left\{\lambda_1, \dots{}, \lambda_{n_\lambda}\right\} \\ 
\cmat{J}_{b, t}^{-1} &= \mathrm{diag}\left\{J(b, t, \lambda_1)^{-1}, \dots{}, J(b, t, \lambda_{n_\lambda})^{-1}\right\} \\
\cmat{\Phi}_{b, t}^{\Delta\alpha, \Delta\delta} &= \mathrm{diag}\left\{e^{-i\frac{2\pi}{\lambda_1}\left(\Delta\alpha{}U(b, t)+\Delta\delta{}V(b, t)\right)},\dots{}\right\}
\end{aligned}
\end{equation}

\subsection{Extracting the astrometry}
 \label{sec:app_astro}
 
At the initial iteration, the planet to star contrast spectrum $\mat{C}$ in Eq.~(\ref{eq:V_cal_mat}) can most generally be replaced by a flat spectrum. In the case of $\beta$ Pictoris b, the temperature and surface gravity of the planet are known from previous work \citep{Chilcote2017}. Thus, the contrast spectrum $\mat{C}$ can be set to a model value. We use a BT-Settl model \citep{Baraffe2015}, at $T=1700~\mathrm{K}$ and $\log{(g/g_0)} = 4.0$ (planet spectrum), divided by a BT-NextGen model \citep{Hauschildt1999} at $T=8000~\mathrm{K}$ and $\log(g/g_0) = 4.0$ (the star).

The contrast spectrum being set to a pre-determined value, Eq.~(\ref{eq:V_cal_mat}) becomes a model at $(m+1)\times{}n_b n_{\mathrm{DIT}}$ complex parameters (the $\complex{a}_k$s), and 2 real parameters ($\Delta\alpha$ and $\Delta\delta$).

Interestingly, this model is linear in all the $\complex{a}_k$, and nonlinear in $\Delta\alpha, \Delta\delta$. To fully benefit from this for the model-fitting, Eq.~(\ref{eq:V_cal_mat}) can be re-arranged in a pseudo matrix form, with real parameters.

\noindent{}Introducing:
\begin{equation}
\mat{x}_{b, t} = \begin{pmatrix}\real{\complex{a}_0} \\ \imag{\complex{a}_0} \\ \vdots{} \\ \real{\complex{a}_m} \\ \imag{\complex{a}_m}\end{pmatrix}
\end{equation}

\noindent{}And $\cmat{A}_{b, t}^{\Delta\alpha, \Delta\delta}$ defined column by column:
\begin{small}
\begin{equation}
\cmat{A}_{b, t}^{\Delta\alpha, \Delta\delta} = 
\left(\mat{\Lambda}^0\mathds{1}, i\mat{\Lambda}^0\mathds{1}, \dots{}, \mat{\Lambda}^m\mathds{1}, i\mat{\Lambda}^m\mathds{1}, \cmat{\Phi}_{b, t}^{\Delta\alpha, \Delta\delta}\mat{C}\right)
\label{eq:A}
\end{equation}
\end{small}

\noindent{}We have:
\begin{equation}
\cmat{U}_{b, t} = \cmat{A}_{b, t}^{\Delta\alpha, \Delta\delta}\mat{x}_{b, t}
\end{equation}

For a given $\Delta\alpha$ and $\Delta\delta$, the corresponding best estimate of $\mat{x}_{b, t}$ (in the sense of the maximum likelihood) is given by:
\begin{equation}
\begin{small}
\hat{\mat{x}}_{b, t} = \transp{{\left(\real{\adj{\cmat{U}_{b, t, 2}}\cmat{W}_2^{-1}\cmat{A}_{b, t, 2}}\left[\real{\adj{\cmat{A}_{b, t, 2}}\cmat{W}_2^{-1}\cmat{A}_{b, t, 2}}\right]^{-1}\right)}}
\label{eq:xhat}
\end{small}
\end{equation}
where $\cmat{W}_2$ is a matrix composed of the covariance and pseudo covariance matrices of $\cmat{U}_{b, t}$ as defined in Eq.~(\ref{eq:covaMatrix}).

The log-likelihood $\log{\mathcal{L}_{b, t}}(\Delta\alpha, \Delta\beta)$, restricted to baseline $b$, DIT $t$, and to the nonlinear parameter $\Delta\alpha$ and $\Delta\delta$ is then given by the following equation, in which the dependance in $\Delta\alpha, \Delta\delta$ of the right-hand side is implicit in the definition of $\cmat{A}$ and $\hat{\mat{x}}$.
\begin{equation}
-\log{\mathcal{L}_{b, t}} = \adj{\left[\cmat{U}_{b, t}-\cmat{A}_{b, t}\hat{\mat{x}}_{b, t}\right]_2}\cmat{W}_{b, t, 2}^{-1}\left[\cmat{U}_{b, t}-\cmat{A}_{b, t}\hat{\mat{x}}_{b, t}\right]_2
\end{equation}

The total log-likelihood can be obtained by summing over all baselines $b$ and DITs $t$:
\begin{equation}
-\log{\mathcal{L}}(\Delta\alpha, \Delta\delta) = -\sum_{b, t}\log{\mathcal{L}_{b, t}} (\Delta\alpha, \Delta\delta) 
\end{equation}
With the expression of $\cmat{A}_{b, t}$ from Eq.~(\ref{eq:A}) and $\hat{\mat{x}}$ from Eq.~(\ref{eq:xhat}), this gives a closed-form expression of the log-likelihood in $\Delta\alpha$, $\Delta\delta$, from which a map can be calculated, in order to extract the best estimate with the associated error bars.

\subsection{Extracting the spectrum}
 \label{sec:app_spectrum}

Extracting the spectrum from the GRAVITY observations is more difficult than extracting the astrometry for two reasons: first, due to the dimensionality of the problem ($n_\lambda > 200$), a $\log{\mathcal{L}}$ map approach is impractical; second, the stellar residuals affecting the on-planet visibility can lead to a degenerated solution for the contrast spectrum $\mat{C}$.

The impact of the stellar residuals on the contrast spectrum can be quantified by using Eq.~(\ref{eq:V_cal_mat}) again. The calculations in the rest of this section are simpler when considering all visibilities shifted to the planet position. To do so, we multiply both sides of Eq.~(\ref{eq:V_cal_mat}) by the inverse of $\cmat{\Phi}_{b, t}^{\Delta\alpha, \Delta\delta}$. From now on, we will denote $\tilde{\cmat{V}}_{b, t}$ any ``shifted'' visibility $\cmat{V}_{b, t}$. Eq.~(\ref{eq:V_cal_mat}) becomes:
\begin{equation}
\tilde{\cmat{U}}_{b, t} = \sum_{k=0}^{m}\complex{a}_{b, t, k}\mat{\Lambda}^k\cmat{\tilde{\mathds{1}}} + \mat{J}_{b, t}^{-1}\mat{C}
\end{equation}

From there, we can introduce the subspace $\mathbb{C}^m[\mat{\Lambda}]\cmat{\tilde{\mathds{1}}}$ of $\mathbb{C}^{n_\lambda}$, defined as the subspace generated by the family of $m+1$ vectors $\mat{\Lambda}^0\cmat{\tilde{\mathds{1}}}, \dots{}, \mat{\Lambda}^m\cmat{\tilde{\mathds{1}}}$ (i.e., the subspace of vectors which are linear combinations of these $m+1$ vectors). Introducing this subspace if of course motivated by the fact that the stellar residual term is part of it. We can then introduce the projector matrix orthogonal to this subspace, which we denote\footnote{The vector $\tilde{\cmat{\mathds{1}}}$ hides a dependancy in $b, t$ through the matrix $\cmat{\Phi}_{b, t}^{\Delta\alpha, \Delta\delta}$ used to define the tilded vector. As this dependency is important in the following calculation, it is made explicit by using $\tilde{\cmat{\mathds{1}}}_{b, t}$ instead} $\cmat{P}_{{\mathbb{C}^m[\mat{\Lambda}]\tilde{\cmat{\mathds{1}}}}_{b,t}}$. Projecting Eq.~(\ref{eq:V_cal_mat}) then gives:
\begin{equation}
\cmat{P}_{{\mathbb{C}^m[\mat{\Lambda}]\cmat{\tilde{\mathds{1}}}}_{b, t}}\cmat{\tilde{U}}_{b, t} = \cmat{P}_{\mathbb{C}^m[\mat{\Lambda}]\tilde{\mathds{1}}_{b, t}}\mat{J}^{-1}_{b, t}\mat{C}
\label{eq:projected_V_cal_mat}
\end{equation}

The new Eq.~(\ref{eq:projected_V_cal_mat}) is a representation of the exact information content of Eq.~(\ref{eq:V_cal_mat}) regarding the contrast spectrum. Since $\cmat{P}_{\mathbb{C}^m[\Lambda]\cmat{\tilde{\mathds{1}}}}$ is a projector matrix, it is necessarily of rank $<n_\lambda$, with an exact value which depends on the dimension of the subspace generated by the $\mat{\Lambda}^k\cmat{\tilde{\mathds{1}}}$. Thus, Eq.~(\ref{eq:projected_V_cal_mat}) is not invertible, and the contrast spectrum cannot be fully recovered from it.

Fortunately, the dataset acquired on $\beta$ Pictoris b contains several baselines and DITs. For each baseline and each DIT, the projector matrix is different, since it depends on the matrix $\cmat{\Phi}_{b, t}$ through the vector $\cmat{\tilde{\mathds{1}}}_{b, t}$. These variations can be leveraged to unambiguously recover the complete contrast spectrum $\mat{C}$.



The proper way to proceed is to start by extracting a set of linearly independant equations from Eq.~(\ref{eq:projected_V_cal_mat}). This can be done by using a diagonal representation of the projector matrix, for example using a singular value decomposition. We can introduce an hermitian matrix $\cmat{H}_{b, t}$ and a diagonal matrix $\mat{D}_{b, t}$ such that:
\begin{align}
{\cmat{H}_{b, t}}\cmat{P}_{{\mathbb{C}^m[\Lambda]\cmat{\tilde{\mathds{1}}}}_{b, t}}\adj{\cmat{H}_{b, t}} &= \mat{D}_{b, t}\\
\cmat{H}_{b, t}\adj{\cmat{H}_{b, t}} &= \mat{I}
\end{align}
\noindent{}Since $\cmat{P}_{{\mathbb{C}^m[\Lambda]\cmat{\tilde{\mathds{1}}}}_{b, t}}$ is a projector matrix, its eigenvalues are either $1$ or $0$. We can assume $\mat{D}$ to be of the form:
\begin{equation}
\mat{D}_{b, t} = \begin{pmatrix}
\mat{I}_{r(b, t)} & 0 \\
0 & 0
\end{pmatrix}
\end{equation}
\noindent{}where $\mat{I}_r(b, t)$ is the identity matrix of size given by the rank of the projector: $r(b, t) = \mathrm{rank}(\cmat{P}_{{\mathbb{C}^m[\Lambda]\cmat{\tilde{\mathds{1}}}}_{b, t}})$.

The matrix $\cmat{H}_{b, t}$ can also be written in blocks:
\begin{equation}
\cmat{H}_{b, t} = \begin{pmatrix}
\cmat{H}^{11}_{b, t} & \cmat{H}^{12}_{b, t} \\
\cmat{H}^{21}_{b, t} & \cmat{H}^{22}_{b, t}
\end{pmatrix}
\end{equation}

From there, multiplying both sides of Eq.~(\ref{eq:projected_V_cal_mat}) by $\cmat{H}_{b,t}$, noting that $\cmat{H}_{b, t}\cmat{P}_{{\mathbb{C}^m[\Lambda]\cmat{\tilde{\mathds{1}}}}_{b, t}} = \mat{D}_{b, t}\cmat{H}_{b, t}$, and using a block calculation gives:
\begin{equation}
\begin{bmatrix}\cmat{H}^{11}_{b, t} & \cmat{H}^{12}_{b, t}\end{bmatrix}\tilde{\cmat{U}}_{b, t} = \begin{bmatrix}\cmat{H}^{11}_{b, t} & \cmat{H}^{12}_{b, t}\end{bmatrix}\mat{J}_{b, t}\mat{C}
\label{eq:HU=HC}
\end{equation}
\noindent{}and a dummy equation $0 = 0$.

Since $r(b, t) < n_\lambda$, the linearly independent system defined by Eq.~(\ref{eq:HU=HC}) is underdetermined (the matrix $\begin{bmatrix}\cmat{H}_{b, t}^{11} & \cmat{H}_{b, t}^{12}\end{bmatrix}$ has more columns than rows).

To invert the problem, it is necessary to combine all the equations obtained for the different baselines $b$ and $t$. In matrix notation, this is just a matter of concatenating all the sub-matrices:
\begin{small}
\begin{equation}
\cmat{H} = \begin{pmatrix}
\begin{bmatrix}\cmat{H}_{1, 1}^{11} & \cmat{H}_{1, 1}^{12}\end{bmatrix} & & 0 \\
 & \ddots{} &  \\
0 & & \begin{bmatrix}\cmat{H}_{n_b, n_\mathrm{DIT}}^{11} & \cmat{H}_{n_b, n_\mathrm{DIT}}^{12}\end{bmatrix}
\end{pmatrix}
\end{equation}
\end{small}

\begin{equation}
\tilde{\cmathbb{U}} = \begin{pmatrix}
\tilde{\cmat{U}}_{1, 1} \\
\tilde{\cmat{U}}_{2, 1} \\
\vdots{} \\
\tilde{\cmat{U}}_{n_b, n_\lambda}
\end{pmatrix}
\end{equation}

\begin{equation}
\mathbb{J} = \begin{pmatrix}
{\mat{J}_{1,1}}^{-1} \\
{\mat{J}_{2,1}}^{-1} \\
\vdots{} \\
{\mat{J}_{n_b,n_\lambda}}^{-1} \\
\end{pmatrix}
\end{equation}

\noindent{}This gives:
\begin{equation}
\cmat{H}\tilde{\cmathbb{U}} = \cmat{H}\mathbb{J}\mat{C}
\end{equation}

\noindent{}Which has the form of a linear problem:
\begin{equation}
\cmathbb{Y} = \cmathbb{H}\mat{C}
\label{eq:Y=HC}
\end{equation}

\noindent{}Where $\cmathbb{Y}$ is a linear transformation of the calibrated visibilities defied by:
\begin{equation}
\mathbb{Y} = \cmat{H}\tilde{\cmathbb{U}}
\end{equation}

\noindent{}And $\cmathbb{H}$ is the collapsed matrix:
\begin{equation}
\cmathbb{H} = \begin{pmatrix}
\begin{bmatrix}\cmat{H}_{1, 1}^{11} & \cmat{H}_{1, 1}^{12}\end{bmatrix}\mat{J}^{-1}_{1, 1} \\
\begin{bmatrix}\cmat{H}_{2, 1}^{11} & \cmat{H}_{2, 1}^{12}\end{bmatrix}\mat{J}^{-1}_{2, 1} \\
\vdots{}\\
\begin{bmatrix}\cmat{H}_{n_b, n_\mathrm{DIT}}^{11} & \cmat{H}_{n_b, n_\mathrm{DIT}}^{12}\end{bmatrix}\mat{J}^{-1}_{n_b, n_\mathrm{DIT}}
\end{pmatrix}
\end{equation}

The problem defined by Eq.~(\ref{eq:Y=HC}) can be solved using the maximum likelihood formalism, adapted to complex random variables. The expression of the maximum likelihood solution is given by Eq.~(\ref{eq:solutionComplex}). For the contrast spectrum, we have:
\begin{equation}
\hat{\mat{C}} = \transp{\left(\real{\adj{\tilde{\mathbb{Y}}}_2\cmat{W}_2^{-1}\cmathbb{H}_2}\left[\real{\adj{\cmathbb{H}}_2\cmat{W}_2^{-1}\cmathbb{H}_2}\right]^{-1}\right)}
\end{equation}
\noindent{}where the extended vectors and matrices $\tilde{\cmat{Y}}_2$, $\cmathbb{H}_2$, and $\cmat{W}_2$, are defined by:
\begin{equation}
\tilde{\cmat{Y}_2} = \begin{pmatrix}\tilde{\cmat{Y}} \\ \conj{\tilde{\cmat{Y}}} \end{pmatrix}
\quad
{\cmathbb{H}_2} = \begin{pmatrix}{\cmathbb{H}} \\ \conj{{\cmathbb{H}}} \end{pmatrix}
\quad
{\cmat{W}_2} = \begin{pmatrix}{\cmat{W}} & {\cmat{Z}}\\ \adj{{\cmat{Z}}} & \conj{{\cmat{W}}} \end{pmatrix}
\end{equation}

The uncertainty on this best estimate of the contrast spectrum can be obtained with a direct error propagation all the way to the real covariance matrix on $\hat{\mat{C}}$. The covariance and pseudo covariance matrices of $\adj{\cmathbb{Y}_2}\cmat{W}_2^{-1}\cmathbb{H}_2$ are given by:
\begin{align}
\cov{\adj{\cmathbb{Y}_2}\cmat{W}_2^{-1}\cmathbb{H}_2} &= \adj{\cmathbb{H}_2}\adj{{\cmat{W}_2^{-1}}}\cov{\cmathbb{Y}_2}\cmat{W}_2^{-1}\cmathbb{H}_2\\
\pcov{\adj{\cmathbb{Y}_2}\cmat{W}_2^{-1}\cmathbb{H}_2} &= 
\transp{\cmathbb{H}_2}\transp{{\cmat{W}_2^{-1}}}\pcov{\cmathbb{Y}_2}\cmat{W}_2^{-1}\cmathbb{H}_2
\end{align}

\noindent{}A proper combination of these covariance and pseudo-covariance matrices gives the covariance matrix of the real part of ${{{\cmathbb{Y}_2}\cmat{W}_2^{-1}\cmathbb{H}_2}}$:
\begin{equation}
\begin{aligned}
& \mathrm{cov}\left\{\real{{\adj{\cmathbb{Y}_2}\cmat{W}_2^{-1}\cmathbb{H}_2}}\right\} = \\
&\frac{1}{2}\real{\cov{\adj{\cmathbb{Y}_2}\cmat{W}_2^{-1}\cmathbb{H}_2}
+ \pcov{\adj{\cmathbb{Y}_2}\cmat{W}_2^{-1}\cmathbb{H}_2}}
\end{aligned}
\end{equation}

\noindent{}Which can then be propagated to give the final full covariance matrix on $\hat{\mat{C}}$:
\begin{footnotesize}
\begin{equation}
\begin{aligned}
& \cov{\hat{\mat{C}}} = \\
& \transp{{\real{\adj{\cmathbb{H}_2}\cmat{W}_2^{-1}\cmathbb{H}_2}^{-1}}}\cov{\real{\adj{\tilde{\mathbb{Y}}}_2\cmat{W}_2^{-1}\cmathbb{H}_2}}{\real{\adj{\cmathbb{H}_2}\cmat{W}_2^{-1}\cmathbb{H}_2}}^{-1}
\end{aligned}
\end{equation}
\end{footnotesize}

\section{Posterior of the petitRADTRANS fit}
\label{sec:pRT_posterior_appendix}
\begin{figure*}
  \includegraphics[width=\linewidth]{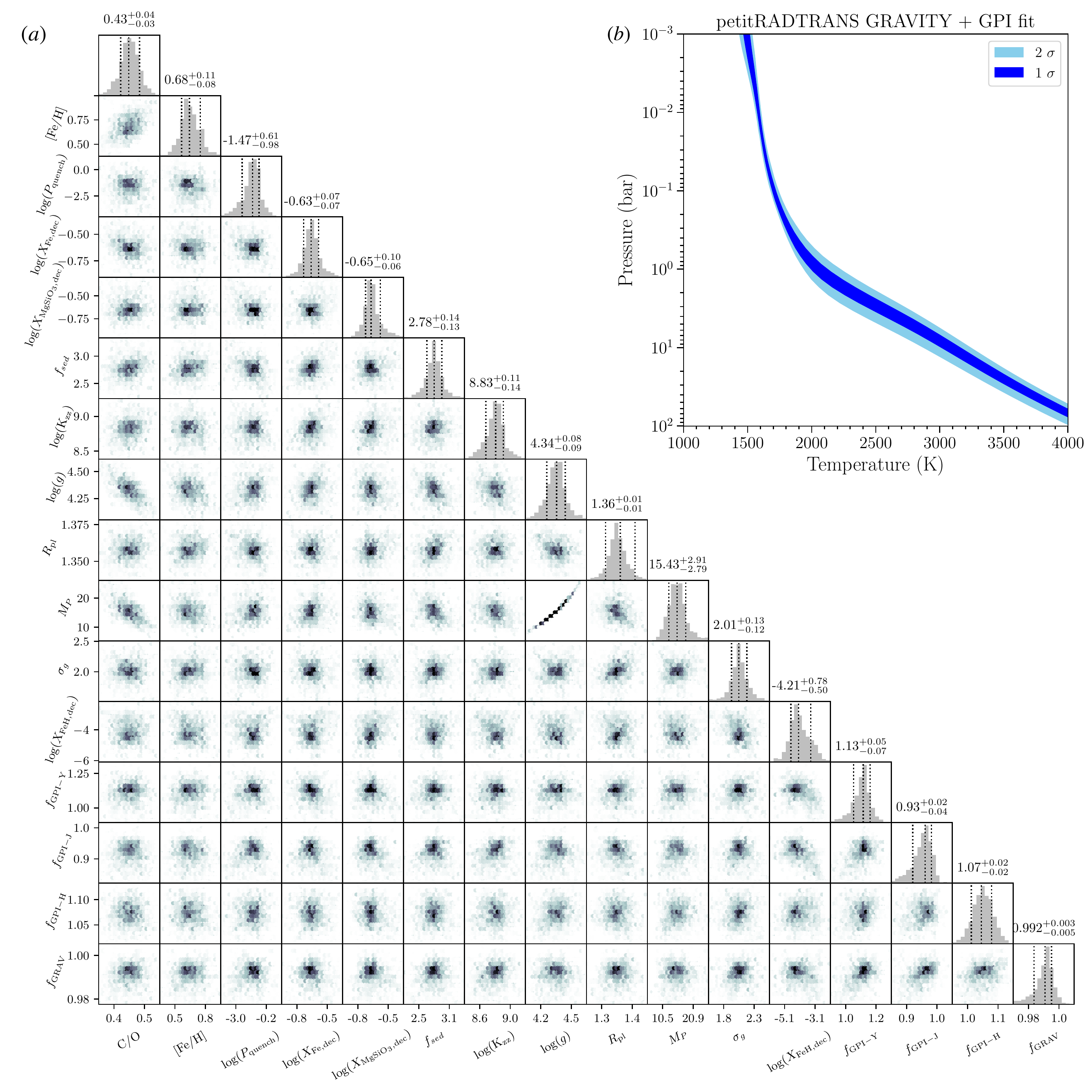}
  \caption{{\it Panel (a)}: projected 2-d posterior of the GRAVITY+GPI fit with petitRADTRANS (spectrum shown in Figure \ref{fig:pRT_fit_spec}), described in Section \ref{sec:pRT_fit}. See the text in 
  Section \ref{sec:pRT_posterior_appendix} for a description of the parameters. {\it Panel (b)}: pressure-temperature envelopes obtained for the same retrieval. At every pressure, we plot the 16 to 84-percentile envelopes in dark blue, and the 2.5 to 97.5~percentile envelopes in light blue. If the temperature values were following a Gauss distribution, this would correspond to the 1 and 2 $\sigma$ envelopes, respectively.}
  \label{fig:pRT_posteriors}
\end{figure*}

In Figure \ref{fig:pRT_posteriors} we show the posteriors of the GRAVITY+GPI fit with petitRADTRANS, described in Section \ref{sec:pRT_fit}. Panel $(a)$ shows the corner plot for all but the temperature nuisance parameters. Panel $(b)$ shows the retrieved temperature uncertainty envelopes.

The parameters shown are the following: the C/O, adjusted by changing the oxygen abundance at a given [Fe/H]. The metallicity [Fe/H], which was used to scale the number fraction of all atomic elements (except H and He) by $10^{\rm [Fe/H]}$. The quench pressure $P_{\rm quench}$ of the atmosphere, here converged to a low enough value such that non-equilibrium chemistry effects are negligible. The mass fractions of Fe and MgSiO$_3$ at the cloud base. Here they are expressed in units of a log-ed decrease factor, which gets multiplied with the maximally allowed mass fraction, based on the elemental composition of the atmosphere. The stoichiometric factors of MgSiO$_3$ are used for finding this upper limit. The cloud settling parameter $f_{\rm sed}$, as described in Section \ref{sec:pRT_fit}. The log-ed eddy diffusion coefficient $K_{zz}$, in units of cm$^2$~s$^{-1}$. This is used for calculating the cloud particle size, as described in Section \ref{sec:pRT_fit}. The planet's surface gravity $\log(g)$. The planetary radius $R_{\rm P}$, in units of Jupiter radii. The planetary mass $M_{\rm P}$, in units of Jupiter masses. This is calculated using the sampled $\log(g)$ and $R_{\rm P}$ values. The width of the log-normal cloud particle size distribution $\sigma_g$, as described in Section \ref{sec:pRT_fit}. The vertically constant mass fraction of FeH, expressed in units of a log-ed decrease factor. This factor gets multiplied with the maximally allowed mass fraction of FeH, based on the elemental composition of the atmosphere and the Fe atoms not yet incorporated into the Fe clouds. Finally the $f_{\rm GPI-Y}$, $f_{\rm GPI-J}$, $f_{\rm GPI-H}$ and $f_{\rm GRAV}$ factors describe the multiplicative scaling of the individual bands, which were allowed to vary by 50~\% in the case of GPI, and by 5~\% in the case of GRAVITY.

The temperature envelopes in Panel $(b)$ of Figure \ref{fig:pRT_posteriors} are obtained by plotting, at every pressure, the 16 to 84-percentile envelopes, and the 2.5 to 97.5~percentile envelopes. If the temperature values at a given pressure layer were following a normal distribution, this would correspond to the 1 and 2 $\sigma$ envelopes, respectively. 

\bibliographystyle{aa}
\bibliography{biblio}

\begin{thebibliography}{106}
\expandafter\ifx\csname natexlab\endcsname\relax\def\natexlab#1{#1}\fi

\bibitem[{{Ackerman} \& {Marley}(2001)}]{Ackerman2001}
{Ackerman}, A.~S. \& {Marley}, M.~S. 2001, \apj, 556, 872

\bibitem[{{Ali-Dib} {et~al.}(2014){Ali-Dib}, {Mousis}, {Petit}, \&
  {Lunine}}]{Ali-Dib2014}
{Ali-Dib}, M., {Mousis}, O., {Petit}, J.-M., \& {Lunine}, J.~I. 2014, \apj,
  785, 125

\bibitem[{{Alibert} {et~al.}(2005){Alibert}, {Mousis}, {Mordasini}, \&
  {Benz}}]{Alibert2005}
{Alibert}, Y., {Mousis}, O., {Mordasini}, C., \& {Benz}, W. 2005, \apjl, 626,
  L57

\bibitem[{{Allard} {et~al.}(2001){Allard}, {Hauschildt}, {Alexander},
  {Tamanai}, \& {Schweitzer}}]{Allard2001}
{Allard}, F., {Hauschildt}, P.~H., {Alexander}, D.~R., {Tamanai}, A., \&
  {Schweitzer}, A. 2001, \apj, 556, 357

\bibitem[{{Allard} {et~al.}(2012){Allard}, {Homeier}, \&
  {Freytag}}]{Allard2012}
{Allard}, F., {Homeier}, D., \& {Freytag}, B. 2012, Philosophical Transactions
  of the Royal Society of London Series A, 370, 2765

\bibitem[{{Andrews} \& {Williams}(2005)}]{Andrews2005}
{Andrews}, S.~M. \& {Williams}, J.~P. 2005, \apjl, 619, L175

\bibitem[{{Ansdell} {et~al.}(2017){Ansdell}, {Williams}, {Manara}, {Miotello},
  {Facchini}, {van der Marel}, {Testi}, \& {van Dishoeck}}]{Ansdell2017}
{Ansdell}, M., {Williams}, J.~P., {Manara}, C.~F., {et~al.} 2017, \aj, 153, 240

\bibitem[{{Asplund}(2005)}]{Asplund2005}
{Asplund}, M. 2005, \araa, 43, 481

\bibitem[{{Asplund} {et~al.}(2009){Asplund}, {Grevesse}, {Sauval}, \&
  {Scott}}]{Asplund2009}
{Asplund}, M., {Grevesse}, N., {Sauval}, A.~J., \& {Scott}, P. 2009, \araa, 47,
  481

\bibitem[{{Baraffe} {et~al.}(2003){Baraffe}, {Chabrier}, {Barman}, {Allard}, \&
  {Hauschildt}}]{Baraffe2003}
{Baraffe}, I., {Chabrier}, G., {Barman}, T.~S., {Allard}, F., \& {Hauschildt},
  P.~H. 2003, \aap, 402, 701

\bibitem[{{Baraffe} {et~al.}(2015){Baraffe}, {Homeier}, {Allard}, \&
  {Chabrier}}]{Baraffe2015}
{Baraffe}, I., {Homeier}, D., {Allard}, F., \& {Chabrier}, G. 2015, \aap, 577,
  A42

\bibitem[{{Baudino} {et~al.}(2015){Baudino}, {B{\'e}zard}, {Boccaletti},
  {Bonnefoy}, {Lagrange}, \& {Galicher}}]{Baudino2015}
{Baudino}, J.-L., {B{\'e}zard}, B., {Boccaletti}, A., {et~al.} 2015, \aap, 582,
  A83

\bibitem[{{Baudino} {et~al.}(2017){Baudino}, {Molli{\`e}re}, {Venot},
  {Tremblin}, {B{\'e}zard}, \& {Lagage}}]{Baudino2017}
{Baudino}, J.-L., {Molli{\`e}re}, P., {Venot}, O., {et~al.} 2017, \apj, 850,
  150

\bibitem[{{B{\'e}jar} \& {Mart{\'\i}n}(2018)}]{Bejar2018}
{B{\'e}jar}, V.~J.~S. \& {Mart{\'\i}n}, E.~L. 2018, {Brown Dwarfs and
  Free-Floating Planets in Young Stellar Clusters}, 92

\bibitem[{{Bell} {et~al.}(2015){Bell}, {Mamajek}, \& {Naylor}}]{Bell2015}
{Bell}, C. P.~M., {Mamajek}, E.~E., \& {Naylor}, T. 2015, \mnras, 454, 593

\bibitem[{{Blunt} {et~al.}(2019){Blunt}, {Wang}, {Angelo}, {Ngo}, {Cody}, {De
  Rosa}, {Graham}, {Hirsch}, {Nagpal}, {Nielsen}, {Pearce}, {Rice}, \&
  {Tejada}}]{Blunt2019}
{Blunt}, S., {Wang}, J., {Angelo}, I., {et~al.} 2019, arXiv e-prints,
  arXiv:1910.01756

\bibitem[{{Bodenheimer}(1974)}]{Bodenheimer1974}
{Bodenheimer}, P. 1974, \icarus, 23, 319

\bibitem[{{Bodenheimer} {et~al.}(1980){Bodenheimer}, {Grossman}, {Decampli},
  {Marcy}, \& {Pollack}}]{Bodenheimer1980}
{Bodenheimer}, P., {Grossman}, A.~S., {Decampli}, W.~M., {Marcy}, G., \&
  {Pollack}, J.~B. 1980, \icarus, 41, 293

\bibitem[{{Bonnefoy} {et~al.}(2013){Bonnefoy}, {Boccaletti}, {Lagrange},
  {Allard}, {Mordasini}, {Beust}, {Chauvin}, {Girard}, {Homeier}, {Apai},
  {Lacour}, \& {Rouan}}]{Bonnefoy2013}
{Bonnefoy}, M., {Boccaletti}, A., {Lagrange}, A.-M., {et~al.} 2013, \aap, 555,
  A107

\bibitem[{{Bonnefoy} {et~al.}(2014){Bonnefoy}, {Marleau}, {Galicher}, {Beust},
  {Lagrange}, {Baudino}, {Chauvin}, {Borgniet}, {Meunier}, {Rameau},
  {Boccaletti}, {Cumming}, {Helling}, {Homeier}, {Allard}, \&
  {Delorme}}]{Bonnefoy2014}
{Bonnefoy}, M., {Marleau}, G.-D., {Galicher}, R., {et~al.} 2014, \aap, 567, L9

\bibitem[{{Brandt}(2018)}]{Brandt2018}
{Brandt}, T.~D. 2018, \apjs, 239, 31

\bibitem[{{Burningham} {et~al.}(2017){Burningham}, {Marley}, {Line}, {Lupu},
  {Visscher}, {Morley}, {Saumon}, \& {Freedman}}]{Burningham2017}
{Burningham}, B., {Marley}, M.~S., {Line}, M.~R., {et~al.} 2017, \mnras, 470,
  1177

\bibitem[{{Chabrier} {et~al.}(2000){Chabrier}, {Baraffe}, {Allard}, \&
  {Hauschildt}}]{Chabrier2000}
{Chabrier}, G., {Baraffe}, I., {Allard}, F., \& {Hauschildt}, P. 2000, \apj,
  542, 464

\bibitem[{{Charnay} {et~al.}(2018){Charnay}, {B{\'e}zard}, {Baudino},
  {Bonnefoy}, {Boccaletti}, \& {Galicher}}]{Charnay2018}
{Charnay}, B., {B{\'e}zard}, B., {Baudino}, J.-L., {et~al.} 2018, \apj, 854,
  172

\bibitem[{{Chauvin} {et~al.}(2012){Chauvin}, {Lagrange}, {Beust}, {Bonnefoy},
  {Boccaletti}, {Apai}, {Allard}, {Ehrenreich}, {Girard}, {Mouillet}, \&
  {Rouan}}]{Chauvin2012}
{Chauvin}, G., {Lagrange}, A.~M., {Beust}, H., {et~al.} 2012, \aap, 542, A41

\bibitem[{{Chilcote} {et~al.}(2015){Chilcote}, {Barman}, {Fitzgerald},
  {Graham}, {Larkin}, {Macintosh}, {Bauman}, {Burrows}, {Cardwell}, {De Rosa},
  {Dillon}, {Doyon}, {Dunn}, {Erikson}, {Gavel}, {Goodsell}, {Hartung},
  {Hibon}, {Ingraham}, {Kalas}, {Konopacky}, {Maire}, {Marchis}, {Marley},
  {Marois}, {Millar-Blanchaer}, {Morzinski}, {Norton}, {Oppenheimer}, {Palmer},
  {Patience}, {Perrin}, {Poyneer}, {Pueyo}, {Rantakyr{\"o}}, {Sadakuni},
  {Saddlemyer}, {Savransky}, {Serio}, {Sivaramakrishnan}, {Song}, {Soummer},
  {Thomas}, {Wallace}, {Wiktorowicz}, \& {Wolff}}]{Chilcote2015}
{Chilcote}, J., {Barman}, T., {Fitzgerald}, M.~P., {et~al.} 2015, \apjl, 798,
  L3

\bibitem[{{Chilcote} {et~al.}(2017){Chilcote}, {Pueyo}, {De Rosa}, {Vargas},
  {Macintosh}, {Bailey}, {Barman}, {Bauman}, {Bruzzone}, {Bulger}, {Burrows},
  {Cardwell}, {Chen}, {Cotten}, {Dillon}, {Doyon}, {Draper}, {Duch{\^e}ne},
  {Dunn}, {Erikson}, {Fitzgerald}, {Follette}, {Gavel}, {Goodsell}, {Graham},
  {Greenbaum}, {Hartung}, {Hibon}, {Hung}, {Ingraham}, {Kalas}, {Konopacky},
  {Larkin}, {Maire}, {Marchis}, {Marley}, {Marois}, {Metchev},
  {Millar-Blanchaer}, {Morzinski}, {Nielsen}, {Norton}, {Oppenheimer},
  {Palmer}, {Patience}, {Perrin}, {Poyneer}, {Rajan}, {Rameau},
  {Rantakyr{\"o}}, {Sadakuni}, {Saddlemyer}, {Savransky}, {Schneider}, {Serio},
  {Sivaramakrishnan}, {Song}, {Soummer}, {Thomas}, {Wallace}, {Wang},
  {Ward-Duong}, {Wiktorowicz}, \& {Wolff}}]{Chilcote2017}
{Chilcote}, J., {Pueyo}, L., {De Rosa}, R.~J., {et~al.} 2017, \aj, 153, 182

\bibitem[{{Cridland} {et~al.}(2016){Cridland}, {Pudritz}, \&
  {Alessi}}]{Cridland2016}
{Cridland}, A.~J., {Pudritz}, R.~E., \& {Alessi}, M. 2016, \mnras, 461, 3274

\bibitem[{{Decampli} \& {Cameron}(1979)}]{DeCampli1979}
{Decampli}, W.~M. \& {Cameron}, A.~G.~W. 1979, \icarus, 38, 367

\bibitem[{{Dupuy} {et~al.}(2019){Dupuy}, {Brandt}, {Kratter}, \&
  {Bowler}}]{Dupuy2019}
{Dupuy}, T.~J., {Brandt}, T.~D., {Kratter}, K.~M., \& {Bowler}, B.~P. 2019,
  \apjl, 871, L4

\bibitem[{{Eistrup} {et~al.}(2016){Eistrup}, {Walsh}, \& {van
  Dishoeck}}]{Eistrup2016}
{Eistrup}, C., {Walsh}, C., \& {van Dishoeck}, E.~F. 2016, \aap, 595, A83

\bibitem[{{Eistrup} {et~al.}(2018){Eistrup}, {Walsh}, \& {van
  Dishoeck}}]{Eistrup2018}
{Eistrup}, C., {Walsh}, C., \& {van Dishoeck}, E.~F. 2018, \aap, 613, A14

\bibitem[{{Feroz} \& {Hobson}(2008)}]{Feroz2008}
{Feroz}, F. \& {Hobson}, M.~P. 2008, \mnras, 384, 449

\bibitem[{{Feroz} {et~al.}(2009){Feroz}, {Hobson}, \& {Bridges}}]{Feroz2009}
{Feroz}, F., {Hobson}, M.~P., \& {Bridges}, M. 2009, \mnras, 398, 1601

\bibitem[{{Foreman-Mackey} {et~al.}(2013){Foreman-Mackey}, {Hogg}, {Lang}, \&
  {Goodman}}]{Foreman-Mackey2013}
{Foreman-Mackey}, D., {Hogg}, D.~W., {Lang}, D., \& {Goodman}, J. 2013, \pasp,
  125, 306

\bibitem[{{Gaia Collaboration} {et~al.}(2018){Gaia Collaboration}, {Brown},
  {Vallenari}, {Prusti}, {de Bruijne}, {Babusiaux}, {Bailer-Jones}, {Biermann},
  {Evans}, {Eyer}, {Jansen}, {Jordi}, {Klioner}, {Lammers}, {Lindegren},
  {Luri}, {Mignard}, {Panem}, {Pourbaix}, {Randich}, {Sartoretti}, {Siddiqui},
  {Soubiran}, {van Leeuwen}, {Walton}, {Arenou}, {Bastian}, {Cropper},
  {Drimmel}, {Katz}, {Lattanzi}, {Bakker}, {Cacciari}, {Casta{\~n}eda},
  {Chaoul}, {Cheek}, {De Angeli}, {Fabricius}, {Guerra}, {Holl}, {Masana},
  {Messineo}, {Mowlavi}, {Nienartowicz}, {Panuzzo}, {Portell}, {Riello},
  {Seabroke}, {Tanga}, {Th{\'e}venin}, {Gracia-Abril}, {Comoretto},
  {Garcia-Reinaldos}, {Teyssier}, {Altmann}, {Andrae}, {Audard},
  {Bellas-Velidis}, {Benson}, {Berthier}, {Blomme}, {Burgess}, {Busso},
  {Carry}, {Cellino}, {Clementini}, {Clotet}, {Creevey}, {Davidson}, {De
  Ridder}, {Delchambre}, {Dell'Oro}, {Ducourant},
  {Fern{\'a}ndez-Hern{\'a}ndez}, {Fouesneau}, {Fr{\'e}mat}, {Galluccio},
  {Garc{\'\i}a-Torres}, {Gonz{\'a}lez-N{\'u}{\~n}ez}, {Gonz{\'a}lez-Vidal},
  {Gosset}, {Guy}, {Halbwachs}, {Hambly}, {Harrison}, {Hern{\'a}ndez},
  {Hestroffer}, {Hodgkin}, {Hutton}, {Jasniewicz}, {Jean-Antoine-Piccolo},
  {Jordan}, {Korn}, {Krone-Martins}, {Lanzafame}, {Lebzelter}, {L{\"o}ffler},
  {Manteiga}, {Marrese}, {Mart{\'\i}n-Fleitas}, {Moitinho}, {Mora}, {Muinonen},
  {Osinde}, {Pancino}, {Pauwels}, {Petit}, {Recio-Blanco}, {Richards},
  {Rimoldini}, {Robin}, {Sarro}, {Siopis}, {Smith}, {Sozzetti}, {S{\"u}veges},
  {Torra}, {van Reeven}, {Abbas}, {Abreu Aramburu}, {Accart}, {Aerts},
  {Altavilla}, {{\'A}lvarez}, {Alvarez}, {Alves}, {Anderson}, {Andrei},
  {Anglada Varela}, {Antiche}, {Antoja}, {Arcay}, {Astraatmadja}, {Bach},
  {Baker}, {Balaguer-N{\'u}{\~n}ez}, {Balm}, {Barache}, {Barata}, {Barbato},
  {Barblan}, {Barklem}, {Barrado}, {Barros}, {Barstow}, {Bartholom{\'e}
  Mu{\~n}oz}, {Bassilana}, {Becciani}, {Bellazzini}, {Berihuete}, {Bertone},
  {Bianchi}, {Bienaym{\'e}}, {Blanco-Cuaresma}, {Boch}, {Boeche}, {Bombrun},
  {Borrachero}, {Bossini}, {Bouquillon}, {Bourda}, {Bragaglia}, {Bramante},
  {Breddels}, {Bressan}, {Brouillet}, {Br{\"u}semeister}, {Brugaletta},
  {Bucciarelli}, {Burlacu}, {Busonero}, {Butkevich}, {Buzzi}, {Caffau},
  {Cancelliere}, {Cannizzaro}, {Cantat-Gaudin}, {Carballo}, {Carlucci},
  {Carrasco}, {Casamiquela}, {Castellani}, {Castro-Ginard}, {Charlot},
  {Chemin}, {Chiavassa}, {Cocozza}, {Costigan}, {Cowell}, {Crifo}, {Crosta},
  {Crowley}, {Cuypers}, {Dafonte}, {Damerdji}, {Dapergolas}, {David}, {David},
  {de Laverny}, {De Luise}, {De March}, {de Martino}, {de Souza}, {de Torres},
  {Debosscher}, {del Pozo}, {Delbo}, {Delgado}, {Delgado}, {Di Matteo},
  {Diakite}, {Diener}, {Distefano}, {Dolding}, {Drazinos}, {Dur{\'a}n},
  {Edvardsson}, {Enke}, {Eriksson}, {Esquej}, {Eynard Bontemps}, {Fabre},
  {Fabrizio}, {Faigler}, {Falc{\~a}o}, {Farr{\`a}s Casas}, {Federici},
  {Fedorets}, {Fernique}, {Figueras}, {Filippi}, {Findeisen}, {Fonti},
  {Fraile}, {Fraser}, {Fr{\'e}zouls}, {Gai}, {Galleti}, {Garabato},
  {Garc{\'\i}a-Sedano}, {Garofalo}, {Garralda}, {Gavel}, {Gavras}, {Gerssen},
  {Geyer}, {Giacobbe}, {Gilmore}, {Girona}, {Giuffrida}, {Glass}, {Gomes},
  {Granvik}, {Gueguen}, {Guerrier}, {Guiraud}, {Guti{\'e}rrez-S{\'a}nchez},
  {Haigron}, {Hatzidimitriou}, {Hauser}, {Haywood}, {Heiter}, {Helmi}, {Heu},
  {Hilger}, {Hobbs}, {Hofmann}, {Holland}, {Huckle}, {Hypki}, {Icardi},
  {Jan{\ss}en}, {Jevardat de Fombelle}, {Jonker}, {Juh{\'a}sz}, {Julbe},
  {Karampelas}, {Kewley}, {Klar}, {Kochoska}, {Kohley}, {Kolenberg},
  {Kontizas}, {Kontizas}, {Koposov}, {Kordopatis}, {Kostrzewa-Rutkowska},
  {Koubsky}, {Lambert}, {Lanza}, {Lasne}, {Lavigne}, {Le Fustec}, {Le
  Poncin-Lafitte}, {Lebreton}, {Leccia}, {Leclerc}, {Lecoeur-Taibi},
  {Lenhardt}, {Leroux}, {Liao}, {Licata}, {Lindstr{\o}m}, {Lister}, {Livanou},
  {Lobel}, {L{\'o}pez}, {Managau}, {Mann}, {Mantelet}, {Marchal}, {Marchant},
  {Marconi}, {Marinoni}, {Marschalk{\'o}}, {Marshall}, {Martino}, {Marton},
  {Mary}, {Massari}, {Matijevi{\v{c}}}, {Mazeh}, {McMillan}, {Messina},
  {Michalik}, {Millar}, {Molina}, {Molinaro}, {Moln{\'a}r}, {Montegriffo},
  {Mor}, {Morbidelli}, {Morel}, {Morris}, {Mulone}, {Muraveva}, {Musella},
  {Nelemans}, {Nicastro}, {Noval}, {O'Mullane}, {Ord{\'e}novic},
  {Ord{\'o}{\~n}ez-Blanco}, {Osborne}, {Pagani}, {Pagano}, {Pailler},
  {Palacin}, {Palaversa}, {Panahi}, {Pawlak}, {Piersimoni}, {Pineau}, {Plachy},
  {Plum}, {Poggio}, {Poujoulet}, {Pr{\v{s}}a}, {Pulone}, {Racero}, {Ragaini},
  {Rambaux}, {Ramos-Lerate}, {Regibo}, {Reyl{\'e}}, {Riclet}, {Ripepi}, {Riva},
  {Rivard}, {Rixon}, {Roegiers}, {Roelens}, {Romero-G{\'o}mez}, {Rowell},
  {Royer}, {Ruiz-Dern}, {Sadowski}, {Sagrist{\`a} Sell{\'e}s}, {Sahlmann},
  {Salgado}, {Salguero}, {Sanna}, {Santana-Ros}, {Sarasso}, {Savietto},
  {Schultheis}, {Sciacca}, {Segol}, {Segovia}, {S{\'e}gransan}, {Shih},
  {Siltala}, {Silva}, {Smart}, {Smith}, {Solano}, {Solitro}, {Sordo}, {Soria
  Nieto}, {Souchay}, {Spagna}, {Spoto}, {Stampa}, {Steele},
  {Steidelm{\"u}ller}, {Stephenson}, {Stoev}, {Suess}, {Surdej}, {Szabados},
  {Szegedi-Elek}, {Tapiador}, {Taris}, {Tauran}, {Taylor}, {Teixeira},
  {Terrett}, {Teyssand ier}, {Thuillot}, {Titarenko}, {Torra Clotet}, {Turon},
  {Ulla}, {Utrilla}, {Uzzi}, {Vaillant}, {Valentini}, {Valette}, {van Elteren},
  {Van Hemelryck}, {van Leeuwen}, {Vaschetto}, {Vecchiato}, {Veljanoski},
  {Viala}, {Vicente}, {Vogt}, {von Essen}, {Voss}, {Votruba}, {Voutsinas},
  {Walmsley}, {Weiler}, {Wertz}, {Wevers}, {Wyrzykowski}, {Yoldas},
  {{\v{Z}}erjal}, {Ziaeepour}, {Zorec}, {Zschocke}, {Zucker}, {Zurbach}, \&
  {Zwitter}}]{Gaia2018}
{Gaia Collaboration}, {Brown}, A.~G.~A., {Vallenari}, A., {et~al.} 2018, \aap,
  616, A1

\bibitem[{{Gravity Collaboration} {et~al.}(2017){Gravity Collaboration},
  {Abuter}, {Accardo}, {Amorim}, {Anugu}, {{\'A}vila}, {Azouaoui}, {Benisty},
  {Berger}, {Blind}, {Bonnet}, {Bourget}, {Brandner}, {Brast}, {Buron},
  {Burtscher}, {Cassaing}, {Chapron}, {Choquet}, {Cl{\'e}net}, {Collin},
  {Coud{\'e} Du Foresto}, {de Wit}, {de Zeeuw}, {Deen},
  {Delplancke-Str{\"o}bele}, {Dembet}, {Derie}, {Dexter}, {Duvert}, {Ebert},
  {Eckart}, {Eisenhauer}, {Esselborn}, {F{\'e}dou}, {Finger}, {Garcia}, {Garcia
  Dabo}, {Garcia Lopez}, {Gendron}, {Genzel}, {Gillessen}, {Gonte}, {Gordo},
  {Grould}, {Gr{\"o}zinger}, {Guieu}, {Haguenauer}, {Hans}, {Haubois}, {Haug},
  {Haussmann}, {Henning}, {Hippler}, {Horrobin}, {Huber}, {Hubert}, {Hubin},
  {Hummel}, {Jakob}, {Janssen}, {Jochum}, {Jocou}, {Kaufer}, {Kellner},
  {Kendrew}, {Kern}, {Kervella}, {Kiekebusch}, {Klein}, {Kok}, {Kolb}, {Kulas},
  {Lacour}, {Lapeyr{\`e}re}, {Lazareff}, {Le Bouquin}, {L{\`e}na}, {Lenzen},
  {L{\'e}v{\^e}que}, {Lippa}, {Magnard}, {Mehrgan}, {Mellein}, {M{\'e}rand},
  {Moreno-Ventas}, {Moulin}, {M{\"u}ller}, {M{\"u}ller}, {Neumann}, {Oberti},
  {Ott}, {Pallanca}, {Panduro}, {Pasquini}, {Paumard}, {Percheron}, {Perraut},
  {Perrin}, {Pfl{\"u}ger}, {Pfuhl}, {Phan Duc}, {Plewa}, {Popovic}, {Rabien},
  {Ram{\'{\i}}rez}, {Ramos}, {Rau}, {Riquelme}, {Rohloff}, {Rousset},
  {Sanchez-Bermudez}, {Scheithauer}, {Sch{\"o}ller}, {Schuhler}, {Spyromilio},
  {Straubmeier}, {Sturm}, {Suarez}, {Tristram}, {Ventura}, {Vincent},
  {Waisberg}, {Wank}, {Weber}, {Wieprecht}, {Wiest}, {Wiezorrek}, {Wittkowski},
  {Woillez}, {Wolff}, {Yazici}, {Ziegler}, \& {Zins}}]{GravityFirstLight}
{Gravity Collaboration}, {Abuter}, R., {Accardo}, M., {et~al.} 2017, \aap, 602,
  A94

\bibitem[{{Gravity Collaboration} {et~al.}(2018){Gravity Collaboration},
  {Abuter}, {Amorim}, {Baub{\"o}ck}, {Berger}, {Bonnet}, {Brand ner},
  {Cl{\'e}net}, {Coud{\'e} Du Foresto}, {de Zeeuw}, {Deen}, {Dexter}, {Duvert},
  {Eckart}, {Eisenhauer}, {F{\"o}rster Schreiber}, {Garcia}, {Gao}, {Gendron},
  {Genzel}, {Gillessen}, {Guajardo}, {Habibi}, {Haubois}, {Henning}, {Hippler},
  {Horrobin}, {Huber}, {Jim{\'e}nez-Rosales}, {Jocou}, {Kervella}, {Lacour},
  {Lapeyr{\`e}re}, {Lazareff}, {Le Bouquin}, {L{\'e}na}, {Lippa}, {Ott},
  {Panduro}, {Paumard}, {Perraut}, {Perrin}, {Pfuhl}, {Plewa}, {Rabien},
  {Rodr{\'\i}guez-Coira}, {Rousset}, {Sternberg}, {Straub}, {Straubmeier},
  {Sturm}, {Tacconi}, {Vincent}, {von Fellenberg}, {Waisberg}, {Widmann},
  {Wieprecht}, {Wiezorrek}, {Woillez}, \& {Yazici}}]{GravityCollaboration2018}
{Gravity Collaboration}, {Abuter}, R., {Amorim}, A., {et~al.} 2018, \aap, 618,
  L10

\bibitem[{{Gravity Collaboration} {et~al.}(2019){Gravity Collaboration},
  {Lacour}, {Nowak}, {Wang}, {Pfuhl}, {Eisenhauer}, {Abuter}, {Amorim},
  {Anugu}, {Benisty}, {Berger}, {Beust}, {Blind}, {Bonnefoy}, {Bonnet},
  {Bourget}, {Brandner}, {Buron}, {Collin}, {Charnay}, {Chapron}, {Cl{\'e}net},
  {Coud{\'e} Du Foresto}, {de Zeeuw}, {Deen}, {Dembet}, {Dexter}, {Duvert},
  {Eckart}, {F{\"o}rster Schreiber}, {F{\'e}dou}, {Garcia}, {Garcia Lopez},
  {Gao}, {Gendron}, {Genzel}, {Gillessen}, {Gordo}, {Greenbaum}, {Habibi},
  {Haubois}, {Hau{\ss}mann}, {Henning}, {Hippler}, {Horrobin}, {Hubert},
  {Jimenez Rosales}, {Jocou}, {Kendrew}, {Kervella}, {Kolb}, {Lagrange},
  {Lapeyr{\`e}re}, {Le Bouquin}, {L{\'e}na}, {Lippa}, {Lenzen}, {Maire},
  {Molli{\`e}re}, {Ott}, {Paumard}, {Perraut}, {Perrin}, {Pueyo}, {Rabien},
  {Ram{\'{\i}}rez}, {Rau}, {Rodr{\'{\i}}guez-Coira}, {Rousset},
  {Sanchez-Bermudez}, {Scheithauer}, {Schuhler}, {Straub}, {Straubmeier},
  {Sturm}, {Tacconi}, {Vincent}, {van Dishoeck}, {von Fellenberg}, {Wank},
  {Waisberg}, {Widmann}, {Wieprecht}, {Wiest}, {Wiezorrek}, {Woillez},
  {Yazici}, {Ziegler}, \& {Zins}}]{GravityLacour2019}
{Gravity Collaboration}, {Lacour}, S., {Nowak}, M., {et~al.} 2019, \aap, 623,
  L11

\bibitem[{{Gray} {et~al.}(2006){Gray}, {Corbally}, {Garrison}, {McFadden},
  {Bubar}, {McGahee}, {O'Donoghue}, \& {Knox}}]{Gray2006}
{Gray}, R.~O., {Corbally}, C.~J., {Garrison}, R.~F., {et~al.} 2006, \aj, 132,
  161

\bibitem[{{Hansen}(1971)}]{Hansen1971}
{Hansen}, J.~E. 1971, Journal of Atmospheric Sciences, 28, 1400

\bibitem[{{Hauschildt} {et~al.}(1999){Hauschildt}, {Allard}, \&
  {Baron}}]{Hauschildt1999}
{Hauschildt}, P.~H., {Allard}, F., \& {Baron}, E. 1999, \apj, 512, 377

\bibitem[{{Helled} {et~al.}(2006){Helled}, {Podolak}, \& {Kovetz}}]{Helled2006}
{Helled}, R., {Podolak}, M., \& {Kovetz}, A. 2006, \icarus, 185, 64

\bibitem[{{Helled} \& {Schubert}(2008)}]{Helled2008}
{Helled}, R. \& {Schubert}, G. 2008, \icarus, 198, 156

\bibitem[{{Helled} \& {Schubert}(2009)}]{Helled2009}
{Helled}, R. \& {Schubert}, G. 2009, \apj, 697, 1256

\bibitem[{{Helling} {et~al.}(2008){Helling}, {Dehn}, {Woitke}, \&
  {Hauschildt}}]{Helling2008}
{Helling}, C., {Dehn}, M., {Woitke}, P., \& {Hauschildt}, P.~H. 2008, \apjl,
  675, L105

\bibitem[{{Helling} \& {Woitke}(2006)}]{Helling2006}
{Helling}, C. \& {Woitke}, P. 2006, \aap, 455, 325

\bibitem[{{Helling} {et~al.}(2014){Helling}, {Woitke}, {Rimmer}, {Kamp}, {Thi},
  \& {Meijerink}}]{Helling2014}
{Helling}, C., {Woitke}, P., {Rimmer}, P.~B., {et~al.} 2014, Life, 4, 142

\bibitem[{{Holweger} {et~al.}(1997){Holweger}, {Hempel}, {van Thiel}, \&
  {Kaufer}}]{Holweger1997}
{Holweger}, H., {Hempel}, M., {van Thiel}, T., \& {Kaufer}, A. 1997, \aap, 320,
  L49

\bibitem[{{Kervella} {et~al.}(2019){Kervella}, {Arenou}, {Mignard}, \&
  {Th{\'e}venin}}]{Kervela2019}
{Kervella}, P., {Arenou}, F., {Mignard}, F., \& {Th{\'e}venin}, F. 2019, \aap,
  623, A72

\bibitem[{{Konopacky} {et~al.}(2013){Konopacky}, {Barman}, {Macintosh}, \&
  {Marois}}]{Konopacky2013}
{Konopacky}, Q.~M., {Barman}, T.~S., {Macintosh}, B.~A., \& {Marois}, C. 2013,
  Science, 339, 1398

\bibitem[{{Kreidberg} {et~al.}(2015){Kreidberg}, {Line}, {Bean}, {Stevenson},
  {D{\'e}sert}, {Madhusudhan}, {Fortney}, {Barstow}, {Henry}, {Williamson}, \&
  {Showman}}]{Kreidberg2016}
{Kreidberg}, L., {Line}, M.~R., {Bean}, J.~L., {et~al.} 2015, \apj, 814, 66

\bibitem[{{Lacour} {et~al.}(2019){Lacour}, {Dembet}, {Abuter}, {F{\'e}dou},
  {Perrin}, {Choquet}, {Pfuhl}, {Eisenhauer}, {Woillez}, {Cassaing},
  {Wieprecht}, {Ott}, {Wiezorrek}, {Tristram}, {Wolff}, {Ram{\'\i}rez},
  {Haubois}, {Perraut}, {Straubmeier}, {Brandner}, \& {Amorim}}]{Lacour2019}
{Lacour}, S., {Dembet}, R., {Abuter}, R., {et~al.} 2019, \aap, 624, A99

\bibitem[{{Lagrange} {et~al.}(2019{\natexlab{a}}){Lagrange}, {Boccaletti},
  {Langlois}, {Chauvin}, {Gratton}, {Beust}, {Desidera}, {Milli}, {Bonnefoy},
  {Cheetham}, {Feldt}, {Meyer}, {Vigan}, {Biller}, {Bonavita}, {Baudino},
  {Cantalloube}, {Cudel}, {Daemgen}, {Delorme}, {D'Orazi}, {Girard},
  {Fontanive}, {Hagelberg}, {Janson}, {Keppler}, {Koypitova}, {Galicher},
  {Lannier}, {Le Coroller}, {Ligi}, {Maire}, {Mesa}, {Messina}, {M{\"u}eller},
  {Peretti}, {Perrot}, {Rouan}, {Salter}, {Samland}, {Schmidt}, {Sissa},
  {Zurlo}, {Beuzit}, {Mouillet}, {Dominik}, {Henning}, {Lagadec}, {M{\'e}nard},
  {Schmid}, {Turatto}, {Udry}, {Bohn}, {Charnay}, {Gomez Gonzales}, {Gry},
  {Kenworthy}, {Kral}, {Mordasini}, {Moutou}, {van der Plas}, {Schlieder},
  {Abe}, {Antichi}, {Baruffolo}, {Baudoz}, {Baudrand}, {Blanchard}, {Bazzon},
  {Buey}, {Carbillet}, {Carle}, {Charton}, {Cascone}, {Claudi}, {Costille},
  {Deboulbe}, {De Caprio}, {Dohlen}, {Fantinel}, {Feautrier}, {Fusco}, {Gigan},
  {Giro}, {Gisler}, {Gluck}, {Hubin}, {Hugot}, {Jaquet}, {Kasper}, {Madec},
  {Magnard}, {Martinez}, {Maurel}, {Le Mignant}, {M{\"o}ller-Nilsson},
  {Llored}, {Moulin}, {Orign{\'e}}, {Pavlov}, {Perret}, {Petit}, {Pragt},
  {Szulagyi}, \& {Wildi}}]{Lagrange2019}
{Lagrange}, A.-M., {Boccaletti}, A., {Langlois}, M., {et~al.}
  2019{\natexlab{a}}, \aap, 621, L8

\bibitem[{{Lagrange} {et~al.}(2012){Lagrange}, {De Bondt}, {Meunier},
  {Sterzik}, {Beust}, \& {Galland}}]{Lagrange2012}
{Lagrange}, A.~M., {De Bondt}, K., {Meunier}, N., {et~al.} 2012, \aap, 542, A18

\bibitem[{{Lagrange} {et~al.}(2018){Lagrange}, {Keppler}, {Meunier}, {Lannier},
  {Beust}, {Milli}, {Bonnavita}, {Bonnefoy}, {Borgniet}, {Chauvin}, {Delorme},
  {Galland}, {Iglesias}, {Kiefer}, {Messina}, {Vidal-Madjar}, \&
  {Wilson}}]{Lagrange2018}
{Lagrange}, A.-M., {Keppler}, M., {Meunier}, N., {et~al.} 2018, \aap, 612, A108

\bibitem[{{Lagrange} {et~al.}(2019{\natexlab{b}}){Lagrange}, {Meunier},
  {Rubini}, {Keppler}, {Galland}, {Chapellier}, {Michel}, {Balona}, {Beust},
  {Guillot}, {Grandjean}, {Borgniet}, {M{\'e}karnia}, {Wilson}, {Kiefer},
  {Bonnefoy}, {Lillo-Box}, {Pantoja}, {Jones}, {Iglesias}, {Rodet}, {Diaz},
  {Zapata}, {Abe}, \& {Schmider}}]{Lagrange2019b}
{Lagrange}, A.~M., {Meunier}, N., {Rubini}, P., {et~al.} 2019{\natexlab{b}},
  Nature Astronomy, 421

\bibitem[{{Lanz} {et~al.}(1995){Lanz}, {Heap}, \& {Hubeny}}]{Lanz1995}
{Lanz}, T., {Heap}, S.~R., \& {Hubeny}, I. 1995, \apjl, 447, L41

\bibitem[{{Lapeyrere} {et~al.}(2014){Lapeyrere}, {Kervella}, {Lacour},
  {Azouaoui}, {Garcia-Dabo}, {Perrin}, {Eisenhauer}, {Perraut}, {Straubmeier},
  {Amorim}, \& {Brandner}}]{Lapeyrere2014}
{Lapeyrere}, V., {Kervella}, P., {Lacour}, S., {et~al.} 2014, in \procspie,
  Vol. 9146, Optical and Infrared Interferometry IV, 91462D

\bibitem[{{Lavie} {et~al.}(2017){Lavie}, {Mendon{\c c}a}, {Mordasini}, {Malik},
  {Bonnefoy}, {Demory}, {Oreshenko}, {Grimm}, {Ehrenreich}, \&
  {Heng}}]{Lavie2017}
{Lavie}, B., {Mendon{\c c}a}, J.~M., {Mordasini}, C., {et~al.} 2017, \aj, 154,
  91

\bibitem[{{Line} {et~al.}(2017){Line}, {Marley}, {Liu}, {Burningham}, {Morley},
  {Hinkel}, {Teske}, {Fortney}, {Freedman}, \& {Lupu}}]{Line2017}
{Line}, M.~R., {Marley}, M.~S., {Liu}, M.~C., {et~al.} 2017, \apj, 848, 83

\bibitem[{{Line} {et~al.}(2015){Line}, {Teske}, {Burningham}, {Fortney}, \&
  {Marley}}]{Line2015}
{Line}, M.~R., {Teske}, J., {Burningham}, B., {Fortney}, J.~J., \& {Marley},
  M.~S. 2015, \apj, 807, 183

\bibitem[{{Lissauer} \& Stevenson(2007)}]{GiantPlanetFormation}
{Lissauer}, J.~J. \& Stevenson, D.~J. 2007, in Protostars and Planets V, ed.
  B.~{Reipurth}, D.~{Jewitt}, \& K.~{Keil}, 591--606

\bibitem[{{Madhusudhan} {et~al.}(2014){Madhusudhan}, {Amin}, \&
  {Kennedy}}]{Madhusudhan2014}
{Madhusudhan}, N., {Amin}, M.~A., \& {Kennedy}, G.~M. 2014, \apjl, 794, L12

\bibitem[{{Madhusudhan} {et~al.}(2017){Madhusudhan}, {Bitsch}, {Johansen}, \&
  {Eriksson}}]{Madhusudhan2017}
{Madhusudhan}, N., {Bitsch}, B., {Johansen}, A., \& {Eriksson}, L. 2017,
  \mnras, 469, 4102

\bibitem[{{Madhusudhan} {et~al.}(2011){Madhusudhan}, {Harrington}, {Stevenson},
  {Nymeyer}, {Campo}, {Wheatley}, {Deming}, {Blecic}, {Hardy}, {Lust},
  {Anderson}, {Collier-Cameron}, {Britt}, {Bowman}, {Hebb}, {Hellier},
  {Maxted}, {Pollacco}, \& {West}}]{Madhusudhan2011}
{Madhusudhan}, N., {Harrington}, J., {Stevenson}, K.~B., {et~al.} 2011, \nat,
  469, 64

\bibitem[{{Madhusudhan} \& {Seager}(2009)}]{Madhusudhan2009}
{Madhusudhan}, N. \& {Seager}, S. 2009, \apj, 707, 24

\bibitem[{{Marboeuf} {et~al.}(2014{\natexlab{a}}){Marboeuf}, {Thiabaud},
  {Alibert}, {Cabral}, \& {Benz}}]{Marboeuf2014b}
{Marboeuf}, U., {Thiabaud}, A., {Alibert}, Y., {Cabral}, N., \& {Benz}, W.
  2014{\natexlab{a}}, \aap, 570, A36

\bibitem[{{Marboeuf} {et~al.}(2014{\natexlab{b}}){Marboeuf}, {Thiabaud},
  {Alibert}, {Cabral}, \& {Benz}}]{Marboeuf2014}
{Marboeuf}, U., {Thiabaud}, A., {Alibert}, Y., {Cabral}, N., \& {Benz}, W.
  2014{\natexlab{b}}, \aap, 570, A35

\bibitem[{{Marley} {et~al.}(2007){Marley}, {Fortney}, {Hubickyj},
  {Bodenheimer}, \& {Lissauer}}]{Marley2007}
{Marley}, M.~S., {Fortney}, J.~J., {Hubickyj}, O., {Bodenheimer}, P., \&
  {Lissauer}, J.~J. 2007, \apj, 655, 541

\bibitem[{{Marley} {et~al.}(2012){Marley}, {Saumon}, {Cushing}, {Ackerman},
  {Fortney}, \& {Freedman}}]{Marley2012}
{Marley}, M.~S., {Saumon}, D., {Cushing}, M., {et~al.} 2012, \apj, 754, 135

\bibitem[{{Millar-Blanchaer} {et~al.}(2015){Millar-Blanchaer}, {Graham},
  {Pueyo}, {Kalas}, {Dawson}, {Wang}, {Perrin}, {moon}, {Macintosh}, {Ammons},
  {Barman}, {Cardwell}, {Chen}, {Chiang}, {Chilcote}, {Cotten}, {De Rosa},
  {Draper}, {Dunn}, {Duch{\^e}ne}, {Esposito}, {Fitzgerald}, {Follette},
  {Goodsell}, {Greenbaum}, {Hartung}, {Hibon}, {Hinkley}, {Ingraham},
  {Jensen-Clem}, {Konopacky}, {Larkin}, {Long}, {Maire}, {Marchis}, {Marley},
  {Marois}, {Morzinski}, {Nielsen}, {Palmer}, {Oppenheimer}, {Poyneer},
  {Rajan}, {Rantakyr{\"o}}, {Ruffio}, {Sadakuni}, {Saddlemyer}, {Schneider},
  {Sivaramakrishnan}, {Soummer}, {Thomas}, {Vasisht}, {Vega}, {Wallace},
  {Ward-Duong}, {Wiktorowicz}, \& {Wolff}}]{MillarBlanchaer2015}
{Millar-Blanchaer}, M.~A., {Graham}, J.~R., {Pueyo}, L., {et~al.} 2015, \apj,
  811, 18

\bibitem[{{Molli{\`e}re} \& {Mordasini}(2012)}]{Molliere2012}
{Molli{\`e}re}, P. \& {Mordasini}, C. 2012, \aap, 547, A105

\bibitem[{{Molli{\`e}re} \& {Snellen}(2019)}]{Molliere2019}
{Molli{\`e}re}, P. \& {Snellen}, I.~A.~G. 2019, \aap, 622, A139

\bibitem[{{Molli{\`e}re} {et~al.}(2017){Molli{\`e}re}, {van Boekel}, {Bouwman},
  {Henning}, {Lagage}, \& {Min}}]{Molliere2017}
{Molli{\`e}re}, P., {van Boekel}, R., {Bouwman}, J., {et~al.} 2017, \aap, 600,
  A10

\bibitem[{{Molli{\`e}re} {et~al.}(2015){Molli{\`e}re}, {van Boekel},
  {Dullemond}, {Henning}, \& {Mordasini}}]{Molliere2015}
{Molli{\`e}re}, P., {van Boekel}, R., {Dullemond}, C., {Henning}, T., \&
  {Mordasini}, C. 2015, \apj, 813, 47

\bibitem[{{Molli{\`e}re} {et~al.}(2019){Molli{\`e}re}, {Wardenier}, {van
  Boekel}, {Henning}, {Molaverdikhani}, \& {Snellen}}]{Molliere2019b}
{Molli{\`e}re}, P., {Wardenier}, J.~P., {van Boekel}, R., {et~al.} 2019, \aap,
  627, A67

\bibitem[{{Mordasini} {et~al.}(2017){Mordasini}, {Marleau}, \&
  {Molli{\`e}re}}]{Mordasini2017}
{Mordasini}, C., {Marleau}, G.-D., \& {Molli{\`e}re}, P. 2017, \aap, 608, A72

\bibitem[{{Mordasini} {et~al.}(2016){Mordasini}, {van Boekel}, {Molli{\`e}re},
  {Henning}, \& {Benneke}}]{Mordasini2016}
{Mordasini}, C., {van Boekel}, R., {Molli{\`e}re}, P., {Henning}, T., \&
  {Benneke}, B. 2016, \apj, 832, 41

\bibitem[{{Morley} {et~al.}(2014){Morley}, {Marley}, {Fortney}, {Lupu},
  {Saumon}, {Greene}, \& {Lodders}}]{Morley2014}
{Morley}, C.~V., {Marley}, M.~S., {Fortney}, J.~J., {et~al.} 2014, \apj, 787,
  78

\bibitem[{{Morzinski} {et~al.}(2015){Morzinski}, {Males}, {Skemer}, {Close},
  {Hinz}, {Rodigas}, {Puglisi}, {Esposito}, {Riccardi}, {Pinna}, {Xompero},
  {Briguglio}, {Bailey}, {Follette}, {Kopon}, {Weinberger}, \&
  {Wu}}]{Morzinski2015}
{Morzinski}, K.~M., {Males}, J.~R., {Skemer}, A.~J., {et~al.} 2015, \apj, 815,
  108

\bibitem[{{Nielsen} {et~al.}(2014){Nielsen}, {Liu}, {Wahhaj}, {Biller},
  {Hayward}, {Males}, {Close}, {Morzinski}, {Skemer}, {Kuchner}, {Rodigas},
  {Hinz}, {Chun}, {Ftaclas}, \& {Toomey}}]{Nielsen2014}
{Nielsen}, E.~L., {Liu}, M.~C., {Wahhaj}, Z., {et~al.} 2014, \apj, 794, 158

\bibitem[{{Notsu} {et~al.}(2019){Notsu}, {Akiyama}, {Booth}, {Nomura}, {Walsh},
  {Hirota}, {Honda}, {Tsukagoshi}, \& {Millar}}]{Notsu2019}
{Notsu}, S., {Akiyama}, E., {Booth}, A., {et~al.} 2019, \apj, 875, 96

\bibitem[{Nowak(2019)}]{NowakPhd}
Nowak, M. 2019, PhD thesis, Université PSL

\bibitem[{{{\"O}berg} \& {Bergin}(2016)}]{Oberg2016}
{{\"O}berg}, K.~I. \& {Bergin}, E.~A. 2016, \apjl, 831, L19

\bibitem[{{{\"O}berg} {et~al.}(2011){{\"O}berg}, {Murray-Clay}, \&
  {Bergin}}]{Oberg2011}
{{\"O}berg}, K.~I., {Murray-Clay}, R., \& {Bergin}, E.~A. 2011, \apjl, 743, L16

\bibitem[{{Owen} {et~al.}(1999){Owen}, {Mahaffy}, {Niemann}, {Atreya},
  {Donahue}, {Bar-Nun}, \& {de Pater}}]{Owen1999}
{Owen}, T., {Mahaffy}, P., {Niemann}, H.~B., {et~al.} 1999, \nat, 402, 269

\bibitem[{{Pascucci} {et~al.}(2016){Pascucci}, {Testi}, {Herczeg}, {Long},
  {Manara}, {Hendler}, {Mulders}, {Krijt}, {Ciesla}, {Henning}, {Mohanty},
  {Drabek-Maunder}, {Apai}, {Sz{\H{u}}cs}, {Sacco}, \&
  {Olofsson}}]{Pascucci2016}
{Pascucci}, I., {Testi}, L., {Herczeg}, G.~J., {et~al.} 2016, \apj, 831, 125

\bibitem[{{Qi} {et~al.}(2015){Qi}, {{\"O}berg}, {Andrews}, {Wilner}, {Bergin},
  {Hughes}, {Hogherheijde}, \& {D'Alessio}}]{Qi2015}
{Qi}, C., {{\"O}berg}, K.~I., {Andrews}, S.~M., {et~al.} 2015, \apj, 813, 128

\bibitem[{{Samland} {et~al.}(2017){Samland}, {Molli{\`e}re}, {Bonnefoy},
  {Maire}, {Cantalloube}, {Cheetham}, {Mesa}, {Gratton}, {Biller}, {Wahhaj},
  {Bouwman}, {Brandner}, {Melnick}, {Carson}, {Janson}, {Henning}, {Homeier},
  {Mordasini}, {Langlois}, {Quanz}, {van Boekel}, {Zurlo}, {Schlieder},
  {Avenhaus}, {Beuzit}, {Boccaletti}, {Bonavita}, {Chauvin}, {Claudi}, {Cudel},
  {Desidera}, {Feldt}, {Fusco}, {Galicher}, {Kopytova}, {Lagrange}, {Le
  Coroller}, {Martinez}, {Moeller-Nilsson}, {Mouillet}, {Mugnier}, {Perrot},
  {Sevin}, {Sissa}, {Vigan}, \& {Weber}}]{Samland2017}
{Samland}, M., {Molli{\`e}re}, P., {Bonnefoy}, M., {et~al.} 2017, \aap, 603,
  A57

\bibitem[{{Snellen} {et~al.}(2014){Snellen}, {Brandl}, {de Kok}, {Brogi},
  {Birkby}, \& {Schwarz}}]{Snellen2014}
{Snellen}, I.~A.~G., {Brandl}, B.~R., {de Kok}, R.~J., {et~al.} 2014, \nat,
  509, 63

\bibitem[{{Snellen} \& {Brown}(2018)}]{Snellen2018}
{Snellen}, I.~A.~G. \& {Brown}, A.~G.~A. 2018, Nature Astronomy, 2, 883

\bibitem[{{Spiegel} {et~al.}(2011){Spiegel}, {Burrows}, \&
  {Milsom}}]{Spiegel2011}
{Spiegel}, D.~S., {Burrows}, A., \& {Milsom}, J.~A. 2011, \apj, 727, 57

\bibitem[{{Tatulli} {et~al.}(2007){Tatulli}, {Millour}, {Chelli}, {Duvert},
  {Acke}, {Hernandez Utrera}, {Hofmann}, {Kraus}, {Malbet}, {M{\`e}ge},
  {Petrov}, {Vannier}, {Zins}, {Antonelli}, {Beckmann}, {Bresson}, {Dugu{\'e}},
  {Gennari}, {Gl{\"u}ck}, {Kern}, {Lagarde}, {Le Coarer}, {Lisi}, {Perraut},
  {Puget}, {Rantakyr{\"o}}, {Robbe-Dubois}, {Roussel}, {Weigelt}, {Accardo},
  {Agabi}, {Altariba}, {Arezki}, {Aristidi}, {Baffa}, {Behrend}, {Bl{\"o}cker},
  {Bonhomme}, {Busoni}, {Cassaing}, {Clausse}, {Colin}, {Connot},
  {Delboulb{\'e}}, {Domiciano de Souza}, {Driebe}, {Feautrier}, {Ferruzzi},
  {Forveille}, {Fossat}, {Foy}, {Fraix-Burnet}, {Gallardo}, {Giani}, {Gil},
  {Glentzlin}, {Heiden}, {Heininger}, {Kamm}, {Kiekebusch}, {Le Contel}, {Le
  Contel}, {Lesourd}, {Lopez}, {Lopez}, {Magnard}, {Marconi}, {Mars},
  {Martinot-Lagarde}, {Mathias}, {Monin}, {Mouillet}, {Mourard}, {Nussbaum},
  {Ohnaka}, {Pacheco}, {Perrier}, {Rabbia}, {Rebattu}, {Reynaud}, {Richichi},
  {Robini}, {Sacchettini}, {Schertl}, {Sch{\"o}ller}, {Solscheid}, {Spang},
  {Stee}, {Stefanini}, {Tallon}, {Tallon-Bosc}, {Tasso}, {Testi}, {Vakili},
  {von der L{\"u}he}, {Valtier}, \& {Ventura}}]{Tatulli2007}
{Tatulli}, E., {Millour}, F., {Chelli}, A., {et~al.} 2007, \aap, 464, 29

\bibitem[{{Th{\'e}bault} \& {Beust}(2001)}]{Thebault2001}
{Th{\'e}bault}, P. \& {Beust}, H. 2001, \aap, 376, 621

\bibitem[{{Thiabaud} {et~al.}(2014){Thiabaud}, {Marboeuf}, {Alibert}, {Cabral},
  {Leya}, \& {Mezger}}]{Thiabaud2014}
{Thiabaud}, A., {Marboeuf}, U., {Alibert}, Y., {et~al.} 2014, \aap, 562, A27

\bibitem[{{van der Bliek} {et~al.}(1996){van der Bliek}, {Manfroid}, \&
  {Bouchet}}]{VanDerBliek1996}
{van der Bliek}, N.~S., {Manfroid}, J., \& {Bouchet}, P. 1996, \aaps, 119, 547

\bibitem[{{van Leeuwen}(2007)}]{VanLeeuwen2007}
{van Leeuwen}, F. 2007, \aap, 474, 653

\bibitem[{{Vousden} {et~al.}(2016){Vousden}, {Farr}, \& {Mandel}}]{Vousden2016}
{Vousden}, W.~D., {Farr}, W.~M., \& {Mandel}, I. 2016, \mnras, 455, 1919

\bibitem[{{Walsh} {et~al.}(2015){Walsh}, {Nomura}, \& {van
  Dishoeck}}]{Walsh2015}
{Walsh}, C., {Nomura}, H., \& {van Dishoeck}, E. 2015, \aap, 582, A88

\bibitem[{{Wang} {et~al.}(2018){Wang}, {Graham}, {Dawson}, {Fabrycky}, {De
  Rosa}, {Pueyo}, {Konopacky}, {Macintosh}, {Marois}, {Chiang}, {Ammons},
  {Arriaga}, {Bailey}, {Barman}, {Bulger}, {Chilcote}, {Cotten}, {Doyon},
  {Duch{\^e}ne}, {Esposito}, {Fitzgerald}, {Follette}, {Gerard}, {Goodsell},
  {Greenbaum}, {Hibon}, {Hung}, {Ingraham}, {Kalas}, {Larkin}, {Maire},
  {Marchis}, {Marley}, {Metchev}, {Millar-Blanchaer}, {Nielsen}, {Oppenheimer},
  {Palmer}, {Patience}, {Perrin}, {Poyneer}, {Rajan}, {Rameau},
  {Rantakyr{\"o}}, {Ruffio}, {Savransky}, {Schneider}, {Sivaramakrishnan},
  {Song}, {Soummer}, {Thomas}, {Wallace}, {Ward-Duong}, {Wiktorowicz}, \&
  {Wolff}}]{Wang2018}
{Wang}, J.~J., {Graham}, J.~R., {Dawson}, R., {et~al.} 2018, \aj, 156, 192

\bibitem[{{Wang} {et~al.}(2016){Wang}, {Graham}, {Pueyo}, {Kalas},
  {Millar-Blanchaer}, {Ruffio}, {De Rosa}, {Ammons}, {Arriaga}, {Bailey},
  {Barman}, {Bulger}, {Burrows}, {Cardwell}, {Chen}, {Chilcote}, {Cotten},
  {Fitzgerald}, {Follette}, {Doyon}, {Duch{\^e}ne}, {Greenbaum}, {Hibon},
  {Hung}, {Ingraham}, {Konopacky}, {Larkin}, {Macintosh}, {Maire}, {Marchis},
  {Marley}, {Marois}, {Metchev}, {Nielsen}, {Oppenheimer}, {Palmer}, {Patel},
  {Patience}, {Perrin}, {Poyneer}, {Rajan}, {Rameau}, {Rantakyr{\"o}},
  {Savransky}, {Sivaramakrishnan}, {Song}, {Soummer}, {Thomas}, {Vasisht},
  {Vega}, {Wallace}, {Ward-Duong}, {Wiktorowicz}, \& {Wolff}}]{Wang2016}
{Wang}, J.~J., {Graham}, J.~R., {Pueyo}, L., {et~al.} 2016, \aj, 152, 97

\bibitem[{{Woitke} \& {Helling}(2003)}]{Woitke2003}
{Woitke}, P. \& {Helling}, C. 2003, \aap, 399, 297

\bibitem[{{Zahnle} \& {Marley}(2014)}]{Zahnle2014}
{Zahnle}, K.~J. \& {Marley}, M.~S. 2014, \apj, 797, 41

\bibitem[{{Zalesky} {et~al.}(2019){Zalesky}, {Line}, {Schneider}, \&
  {Patience}}]{Zalesky2019}
{Zalesky}, J.~A., {Line}, M.~R., {Schneider}, A.~C., \& {Patience}, J. 2019,
  \apj, 877, 24

\bibitem[{{Zieba} {et~al.}(2019){Zieba}, {Zwintz}, {Kenworthy}, \&
  {Kennedy}}]{Zieba2019}
{Zieba}, S., {Zwintz}, K., {Kenworthy}, M.~A., \& {Kennedy}, G.~M. 2019, \aap,
  625, L13

\end{thebibliography}

\end{document}